\documentclass[a4paper,12pt]{article}
\usepackage[a4paper, left=2.25cm, right=2.25cm, top=2.25cm, bottom=2.25cm]{geometry}
\usepackage{pgfplots}
\pgfplotsset{compat=1.18}
\usepackage{amsmath,xfrac}
\usepackage{bm}
\usepackage{bbm}
\usepackage{amsthm}
\usepackage{dsfont}
\usepackage[utf8]{inputenc} 
\usepackage[english]{babel} 
\usepackage{xcolor} 
\usepackage{tikz}
\usepackage{adjustbox} 
\usepackage[onehalfspacing]{setspace}
\usepackage{graphicx}
\usepackage{xcolor, lipsum} 
\usepackage{caption}
\usepackage{mathcomp}
\usepackage{verbatim} 
\usepackage{enumitem}
\usepackage[overload]{abraces}
\usepackage{array}
\usepackage{pict2e} 
\usepackage{amssymb}
\usepackage{epstopdf}
\usepackage{pbox}
\usepackage{mathrsfs}
\usepackage{graphicx,wrapfig,lipsum}
\usepackage{multirow}
\usepackage{subcaption}
\usepackage{colortbl}
\usepackage{mathtools}
\usepackage{multicol}
\usepackage{wasysym}
\usepackage{makecell}
\usepackage{eurosym}
\usepackage{booktabs}

\usepackage{pdflscape}
\usepackage{afterpage}
\usepackage{booktabs,arydshln}
\usepackage{capt-of}
\usepackage{rotating}
\usepackage{longtable}
\usepackage{pdfpages}
\usepackage[hidelinks,
colorlinks=true,
urlcolor=blue,
linkcolor=blue,
citecolor=blue,
hyperfootnotes=true]{hyperref}
\usepackage{natbib}
\usepackage{mathtools}

\usetikzlibrary{decorations.pathreplacing}
\usetikzlibrary{positioning, shapes,arrows.meta}

\renewcommand{\ref}[1]{\hyperref[#1]{\ref{#1}}}

\newtheorem{proposition}{Proposition}
\newtheorem{lemma}{Lemma}
\newtheorem{corollary}{Corollary}

\newtheorem{assumption}{Assumption}

\renewcommand{\thefootnote}{\fnsymbol{footnote}}
\begin{document}
	\onehalfspacing
	\begin{center}
		{\LARGE \textbf{{\LARGE \textbf{Memory Manipulation}}}
		}\\
		\vspace{5mm}
		\textit{Benson Tsz Kin Leung\footnote{Hong Kong Baptist University. Email: \href{mailto:btkleung@hkbu.edu.hk}{\texttt{btkleung@hkbu.edu.hk}}, \href{mailto:ltkbenson@gmail.com}{\texttt{ltkbenson@gmail.com}}. }.}\\
		\textit{\today}\\

	\end{center}
	\vspace{4mm} 
	\begin{abstract}
		We study persuasion that works through memory: a manipulator alters the probabilities with which a Bayesian decision-maker recalls the signals he has received. The decision-maker corrects for the elevated or reduced probabilities of recalling different signals, and this correction disciplines what manipulation can achieve. Manipulation cannot move the average posterior. It backfires on decision-makers with few memories, whose sparse recall is evidence against the favored state, yet it is effective on those with many, who can be led to arbitrarily confident beliefs in the wrong state. It thickens the right tail of the belief distribution and, under Poisson memory, lowers its hazard rate. In a monopoly pricing application, the seller manipulates memory retrieval at high prices, and manipulation in turn raises the optimal price. In a narrative persuasion application, memory manipulation reduces information provision and creates echo chambers.

		\vspace{5mm}


	\end{abstract}
	
	\renewcommand{\thefootnote}{\arabic{footnote}}
	\setcounter{footnote}{0}
	\newpage
	\section{Introduction}
	
	Memory plays a key role in our belief-formation and decision-making (\cite{bordalo2020belief,bordalo2022overreaction,bordalo2023imagining,bordalo2023memory,enke2024associative}). However, given the volatile and imprecise nature of memory, there is room for memory manipulation as a tool of persuasion (\cite{loftus1979malleability,garry1996imagination,kassin1996social,liv2020deep}). This paper develops a model of how manipulation of memory retrieval probabilities affects belief formation, characterizes when such manipulation is effective and when it backfires, and applies the framework to two settings: monopoly pricing and advertising, and narrative persuasion in political communication.
	
	While there are numerous ways to manipulate memory, this paper, as a first and important step in the inquiry, focuses on manipulations of the retrieval probability of memories. The phenomena in which there are varying retrieval probabilities based on the nature of the memories and cues are well-documented. In Economics, \cite{bordalo2023memory} and \cite{enke2024associative} found that subjects recall similar past signals upon receiving a new piece of information, affecting probabilistic judgement, leading to overreaction and extreme market prices. In \cite{enke2024associative}, the overreaction arises when memory holds signals that the new information cues. \cite{graeber2022stories} shows that stories lead to more persistently higher retrieval probabilities than statistics. In Psychology, the classical availability heuristic, as presented in the seminal paper (\cite{tversky1973availability}), illustrates how the varying ease of memory retrieval distorts individuals' perception of frequencies and probabilities. In Marketing, nostalgic marketing campaigns trigger retrieval of brand-related memories, strengthening brand awareness (\cite{zhao2014remembering,fetterman2025topics,thoma2025nostalgic}), and are more effective for consumers with stronger memories of the brand (\cite{muehling2014exploring,thoma2025nostalgic}).

	Given this evidence, this paper analyzes how manipulation of memory retrieval affects the persuadee's belief and the persuader's objective. In the general model (Section~\ref{sec:model}), a decision-maker (DM) probabilistically retrieves signals from memory and forms beliefs using Bayes' rule. In a benchmark, the DM retrieves each signal with equal probability. Memory manipulation involves favoring retrieval of some signals relative to others. Consider, for example, the goal of persuading an individual that immigration is harmful to the local economy. We focus on manipulation of memory retrieval so that the individual recalls signals against immigration with higher probability and signals in favor of immigration with lower probability. We show that such manipulation can backfire: if the individual does not recall any anti-immigration signals despite the increased retrieval probability, he rationally infers that his memory contains few such signals and becomes \textit{more} skeptical that immigration is harmful. This indirect ``ease-of-retrieval'' effect (\cite{schwarz1991ease}) can dominate the direct effect, depending on the individual's memory size, the information structure, and other model parameters.

	As the DM is Bayesian, his average posterior belief is invariant to any memory manipulation: this martingale property is the analogue of the Bayesian feasibility constraint in \cite{kamenica2011bayesian}. The content of our results therefore lies in how manipulation reshapes the \emph{distribution} of posteriors. Our first result presents the effect of manipulation based on memory size, and thus sheds light on the heterogeneous impact on individuals. Manipulation backfires for individuals with small memories, for whom the ease-of-retrieval correction dominates, and is effective for all individuals with large enough memories, for whom the recalled evidence dominates. This pattern is in line with evidence that experts judge by what they recall whereas laymen judge by how difficult recall felt \citep{ofir2000ease,weingarten2018does}, and that the effect of advertising is concentrated among older consumers, plausibly those with more memories of the product \citep{lewis2014advertising}. When manipulation is aggressive enough that the recalled evidence favors the wrong state on average, an individual with a large memory becomes nearly certain of the wrong state---confidently holding a false belief precisely because he has accumulated many, selectively recalled, memories.

	Our second result concerns the tails of the belief distribution: whatever the signal structure, manipulation makes extreme beliefs in the favored direction more likely and extreme beliefs against it less likely, and under Poisson memory it also lowers the hazard rate of the right tail, which is useful when we apply our model in economic problems like monopoly pricing, optimal auctions, and screening. Our third result asks how signal informativeness affects the effectiveness of manipulation. Informative signals cut both ways: each recalled or suppressed signal moves beliefs further, but Bayesian learning also reaches extreme beliefs more easily on its own. Manipulation has no traction when signals are uninformative; under perfectly revealing signals it retains a gain only through the individuals who recall nothing, and that gain vanishes when memories are abundant. 

	These results have broad implications for understanding persuasion in settings where the persuader influences not what information the receiver \textit{has}, but what information he \textit{recalls}. We develop and illustrate these implications in two applied settings.

	The first application (Section~\ref{sec:app3}) studies a monopolist selling a product of uncertain match value to a consumer who has accumulated memories about the product category. The monopolist jointly chooses a price and a memory-manipulation strategy (e.g., through advertising or framing). The heavier right tail result implies that for any sufficiently high price the monopolist strictly prefers to manipulate. Moreover, when the marginal cost, and thus the price, is sufficiently high, the availability of manipulation raises the optimal price. It is driven by the hazard-rate comparison: manipulation amplifies recall of favorable signals, which lowers demand elasticity in the right tail relative to the no-manipulation benchmark. The result is consistent with evidence that emotional rather than informational content in television ads raises sales more for high-price than for low-price cars \citep{guitart2021impact}, and with the narrative storylines of luxury fashion advertising \citep{kim2016narrative}.

	The second application (Section~\ref{sec:app1}) embeds the model in a political persuasion game where a politician chooses between disclosing verifiable evidence and manipulating a voter's memory recall. This captures a trade-off between two styles of persuasion: verifiable evidence such as statistics is more informative, whereas stories are better remembered and bring to mind the audience's own related experiences.\footnote{\cite{graeber2022stories} find that statistics move beliefs more than stories on impact, but a day later stories are recalled far more often and retain more of their effect. \cite{hamby2017reflecting} found that the effect of retrospective reflection, i.e., the recall of memories related to the self and others, is significant and positively related to persuasion-related outcomes. Similarly, \cite{green2004transportation} finds that readers with relevant personal experience, such as having gay friends or relatives when reading a story about a gay man, are more transported into the story, and that more transported readers hold more story-consistent beliefs.} For a demanding voter, the equilibrium is a cutoff in the strength of the politician's evidence: weak evidence is withheld in favor of manipulation and strong evidence is disclosed, and once the voter is demanding enough no evidence is disclosed at all and the politician relies solely on memory manipulation. Thus, the availability of memory manipulation reduces information disclosure. In an extension in which the voter can choose whether to listen to the politician, those who listen are the ones whose own impressions already favor the politician: his audience is an echo chamber.

	\paragraph{Related Literature.}
	This paper contributes to the theoretical economics literature on memory-based belief formation (\cite{bordalo2022overreaction,bordalo2023memory,bordalo2020belief,bordalo2023imagining,leung2024simple}). It departs from these studies in two respects. First, rather than focusing solely on the individual process of memory retrieval and belief formation, we analyze how a strategic agent can manipulate the retrieval process to influence the decision-maker's beliefs. Second, we embed the manipulation in two applied settings, monopoly pricing and political communication, and characterize the resulting equilibria.
	
	The paper also contributes to the literature on communication and persuasion, building on the seminal work on cheap talk and Bayesian persuasion (\cite{crawford1982strategic,kamenica2011bayesian} and many others). A recent strand of this literature explores how persuaders can influence beliefs by manipulating the statistical models that receivers use to make inferences from data (e.g., \cite{eliaz2020model}). Our paper complements this work by studying a distinct channel: manipulation of memory retrieval, rather than manipulation of models or information structures.
	
	Lastly, the paper contributes to the fast-growing economic literature on narratives. \cite{shiller2017narrative} and \cite{andre2024narratives} argue that narratives affect individuals' expectations of macroeconomic indicators and government policies. \cite{roos2024narratives} offers a comprehensive review, documenting rapid growth from roughly five papers per year in the late 1990s to over 60 per year by 2021. The closest papers are \cite{eliaz2020model,schwartzstein2021using,barron2024narrative}, which study narrative persuasion theoretically and experimentally. These papers define narratives as (causal) models that receivers use to interpret data. Our paper complements this work by formalizing narratives as tools that manipulate memory retrieval rather than statistical models, and by deriving equilibrium implications in a political-communication setting.
	
	\section{A Model of Memory Recall and Manipulation}\label{sec:model}
	
	To fix the idea, suppose a voter is forming his belief on whether immigration is beneficial or hurtful to the local economy. The voter first tries to retrieve information from his memory, such as articles he has seen, conversations he has had, or daily observations of local businesses. Memory retrieval is imperfect so the voter retrieves some of his memory but might not retreive all. Lastly, the voter forms his belief with the retrieved memories and takes into account the possibility that he might fail to retrieve some informative memories.
	
	Formally, denote $\omega\in\lbrace 0,1\rbrace$ the state of the world, which is a priori equally likely.\footnote{Note that a non-uniform prior can easily be modeled as existing memories of the individual.} The model is illustrated in Figure~\ref{fig:timeline}. First, nature draws the true state of the world. Second, the DM receives $M$ signals drawn i.i.d.~from a state-dependent distribution $f_\omega$, where $M$ is independently drawn from a state-independent discrete distribution $g$. We assume that no single signal perfectly rules out any state, such that $\inf_{s,\omega\neq\omega'}\frac{f_\omega(s)}{f_{\omega'}(s)}$ is bounded away from $0$. We refer to the collection of the $M$ signals as the DM's memory and denote it as $\mathscr{M}$. Now, in a standard Bayesian model, his posterior belief that state 1 is true will be:
	\begin{equation*}
		p_1=\frac{\prod_{s\in \mathscr{M}}f_1(s)}{\prod_{s\in \mathscr{M}}f_1(s)+\prod_{s\in \mathscr{M}}f_0(s)}.
	\end{equation*}
	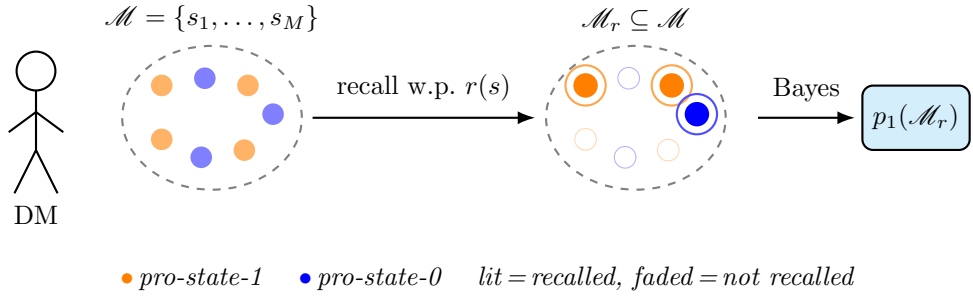
\begin{figure}
		\centering
		\begin{tikzpicture}[scale=0.95, every node/.style={font=\small}]
			\begin{scope}[shift={(-5.4,0)}]
				\draw[thick] (0,0.65) circle (0.27);
				\draw[thick] (0,0.38)--(0,-0.45);
				\draw[thick] (0,0.08)--(-0.38,-0.2);
				\draw[thick] (0,0.08)--(0.38,-0.2);
				\draw[thick] (0,-0.45)--(-0.30,-1.05);
				\draw[thick] (0,-0.45)--(0.30,-1.05);
				\node at (0,-1.35) {\footnotesize DM};
			\end{scope}

			\begin{scope}[shift={(-2.95,0)}]
				\draw[thick, dashed, gray] (0,0) ellipse (1.25 and 1.0);
				\node[anchor=south] at (0,1.05) {\footnotesize $\mathscr{M}=\{s_1,\ldots,s_M\}$};
				\fill[orange!60] (-0.70,0.45) circle (0.15);
				\fill[blue!50]   (-0.10,0.55) circle (0.15);
				\fill[orange!60] ( 0.50,0.45) circle (0.15);
				\fill[blue!50]   ( 0.85,0.05) circle (0.15);
				\fill[orange!60] ( 0.45,-0.45) circle (0.15);
				\fill[blue!50]   (-0.15,-0.55) circle (0.15);
				\fill[orange!60] (-0.70,-0.30) circle (0.15);
			\end{scope}

			\draw[->, thick, >=Latex] (-1.55,0) -- (1.55,0)
				node[midway, above=2pt] {\footnotesize recall w.p.\ $r(s)$};

			\begin{scope}[shift={(2.95,0)}]
				\draw[thick, dashed, gray] (0,0) ellipse (1.25 and 1.0);
				\node[anchor=south] at (0,1.05) {\footnotesize $\mathscr{M}_r\subseteq\mathscr{M}$};
				\fill[orange]     (-0.70,0.45) circle (0.17);
				\draw[orange!70, thick] (-0.70,0.45) circle (0.28);
				\fill[orange]     ( 0.50,0.45) circle (0.17);
				\draw[orange!70, thick] ( 0.50,0.45) circle (0.28);
				\fill[blue]       ( 0.85,0.05) circle (0.17);
				\draw[blue!70, thick]   ( 0.85,0.05) circle (0.28);
				\draw[blue!30,   thin] (-0.10,0.55) circle (0.15);
				\draw[orange!30, thin] ( 0.45,-0.45) circle (0.15);
				\draw[blue!30,   thin] (-0.15,-0.55) circle (0.15);
				\draw[orange!30, thin] (-0.70,-0.30) circle (0.15);
			\end{scope}

			\draw[->, thick, >=Latex] (4.65,0) -- (6.00,0)
				node[midway, above=2pt] {\footnotesize Bayes};

			\node[draw, thick, rounded corners, fill=cyan!15, minimum width=1.4cm, minimum height=0.85cm]
				at (6.85,0) {\footnotesize $p_1(\mathscr{M}_r)$};
		\end{tikzpicture}

		\vspace{0.4em}
		{\footnotesize\itshape \textcolor{orange}{$\bullet$}\,pro-state-1 \quad \textcolor{blue}{$\bullet$}\,pro-state-0 \quad lit\,=\,recalled, faded\,=\,not recalled}
		\caption{The DM has a memory $\mathscr{M}$ of $M$ signals; each signal $s$ is independently recalled with probability $r(s)$, producing the recalled set $\mathscr{M}_r\subseteq\mathscr{M}$ (the ``lit'' signals on the right); the DM then forms the posterior $p_1(\mathscr{M}_r)$ by Bayes' rule, accounting for the possibility that some signals were not recalled.}
		\label{fig:timeline}
	\end{figure}

	In contrast, we assume that the DM forms his belief by retrieving information from his memory, where the retrieval probability can vary across signal realizations. In particular, these retrieval probabilities can be manipulated or influenced by persuasion tactics and advertisement, as analyzed in later sections. The DM recalls each signal $s$ independently with probability $r(s)$. Let
	\begin{equation}\label{eq:Aomega_def}
		\overline{R}_\omega\;\equiv\;\int_S f_\omega(s)\bigl(1-r(s)\bigr)\,ds
	\end{equation}
	denote the probability of failing to recall a signal in state $\omega$. The DM observes only the recalled set $\mathscr{M}_r\subseteq\mathscr{M}$: he knows the primitives of the environment---$f_0$, $f_1$, $g$, and the retrieval function $r(\cdot)$---but observes neither the memory size $M$ nor the signals that went unrecalled. The joint likelihood of observing a recalled set $\mathscr{M}_r$ in state $\omega$ is
	\begin{equation}\label{eq:Lomega_def}
		L_\omega(\mathscr{M}_r)\;=\;\underbrace{\prod_{s\in\mathscr{M}_r}r(s)\,f_\omega(s)}_{\text{recalled signals}}\;\cdot\;\underbrace{E_{n\geq|\mathscr{M}_r|}\!\left[\binom{n}{|\mathscr{M}_r|}\overline{R}_\omega^{\,n-|\mathscr{M}_r|}\right]}_{\text{forgotten signals}},
	\end{equation}
	where $E_{n\geq|\mathscr{M}_r|}$ denotes the expectation over $n\sim g$ conditional on $n\geq|\mathscr{M}_r|$. The first factor is the probability of drawing and recalling exactly the observed signals in state $\omega$; the second factor accounts for the DM's uncertainty about the unknown number of additional unrecalled signals---the binomial coefficient counts which $|\mathscr{M}_r|$ of the $n$ signals were recalled, and each of the remaining $n-|\mathscr{M}_r|$ independently failed to be recalled with probability $\overline{R}_\omega$. Bayes' rule then gives
	\begin{equation}\label{eq:memorybelief_complex}
		p_1(\mathscr{M}_r)\;=\;\frac{L_1(\mathscr{M}_r)}{L_1(\mathscr{M}_r)+L_0(\mathscr{M}_r)}.
	\end{equation}
	For ease of computation and interpretation, we also work on the log-odds from of Equation~\eqref{eq:memorybelief_complex}, which additively separate the two factors:
	\begin{equation}\label{eq:memorybelief_complex1}
		\log\frac{p_1(\mathscr{M}_r)}{1-p_1(\mathscr{M}_r)}\;=\;\sum_{s\in\mathscr{M}_r}\log\frac{f_1(s)}{f_0(s)}\;+\;\log\frac{E_{n\geq|\mathscr{M}_r|}\!\left[\binom{n}{|\mathscr{M}_r|}\overline{R}_1^{\,n-|\mathscr{M}_r|}\right]}{E_{n\geq|\mathscr{M}_r|}\!\left[\binom{n}{|\mathscr{M}_r|}\overline{R}_0^{\,n-|\mathscr{M}_r|}\right]}.
	\end{equation}
	Throughout, $\ell(s)\equiv\log\frac{f_1(s)}{f_0(s)}$ denotes the log-likelihood ratio of signal $s$, so the first term of Equation~\eqref{eq:memorybelief_complex1} is the sum of the recalled signals' log-likelihood ratios.

	In the rest of the paper, we analyze how changes in retrieval probability $r(s)$ affect the DM's posterior belief. Consider a benchmark where $r(s)\equiv r\in[0,1]$ for all $s$, i.e., the retrieval probability remains invariant across different signal realizations. We refer to such benchmark as the case where memory retrieval is not manipulated. For simplicity, we focus on memory retrieval probability manipulation that is in favor of state~1: there exist constants $r_+>r>r_-$ such that $r(s)\geq r_+$ for all $s$ with $f_1(s)>f_0(s)$ and $r(s)\leq r_-$ for all $s$ with $f_1(s)<f_0(s)$.\footnote{We allow for any $r(s'')$ for $f_1(s'')=f_0(s'')$, which does not affect the DM's posterior belief. The bounds $r_+$ and $r_-$ are imposed on all signals for simplicity; our results require them only at the extreme signals---those with sufficiently high and sufficiently low likelihood ratios, respectively---with $r(s)>r>r(s')$ elsewhere.} That is, the DM recalls signals that support state~1 with a probability bounded above the baseline, and signals supporting state~0 with a probability bounded below it. A special case, used repeatedly in the paper, is \emph{bang-bang} manipulation, which pushes these recall probabilities to their extremes: $r(s)=\overline{r}$ for all $s$ with $f_1(s)>f_0(s)$ and $r(s)=\underline{r}$ for all $s$ with $f_1(s)<f_0(s)$, for constants $\overline{r}>r>\underline{r}$.

	Throughout, we maintain a mild restriction on the tail of the memory-size distribution.
	\begin{assumption}[Poisson-type memory]\label{ass:light_tail}
		There are constants $\overline{C}\geq\underline{C}>0$ and $n_0$ such that $(n+1)\,g(n+1)\leq \overline{C}\,g(n)$ for all $n\geq n_0$ and, if the support of $g$ is unbounded, also $(n+1)\,g(n+1)\geq\underline{C}\,g(n)$ for all $n\geq n_0$.
	\end{assumption}
	Assumption~\ref{ass:light_tail} says that, from one size to the next, the memory-size distribution decays at least as fast as a Poisson distribution and, no faster than one when its support is unbounded: it implies $g(n)\leq g(n_0)\,n_0!\,\overline{C}^{\,n-n_0}/n!$ and, for unbounded support, $g(n)\geq g(n_0)\,n_0!\,\underline{C}^{\,n-n_0}/n!$. In terms of the model, Assumption~\ref{ass:light_tail} ensures that the expected number of signals the DM failed to recall, given how many he recalled, stays bounded above and, for unbounded memory, below: the DM neither infers that he has forgotten infinite signals nor becomes certain that nothing was forgotten when $g$ is unbounded. The assumption holds for Poisson memory ($\underline{C}=\overline{C}=\lambda$), for Poisson mixtures with bounded mixing support, and for every finitely supported $g$, in particular a deterministic memory size. It excludes geometric and negative binomial memory, whose tails are heavier, distributions lighter than Poisson, and gaps in the support arbitrarily far out. We discuss relaxing it at the end of Section~\ref{sec:main_results}.


	\section{Simple Example}\label{sec:simpleexample}
	To build intuition, this section presents the minimal example that illustrates the core mechanisms of the model. We assume 1) a binary signal space $S=\{s_0,s_1\}$; 2) the DM has and knows he has only one piece of memory, i.e., $M$ is deterministic and equal to $1$; 3) signal structure is symmetric such that
	\begin{equation*}
		\begin{split}
			f_1(s_1)=f_0(s_0)=\,& \theta\in(\tfrac{1}{2},1),\\
			 f_1(s_0)=f_0(s_1)=\,&1-\theta,
		\end{split}
	\end{equation*}
	so $s_0$ supports state~$0$ and $s_1$ supports state~$1$, and $\theta$ measures the precision of the signal. The signal log-likelihood ratios are then $\ell(s_1)=-\ell(s_0)=\log\frac{\theta}{1-\theta}\equiv\ell>0$. We consider manipulation in favor of state~1 such that
	\[
		r^{\text{manip}}(s_1)\;=\;\overline{r},\qquad r^{\text{manip}}(s_0)\;=\;\underline{r},\qquad\overline{r}>\underline{r},
	\]
	and the no-manipulation benchmark sets $r^{\text{no}}(s)\equiv r\equiv\tfrac{\overline{r}+\underline{r}}{2}$.

	\paragraph{Recalled Memories and Posteriors.} In this simple example, the DM either remembers nothing, a signal supporting state~0 or a signal supporting state~1, i.e., $\mathscr{M}_r$ takes one of three values: $\emptyset$, $\{s_0\}$, or $\{s_1\}$. The log-odds posterior beliefs $L=\log\frac{p_1}{1-p_1}$ in each value of $\mathscr{M}_r$, with and without manipulation, and their associated probabilities are:
	\begin{center}
		\begin{tabular}{lcccc}
			\toprule
			$\mathscr{M}_r$ & $L^{\text{no}}$ & $L^{\text{manip}}$ & $\Pr^{\text{no}}$ & $\Pr^{\text{manip}}$ \\
			\midrule
			$\emptyset$    & $0$      & $\log(\overline{R}_1^{\text{manip}}/\overline{R}_0^{\text{manip}})<0$ & $1-r$    & $1-r$ \\
			$\{s_0\}$      & $-\ell$  & $-\ell$  & $r/2$    & $\underline{r}/2$ \\
			$\{s_1\}$      & $+\ell$  & $+\ell$  & $r/2$    & $\overline{r}/2$ \\
			\bottomrule
		\end{tabular}
	\end{center}
	The following three observations summarize the example.

	\paragraph{(i)~Manipulation sometimes backfires.}
	As shown in the first row of the table, with and without manipulation, the DM sometimes fails to recall any signals (with probability $1-r$). Given that the DM does not recall any signals, without manipulation, the DM's posterior belief stays at $\frac{1}{2}$. In contrast, with manipulation, the DM's posterior belief drops below $\frac{1}{2}$:
	\begin{equation*}
			p_1^{\text{manip}}(\emptyset)\;=\;\frac{\overline{R}_1^{\text{manip}}}{\overline{R}_1^{\text{manip}}+\overline{R}_0^{\text{manip}}}\;<\;\tfrac{1}{2},
	\end{equation*}
	since $\overline{R}_1^{\text{manip}}<\overline{R}_0^{\text{manip}}$. Manipulation backfires when the DM does not recall any signals supporting the favored state of the world. This is reminiscent of the classic ``ease of retrieval'' effect documented in \cite{schwarz1991ease}: ``If I cannot retrieve any signals supporting state~1 given the elevated retrieval probability, it must be more likely that my memory does not contain any signals supporting state~1 and thus state~1 is not true.''.

	\paragraph{(ii)~Manipulation generates a heavier right tail.} The third row of the table shows the main benefit of manipulation: the probability that the DM recalls a pro-state-1 signal increases from $\frac{r}{2}$ to $\frac{\overline{r}}{2}$. As a result, the distribution of the DM's posterior has a heavier right tail under manipulation compared to no manipulation. In this simple example, when the DM recalls a signal, the ``ease of retrieval'' effect vanishes. Manipulation thus has a positive direct effect (a higher probability of recalling pro-state-1 signals) without the offsetting negative indirect effect, so the DM's posterior is pushed in favor of state~1.

	\paragraph{(iii)~Manipulation does not change the average posterior.} Lastly, on average, since the DM is Bayesian and manipulation is merely a transformation of the information structure, the martingale property implies that manipulation does not change the average posterior. Simple algebra verifies this. With manipulation, using $\overline{R}_1^{\text{manip}}+\overline{R}_0^{\text{manip}}=2(1-r)$:
	\begin{equation*}
		\begin{split}
			E[p_1^{\text{manip}}]\;=\;&(1-r)\cdot\frac{\overline{R}_1^{\text{manip}}}{\overline{R}_1^{\text{manip}}+\overline{R}_0^{\text{manip}}} + \frac{\underline{r}}{2}(1-\theta) + \frac{\overline{r}}{2}\theta\\
			=\;&\frac{\overline{R}_1^{\text{manip}}}{2} + \frac{\underline{r}(1-\theta) + \overline{r}\theta}{2}\\
			=\;&\frac{(1-\overline{r})\theta + (1-\underline{r})(1-\theta) + \underline{r}(1-\theta) + \overline{r}\theta}{2}\\
			=\;&\frac{\theta+(1-\theta)}{2}\;=\;\frac{1}{2}.
		\end{split}
	\end{equation*}
	Without manipulation, the empty recall is uninformative ($p_1^{\text{no}}(\emptyset)=\frac{1}{2}$), and:
	\begin{equation*}
		E[p_1^{\text{no}}]\;=\;(1-r)\cdot\tfrac{1}{2} + \tfrac{r}{2}(1-\theta) + \tfrac{r}{2}\theta\;=\;\tfrac{1-r}{2} + \tfrac{r}{2}\;=\;\tfrac{1}{2}.
	\end{equation*}
	Combined with (i) and (ii), this simple example shows that, as in the Bayesian persuasion literature \citep{kamenica2011bayesian}, the key impact of manipulation lies in the details of the distribution of posterior beliefs, i.e., in this model a heavier right tail.


	\section{Main Results}\label{sec:main_results}
	We now turn to a model with general signal and memory-size distributions and study how manipulation in memory retrieval probabilities affects the DM's posterior beliefs.

	First, as in the simple example, manipulation cannot alter the average posterior in favor of or against state~1. 

	\begin{lemma}\label{lemma:martingale}
		For all $r(s)$, the average posterior belief of the DM is equal to his prior belief.
	\end{lemma}


	While this mean-preservation property rules out average belief manipulation, as foreshadowed in the simple example, the key insight of the model lies in the distributional results: memory manipulation reshapes the distribution of posterior beliefs, generating a heavier right tail. Unlike the simple example, the following proposition also presents results conditional on the memory size~$M$.\footnote{Recall that the DM never observes $M$. The memory-size distribution $g$ is held fixed throughout, so conditioning on different values of $M$ compares individuals who happened to accumulate few or many signals.}

	\begin{proposition}\label{prop:general_g}
		Let $g$ be a non-degenerate memory-size distribution with $g(0)>0$, and consider memory manipulation in favor of state~$1$, i.e., $r(s)\geq r_+>r>r_-\geq r(s')$ for all $s,s'$ with $f_1(s)>f_0(s)$ and $f_1(s')<f_0(s')$. 
		\begin{enumerate}[label=(\roman*)]
			\item \textbf{Backfire at small $M$ and effective at large $M$.} $E\left(p_1(\mathscr{M}_r)\mid M\right)<\frac{1}{2}$ for small enough $M$, while $E\left(p_1(\mathscr{M}_r)\mid M\right)>\frac{1}{2}$ for all large enough $M$. Moreover, whenever $\int r(s)f_0(s)\log\frac{f_1(s)}{f_0(s)}\,ds>0$, $p_1(\mathscr{M}_r)\to 1$ almost surely as $M\to\infty$ conditional on $\omega=0$.\footnote{Almost-sure convergence is defined with a nested memories structure: the memory at size $M$ consists of the first $M$ draws of an infinite i.i.d.\ signal sequence. Without it, each $M$ indexes a separate experiment, and thus only distributional convergence are meaningful, i.e., we have $\lim_{M\to\infty}E\left(p_1(\mathscr{M}_r)\mid M,\omega=0\right)=1$.}
			\item \textbf{Robust impact on extreme beliefs.} Denote by $B^{\emph{manip}}$ and $B^{\emph{no}}$ the cumulative distribution functions of the posterior belief $p_1$ with and without manipulation. There exist $\hat{p}_H\in(0,1)$ and $\hat{p}_L\in(0,\hat{p}_H)$ such that $B^{\emph{manip}}(p) < B^{\emph{no}}(p)$ for every $p\in(0,\hat{p}_L)\cup(\hat{p}_H,1)$ with $B^{\emph{no}}(p)\in(0,1)$.
		\end{enumerate}
	\end{proposition}

	Proposition~\ref{prop:general_g} characterizes the impact of memory manipulation along two margins: how it varies with the DM's realized memory size (Part~(i)), and how it reshapes the tails of the posterior distribution (Part~(ii)). Part~(i) shows that manipulation backfires for individuals with small memories yet is effective for those with large ones---the same manipulation moves different individuals' beliefs in opposite directions. The intuition rests on the two opposing forces in Equation~\eqref{eq:memorybelief_complex1}: the \textit{direct effect} (the first term: recalled signals are tilted toward state~1) and the \textit{indirect ease-of-retrieval effect} (the second term: failing to recall state-1-supporting signals despite their boosted retrieval probability is itself evidence against state~1). For an individual with a small memory, little is recalled, so the indirect effect dominates and manipulation backfires. For an individual with a large memory, the law of large numbers delivers many recalled state-1-supporting signals, and the accumulated direct evidence overwhelms the indirect correction---which Assumption~\ref{ass:light_tail} keeps asymptotically negligible. 

	This pattern is in line with existing evidence. In experiments that prompt subjects to retrieve supporting memories, those with little knowledge of the domain judge by how difficult retrieval felt (the indirect effect), whereas experts, who retrieve with ease, judge by what they recall (the direct effect) \citep{ofir2000ease,ofir2008memory}; in a meta-analysis of 263 studies, \cite{weingarten2018does} find that the ease-of-retrieval effect reverses among experts and individuals with firmly held attitudes. In a field experiment, \cite{lewis2014advertising} find that customers aged 65 and above, 5\% of a retailer's customers, account for 40\% of the effect of its advertising on purchases, which suggests that memory size, as proxied by age, plays a role.
	
	Part~(i) also highlights that sufficiently aggressive manipulation can generate confident \emph{false} beliefs: whenever $\int r(s)f_0(s)\log\frac{f_1(s)}{f_0(s)}\,ds>0$---that is, whenever manipulated recall is selective enough that the recalled evidence favors state~1 on average even when the true state is~0---the DM's posterior converges almost surely to certainty in state~1 conditional on state~0 as the realized $M$ goes to $\infty$. This contrasts with the classic result that the Bayesian posterior converges to the truth as evidence accumulates. The failure is not due to naivety: the DM knows the manipulation and corrects for it. The key observation is that the correction is based on the expected number of unrecalled signals, not on the DM's actual memory size. Under Assumption~\ref{ass:light_tail} this expected number grows less than proportionally with the amount recalled,\footnote{Under Poisson memory the correction is a constant: the expected number of unrecalled signals is independent of the amount recalled.} so as $M\to\infty$ the correction is outpaced by the biased direct evidence accumulating without bound. The DM thus becomes certain of the wrong state.

	Part~(ii) characterizes the distributional effects: manipulation generates a strictly \textit{heavier right tail} (more extreme beliefs favoring state~1) and a strictly \textit{lighter left tail} (fewer extreme beliefs favoring state~0). 
	 Both effects share the same mechanism: extreme beliefs typically arise from recalling many signals favoring one state, in which case Assumption~\ref{ass:light_tail} ensures that the indirect effect is dominated by the direct effect. Manipulation tilts each recall toward state~1, making the far right tail much easier to reach and the far left tail much harder. The maintained assumption that no single signal rules out a state ensures that extreme beliefs are reached only through many recalled signals, and the lower bound in Assumption~\ref{ass:light_tail} ensures that one more recalled signal is never prohibitively unlikely, so the fixed indirect effect is always dominated at the tail.

	 Proposition~\ref{prop:general_g}(ii) shows that manipulation pushes more mass past any sufficiently high attainable threshold. The next result imposes extra assumptions and sharpens this to a comparison of \emph{hazard rates}, which are widely used in monopoly pricing (Section~\ref{sec:app3}), optimal auctions, and screening problems \citep{myerson1981optimal,bagnoli2005log}. Write $1-B(\bar{p})=\Pr(p_1>\bar{p})$ for the \emph{survival function} of the posterior and $h(\bar{p})\equiv B'(\bar{p})/[1-B(\bar{p})]$ for its \emph{hazard rate}---the rate at which the survival probability decays as the threshold rises. Manipulation not only raises the survival probability in the right tail but also lowers its hazard rate. Equivalently, the survival-function ratio $[1-B^{\text{manip}}(\bar{p})]/[1-B^{\text{no}}(\bar{p})]$ is increasing in the right tail: manipulation's right-tail advantage increases as the threshold rises.

	\begin{proposition}[Hazard-rate dominance in the right tail]\label{prop:hazard}
		Suppose (i) $M\sim\emph{Poisson}(\lambda)$ with $\lambda>0$; and (ii) the log-likelihood ratio $\ell(s)$ is continuously distributed (admits a density). Then, with $\hat{p}_H$ the threshold defined in Proposition~\ref{prop:general_g}(ii), there exists $p^*\in[\hat{p}_H,1)$ such that the posterior hazard rates satisfy
		\begin{equation}\label{eq:hazard_dominance}
			h^{\emph{manip}}(\bar{p})\;<\;h^{\emph{no}}(\bar{p})\qquad\text{for all }\bar{p}\in[p^*,1),
		\end{equation}
		or equivalently, the survival-function ratio $[1-B^{\emph{manip}}(\bar{p})]/[1-B^{\emph{no}}(\bar{p})]$ is strictly increasing on $[p^*,1)$ and tends to infinity as $\bar{p}\uparrow 1$.
	\end{proposition}

	The two conditions read as follows. Condition~(i) is Poisson memory, under which the ease-of-retrieval correction is a constant shift of the log-odds. Condition~(ii) makes the posterior belief continuously distributed, so that hazard rates are well defined; it excludes discrete signals. 
	
	The proof of Proposition~\ref{prop:hazard} relies on saddlepoint approximations \citep{jensen1995saddlepoint}. A saddlepoint approximation estimates the probability that a sum of many random terms exceeds a high threshold, and it becomes more accurate as the threshold rises. It approximates both the density and the survival probability of the posterior log-odds, and hence their ratio, the hazard rate, with and without manipulation.

	The intuition is as follows. Because $\ell$ is bounded, raising the posterior log-odds by one unit requires at least $1/\sup_s\ell(s)$ more recalled favorable signals. Manipulation raises each favorable signal's recall probability by a factor of at least $r_+/r$. Under Poisson memory, this makes recalling one more favorable signal at least $r_+/r$ times as likely, however many have already been recalled. Hence, given that the log-odds clears a high threshold, the probability that it also clears the threshold plus one is at least $(r_+/r)^{1/\sup_s\ell(s)}$ times higher with manipulation. Equivalently, measured in log-odds, the hazard rate is lower by at least $\log(r_+/r)/\sup_s\ell(s)$. The ease-of-retrieval correction cannot undo this: under Poisson memory it lowers every posterior log-odds by the same constant $\lambda(\overline{R}_0-\overline{R}_1)$, which moves the tail sideways without materially changing how fast it decays far out.

	Propositions~\ref{prop:general_g} and~\ref{prop:hazard} fix the signal structure and vary the memory size (Proposition~\ref{prop:general_g}) or the threshold (Proposition~\ref{prop:hazard}). We close the analysis by varying the \emph{informativeness} of the signal structure itself and asking when manipulation is valuable. More informative signals strengthen both the direct and the indirect effect: each recalled signal moves beliefs further, but the DM also grows more skeptical when he retrieves only a few. For simplicity, we focus on bang-bang manipulation for the following result.

	\begin{proposition}[Informative signals crowd out manipulation when memories are large]\label{prop:crowding}
		Fix a threshold $\bar{p}>\tfrac{1}{2}$ and a memory-size distribution $g$ with $\Pr(M\geq 1)>0$ and generating function $Q_g(y)=E[y^M]$. Consider a family of signal structures $\{f_0^{\gamma},f_1^{\gamma}\}_{\gamma\in[0,1]}$, continuous in $\gamma$ (in total variation) and increasing in the Blackwell order, from uninformative ($\gamma=0$: $f_1^{0}=f_0^{0}$) to perfectly revealing ($\gamma=1$: $f_1^{1}$ and $f_0^{1}$ have disjoint supports), with bang-bang manipulation and baseline recall $r$. Let
		\begin{equation*}
			\Gamma_\gamma(\bar{p})\;\equiv\;\Pr\!\big(p_1^{\emph{manip}}\geq\bar{p}\big)-\Pr\!\big(p_1^{\emph{no}}\geq\bar{p}\big)
		\end{equation*}
		denote the manipulation gain. Then
		\begin{enumerate}[label=(\roman*)]
			\item \textbf{No traction under uninformative information.} $\Gamma_\gamma(\bar{p})\to 0$ as $\gamma\to 0$.
			\item \textbf{Gain under perfect information.} $\Gamma_\gamma(\bar{p})\to\tfrac{1}{2}\big[Q_g(1-r)-Q_g(1-\overline{r})\big]>0$ as $\gamma\to 1$.
			\item \textbf{Crowding out under large memory.} The limit in~(ii) vanishes as $M\to\infty$ in probability (e.g., Poisson memory with $\lambda\to\infty$).
		\end{enumerate}
	\end{proposition}

	The result shows an intricate relationship between informativeness and manipulability. When signals are uninformative, no recalled or suppressed signal moves the DM's belief, so the posterior stays at $\tfrac{1}{2}$ with or without manipulation and manipulation has no traction. Part~(ii) is more surprising: manipulation always gains under perfect information. A single recalled signal then pins down the state, and manipulation's only remaining lever is the empty-recall event: it reduces the probability of recalling nothing in state~1 from $Q_g(1-r)$ to $Q_g(1-\overline{r})$. This gain is substantial for a DM with few memories, but for large memories the empty-recall event is rare and the gain evaporates, which is Part~(iii).

	\subsection{Discussion}\label{sec:main_discussion}

	We close this section with four observations on the scope of the results.

	\paragraph{Unbiased baseline recall.} The assumption that the no-manipulation benchmark uses a constant recall probability $r^{\text{no}}(s)\equiv r$ can be substantially weakened. All parts of Proposition~\ref{prop:general_g} extend, with the same arguments, to any signal-dependent baseline $r^{\text{no}}(s)$ satisfying the \emph{unbiased baseline recall} condition
	\begin{equation}\label{eq:unbiased_baseline}
		\int r^{\text{no}}(s)f_1(s)\,ds\;=\;\int r^{\text{no}}(s)f_0(s)\,ds,
	\end{equation}
	and any manipulation whose recall probability $r^{\text{manip}}(s)$ exceeds $r^{\text{no}}(s)$ wherever $\ell(s)>0$ and falls short of it wherever $\ell(s)<0$, in both cases by a margin bounded away from zero. Propositions~\ref{prop:hazard} and~\ref{prop:crowding} extend as well. Proposition~\ref{prop:hazard} holds as stated; in the intuition above and in the proof, the factor $r_+/r$ becomes $\inf_{s:\,\ell(s)>0}r^{\text{manip}}(s)/r^{\text{no}}(s)$, which exceeds one because $r^{\text{manip}}$ exceeds $r^{\text{no}}$ on these signals by a margin bounded away from zero and $r^{\text{no}}\leq 1$. In Proposition~\ref{prop:crowding}, suppose the baseline is unbiased for every $\gamma$ and its recall probability $\int r^{\text{no}}(s)f_1^{\gamma}(s)\,ds$, which by \eqref{eq:unbiased_baseline} is the same in both states, converges to a positive limit as $\gamma\to 1$. Then Part~(ii) holds with $r$ replaced by this limiting baseline recall and $\overline{r}$ unchanged, and the gain is still positive because $\overline{r}$ exceeds the baseline on favorable signals; Part~(iii) holds because the limiting baseline recall is positive. Condition~\eqref{eq:unbiased_baseline} says the benchmark is equally likely to recall a signal in either state---equivalently, the probability of \emph{not} recalling any given signal is state-independent, $\overline{R}_1^{\text{no}}=\overline{R}_0^{\text{no}}$---which is exactly what preserves the clean benchmark $E[p_1^{\text{no}}\mid M]=\tfrac{1}{2}$.
	 A natural sufficient condition is that signals are symmetric (there exists a measure-preserving involution $\phi$ with $f_1(\phi(s))=f_0(s)$, hence $\ell(\phi(s))=-\ell(s)$) and $r^{\text{no}}$ depends only on $|\ell(s)|$. Economically, this accommodates realistic features such as heightened recall of emotionally salient or extreme signals, as long as the salience is symmetric across states.

	\paragraph{Results conditional on the observable recalled count.} Although Proposition~\ref{prop:general_g}(i) conditions on the latent memory size $M$, the analogue of Part~(i) holds conditional on the \emph{observable} recalled count $|\mathscr{M}_r|$: the benchmark gives $E[p_1^{\text{no}}\mid |\mathscr{M}_r|=k]=\tfrac{1}{2}$ for every $k$ and any $g$, while manipulation backfires at low recalled counts and is effective at every sufficiently high count. The intuition is that $M$ and $|\mathscr{M}_r$ are positively correlated. This formulation enables empirical tests, since the recalled count is observable whereas $M$ is not.

	\paragraph{Manipulation at thresholds below the prior.} Propositions~\ref{prop:hazard} and~\ref{prop:crowding} evaluate manipulation at a threshold $\bar{p}>\tfrac{1}{2}$. For Proposition~\ref{prop:hazard}, although we do not state it formally, the same argument applied to the negative of the posterior log-odds $L=\log\frac{p_1}{1-p_1}$ delivers the mirror statement for the left tail: the CDF ratio $B^{\text{manip}}(\bar{p})/B^{\text{no}}(\bar{p})\to 0$ as $\bar{p}\to 0$, and the \emph{reversed} hazard rate $\frac{\mathrm{d}}{\mathrm{d}\bar{p}}\log B(\bar{p})$ is strictly higher under manipulation in the left tail. This mirror argument follows similar argument as in Propsition~\ref{prop:hazard}.

	For Proposition~\ref{prop:crowding}, the limit in Part~(ii) depends on where the threshold lies relative to the empty-recall posterior. Under manipulation, recalling nothing is evidence against state~1, so the posterior after empty recall falls to some $p^{\varnothing}<\tfrac{1}{2}$. For $\bar{p}\in(p^{\varnothing},\tfrac{1}{2})$, individuals who recall nothing clear the threshold without manipulation but not with it, and the gain converges to $-\tfrac{1}{2}\big[Q_g(1-\overline{r})+Q_g(1-r)\big]<0$. For $\bar{p}<p^{\varnothing}$, they clear it in both cases, and suppressing unfavorable signals makes state-0 individuals more likely to recall nothing; the gain converges to $\tfrac{1}{2}\big[Q_g(1-\underline{r})-Q_g(1-r)\big]>0$. Parts~(i) and~(iii) are unchanged, the latter provided $\underline{r}>0$: every limit above vanishes as memories grow.

	\paragraph{Relaxing Assumption~\ref{ass:light_tail}.} Three results hold for any memory-size distribution: the martingale property of Lemma~\ref{lemma:martingale}; backfire at small $M$, which requires only $g(0)>0$; and effectiveness at \emph{some} memory size, since the conditional expectations $E(p_1\mid M)$ average to $\tfrac{1}{2}$. Assumption~\ref{ass:light_tail} matters at large memories. For example, when $g$ has geometric tails, as for geometric or negative binomial memory, each recalled signal is also evidence of a larger memory, so the ease-of-retrieval effect grows with the number of signals recalled instead of staying bounded, and it competes with the direct evidence. Effectiveness for all large $M$ is then no longer guaranteed; confident false beliefs require the drift $\int r(s)f_0(s)\ell(s)\,ds$ to exceed a positive hurdle rather than zero; and the tail results can fail, because extreme beliefs are reached mainly through unusually large memories, which are mainly determined by $g$ and equally likely with and without manipulation. In numerical examples with geometric memory, manipulation produces a \emph{lighter} right tail. 


	\section{Applications}
	This section develops two applications of the general model. The first studies monopoly pricing; the second analyzes narrative persuasion in political communication. Throughout, we focus on bang-bang manipulation.

	\subsection{Application 1: Monopoly Pricing}\label{sec:app3}

	The first application shows how the heavier right tail result (Proposition~\ref{prop:general_g}) shapes a monopolist's incentive to manipulate consumer memory. We show that a monopolist charging a sufficiently high price strictly prefers to manipulate consumer memory, and manipulation in turn raises the monopoly price.

	\subsubsection{Setup}
	A monopolist sells a product of uncertain match value to a representative consumer. The match value is $v(\omega)\in\{v_L,v_H\}$ with $v_H>v_L\geq 0$, where $\omega\in\{0,1\}$ is the state: $v(1)=v_H$ (good match) and $v(0)=v_L$ (bad match). The consumer's prior is $\Pr(\omega=1)=\frac{1}{2}$. The monopolist produces at constant marginal cost $c\in[v_L,v_H)$ and chooses a price $P\in[c,v_H]$.

	Prior to the purchase decision, the consumer has accumulated memories about the product: a Poisson number of signals $M\sim\text{Poisson}(\lambda)$, each drawn from the state-dependent distribution $f_\omega$ on $S$ with bounded likelihood ratio. The monopolist chooses whether to manipulate memory retrieval (e.g., through advertising, reviews, or framing) such that the consumer recalls favorable signals $s:f_1(s)>f_0(s)$ with probability $\overline{r}$ and non-favorable signals $s':f_1(s')<f_0(s')$ with probability $\underline{r}<\overline{r}$. If the monopolist does not manipulate, the consumer recalls each signal with probability $r\in (\underline{r},\overline{r})$. Given the recalled set $\mathscr{M}_r$, the consumer forms the posterior $p_1=\Pr(\omega=1\mid\mathscr{M}_r)$ and buys if and only if the expected match value weakly exceeds the price:
	\begin{equation}\label{eq:buy_rule}
		v_L+p_1(v_H-v_L)\geq P\quad\Longleftrightarrow\quad p_1\geq \bar{p}(P)\equiv \frac{P-v_L}{v_H-v_L}.
	\end{equation}
	The monopolist's profit is
	\begin{equation}\label{eq:monopoly_profit}
		\Pi(P,r)=(P-c)\cdot\Pr\!\left(p_1(\mathscr{M}_r)\geq \bar{p}(P)\right).
	\end{equation}
	The consumer's buy-or-not decision in \eqref{eq:buy_rule} is a threshold rule, with $\bar{p}(P)$ strictly increasing in $P$ and ranging over $[0,1]$ as $P$ ranges over $[v_L,v_H]$. The monopolist's problem therefore couples a standard pricing trade-off (higher margin v.s. higher demand) with the threshold objective to which Proposition~\ref{prop:general_g} directly applies.

	\subsubsection{The Monopolist's Incentive to Manipulate}
	In this subsection, we first fix a price $P\in(c,v_H]$ and analyze the monopolist's incentive to manipulate. Let $\Pi^{\text{no}}(P)\equiv (P-c)\Pr(p_1^{\text{no}}\geq\bar{p}(P))$ and $\Pi^{\text{manip}}(P,r)\equiv (P-c)\Pr(p_1^{\text{manip}}\geq\bar{p}(P))$ denote the profit under no manipulation and under manipulation $r$, respectively. By Proposition~\ref{prop:general_g}(ii), for every price high enough that $\bar{p}(P)>\hat{p}_H$, the purchase probability strictly increases:
	\[
		\Pr(p_1^{\text{manip}}\geq\bar{p}(P))>\Pr(p_1^{\text{no}}\geq\bar{p}(P)),
	\]
	where $\hat{p}_H\in(0,1)$ is the threshold from Proposition~\ref{prop:general_g}(ii). That proposition compares $1-B(\bar{p})=\Pr(p_1>\bar{p})$, while demand is $\Pr(p_1\geq\bar{p})$; its proof applies verbatim with $\{L\geq x\}$ in place of $\{L>x\}$. Translating this into the price variable via \eqref{eq:buy_rule}:
	\begin{equation}\label{eq:price_threshold}
		\hat{P}\equiv v_L+\hat{p}_H(v_H-v_L)\in(v_L,v_H).
	\end{equation}
	
	\begin{corollary}[Monopoly manipulation at high prices]\label{cor:monopoly_manipulation}
		Let $\hat{P}$ be defined in \eqref{eq:price_threshold}. For every price $P\in(\max\{\hat{P},c\},v_H)$, the monopolist strictly prefers to manipulate memory: the bang-bang manipulation
		\begin{equation}\label{eq:monopoly_bangbang}
			r^*(s)=\begin{cases}\overline{r} & \text{if }\ell(s)>0\quad\text{(favorable signal)}\\ \underline{r} & \text{if }\ell(s)<0\quad\text{(unfavorable signal)}\end{cases}
		\end{equation}
		yields $\Pi^{\emph{manip}}(P,r^*)>\Pi^{\emph{no}}(P)$.
	\end{corollary}

The logic follows directly from Proposition~\ref{prop:general_g}(ii). First, a high price $P$ implies only consumers with extreme posteriors buy. Second, manipulation creates a strictly heavier right tail (Proposition~\ref{prop:general_g}(ii)), which moves additional mass past $\bar{p}$ and increases demand exactly when the price is high enough. 

	In contrast, by the martingale property, $\int_0^1[B^{\text{manip}}-B^{\text{no}}]\,dp=0$ for any manipulation $r$, so manipulation reduces demand at \emph{some} prices.\footnote{The direction of the effect at any particular price below $\hat{P}$ depends on the specific details of the signal distributions and manipulation function $r$. The corollary's contribution is that the heavier right tail robustly guarantees manipulation is profitable at \emph{high} prices regardless of the details of the manipulation and signal distributions.} The model thus predicts that manipulation-driven advertising (narrative framing, testimonials, imagery-rich campaigns) should be concentrated in categories of expensive experience goods, where $P$ is high enough that $\bar{p}(P)$ exceeds the threshold $\hat{p}_H$. In contrast, inexpensive commodity products should feature less narrative manipulation, consistent with the observation that commodities compete on price and informative disclosure while premium experience goods compete on imagery and story. In the car market, for instance, emotional content in television ads raises sales more for high-price than for low-price models, whereas informational content raises sales more for low-price models \citep{guitart2021impact}; and luxury fashion houses such as Herm\`es, Chanel and Louis Vuitton advertise through storylines that transport the viewer into a narrative \citep{kim2016narrative}. Regulation that narrows $\overline{r}-\underline{r}$ (e.g., restrictions on cherry-picked testimonials, mandatory disclosure of adverse outcomes, limits on framing effects) is therefore most valuable for high-priced experience goods.

	\subsubsection{Pricing--Manipulation Complementarity}
	The results above characterize the manipulation decision \emph{conditional on price}. We now ask the converse: does the availability of manipulation change the monopolist's \emph{pricing} decision? We show that when the marginal cost is sufficiently high, so that the product is expensive, the availability of manipulation unambiguously raises the optimal price.

	Let $\bar{p}_c\equiv(c-v_L)/(v_H-v_L)\in[0,1)$ denote the \emph{breakeven belief}: the minimum posterior at which a consumer's expected match value covers the marginal cost. Since $P\geq c$ implies $\bar{p}(P)\geq\bar{p}_c$, the monopolist operates entirely in the region $\bar{p}\in[\bar{p}_c,1]$. Recalling from Section~\ref{sec:main_results} the hazard rate of the posterior, $h(\bar{p})\equiv B'(\bar{p})/[1-B(\bar{p})]$, the first-order condition for the monopolist's optimal price reduces to
	\begin{equation}\label{eq:monopoly_foc}
		(\bar{p}^*-\bar{p}_c)\cdot h(\bar{p}^*)=1,
	\end{equation}
	so a \emph{lower} hazard rate at the same threshold raises the optimal markup $\bar{p}^*-\bar{p}_c$, hence raises the optimal price.\footnote{Differentiating $\Pi(P)=(P-c)[1-B(\bar{p}(P))]$ gives $d\Pi/dP=[1-B(\bar{p})]-(P-c)B'(\bar{p})/(v_H-v_L)=[1-B(\bar{p})][1-(\bar{p}-\bar{p}_c)h(\bar{p})]$. Setting this to zero gives \eqref{eq:monopoly_foc}.}


	\begin{corollary}[Pricing--manipulation complementarity]\label{cor:complementarity}
		Assume the log-likelihood ratio $\ell(s)$ is continuously distributed (admits a density). There exists $p^*\in[\hat{p}_H,1)$ such that if the marginal cost is sufficiently high,
		\begin{equation}\label{eq:high_cost}
			c\;\geq\;v_L+p^*(v_H-v_L),
		\end{equation}
		and the monopolist's profit function under manipulation is quasi-concave in $P$, then the optimal price under (optimal bang-bang) manipulation strictly exceeds the no-manipulation optimal price:
		\[
			P^{\emph{manip},*}\;>\;P^{\emph{no}}.
		\]
	\end{corollary}

	By Proposition~\ref{prop:hazard}, bang-bang manipulation lowers the hazard rate of the posterior in the right tail. Demand is then less elastic at high prices, so the monopolist optimally charges a higher markup. The high-cost condition~\eqref{eq:high_cost} ensures that all feasible prices lie in this right-tail region, which turns the local hazard-rate comparison into a global price comparison. Economically, the result describes markets for expensive experience goods, such as cars and luxury goods, where $c$ is close to $v_H$ and only consumers with high posteriors purchase. In these markets, emotional and narrative advertising content is effective \citep{guitart2021impact,kim2016narrative}, and the model predicts that it also allows sellers to charge higher prices.

	\subsection{Application 2: Narrative Persuasion}\label{sec:app1}
	The Economics literature has mainly studied communication in the form of hard and soft evidence, i.e., Bayesian persuasion and cheap talk (\cite{crawford1982strategic,kamenica2011bayesian}). However, much of political communication goes beyond presenting evidence. Political advertisements routinely use narratives as a tool of persuasion (\cite{vafeiadis2018narratives}). This application develops a formal model of a sender choosing between disclosing informative evidence and engaging in memory manipulation, and characterizes the equilibrium.

	\subsubsection{Setup}\label{sec:app1_setup}

	A sender (politician) wants to persuade a receiver (voter) that state~$\omega=1$ is true. The receiver takes a binary action $a\in\{0,1\}$ and prefers $a=1$ if and only if she believes state~1 is true with a high enough probability. The sender prefers $a=1$ regardless of $\omega$.

	\paragraph{Timeline.}
	\begin{enumerate}
		\item Nature draws $\omega\in\{0,1\}$ with equal probability.
		\item The receiver accumulates $M$ signals drawn i.i.d.\ from $f_\omega$, with $M\sim\text{Poisson}(\lambda)$.
		\item The sender privately observes a signal $s^P$ drawn from a density $f^P_\omega$ in state~$\omega$. Its log-likelihood ratio $\ell^P\equiv\log[f^P_1(s^P)/f^P_0(s^P)]$ has a density that is positive on its support $[\ell_{\min},\ell_{\max}]$, where $\ell_{\min}<0<\ell_{\max}$; write $F^P_\omega(c)\equiv\Pr(\ell^P\leq c\mid\omega)$.
		\item The sender chooses one of two actions:
		\begin{itemize}
			\item \textbf{Disclose:} Verifiably reveal $s^P$ to the receiver. The receiver then recalls her memory with the default retrieval probability $r(s)\equiv r$ for all $s$.
			\item \textbf{Manipulate:} Withhold $s^P$ and instead manipulate the receiver's recall such that receiver recalls signals $s: f_1(s)>f_0(s)$ with probability $\overline{r}>r$ and signals $s: f_1(s)<f_0(s)$ with probability $\underline{r}<r$.
		\end{itemize}
		\item The receiver observes whether the sender disclosed (and, if so, $s^P$) or manipulated. The receiver retrieves signals from memory and forms posterior belief $p_1(\mathscr{M}_r)$ using Bayes' rule. She then takes action $a=\mathds{1}[p_1\geq\bar{p}]$ for some exogenous threshold $\bar{p}$.
	\end{enumerate}

	The assumption that the sender can disclose \emph{or} manipulate, but not both, captures the incompatibility of the two communication strategies: a factual statistical argument and an emotional narrative designed to trigger selective recall involve different rhetorical modes and are difficult to combine effectively within a single message. We further assume that the memory log-likelihood ratio $\ell(s)$ has a bounded continuous density, so that, as in Proposition~\ref{prop:hazard}, the posterior log-odds has a density except for the atom at empty recall.


	\paragraph{Sender's objective.} The sender maximizes the probability that the receiver takes action $a=1$, e.g., that the sender wants to win a vote, secure support for a policy.

	\paragraph{Receiver's belief formation.} If the sender discloses $s^P$, the receiver's log-odds belief is:
	\begin{equation}\label{eq:belief_disclose}
		\log\frac{p_1}{1-p_1} = \underbrace{\log\frac{f^P_1(s^P)}{f^P_0(s^P)}}_{\text{sender's signal}} + \underbrace{\sum_{s\in\mathscr{M}_r}\log\frac{f_1(s)}{f_0(s)}}_{\text{recalled memory}},
	\end{equation}
	where $f^P_\omega$ denotes the distribution of $s^P$ in state $\omega$, and the indirect effect vanishes because $r(s)\equiv r$ implies $\int r(f_1-f_0)\,ds=0$.

	If the sender manipulates with retrieval function $r(\cdot)$, the receiver infers that the sender chose not to disclose. In equilibrium, this inference is itself informative. The log-odds is:
	\begin{equation}\label{eq:belief_manipulate}
		\log\frac{p_1}{1-p_1} = \underbrace{\log\frac{\Pr(\text{manipulate}\mid\omega=1)}{\Pr(\text{manipulate}\mid\omega=0)}}_{\text{inference from non-disclosure}} + \underbrace{\sum_{s\in\mathscr{M}_r}\log\frac{f_1(s)}{f_0(s)}}_{\text{recalled memory}} - \underbrace{\lambda\int_{S}r(s)(f_1(s)-f_0(s))\,ds}_{\text{indirect effect }\lambda\Delta}.
	\end{equation}
	The first term reflects rational updating from the sender's strategy. If the sender manipulates when $\ell^P<c$ and discloses when $\ell^P>c$, it equals $\ell_{\text{inf}}(c)\equiv\log[F^P_1(c)/F^P_0(c)]$, which is increasing in $c$, equals $0$ at $c=\ell_{\max}$, and tends to $\ell_{\min}$ as $c\downarrow\ell_{\min}$. The third term is the indirect effect under manipulation, with $\Delta\equiv\int r(s)(f_1-f_0)\,ds>0$.

	\subsubsection{Equilibrium with Memory Manipulation}

	We now analyze the game where the sender choose whether to manipulate memory retrieval. A perfect Bayesian equilibrium (PBE) consists of a disclosure decision for each sender type and the receiver's beliefs after each history, such that each sender type maximizes the probability of $a=1$ given the receiver's beliefs, the receiver acts by the threshold rule at her posterior, and beliefs follow Bayes' rule on the equilibrium path. After a disclosed signal, the receiver's log-odds include $\ell^P$ as in Equation~\eqref{eq:belief_disclose} whether or not disclosure is on the path, because the signal is verifiable. After non-disclosure on the path, the inference term in Equation~\eqref{eq:belief_manipulate} is the log-likelihood ratio of the event that the sender manipulates. Off the path, that is, when every type discloses, we allow the receiver to attribute non-disclosure to any distribution over the sender's signals. On and off the path, the inference term therefore lies in $[\ell_{\min},\ell_{\max}]$, and the results below hold for every such off-path belief. We focus on thresholds $\bar{p}$ close to $1$, where the tail results of the general model apply.

\begin{proposition}\label{prop:equilibrium_threshold}
		There exists $\bar{p}_0<1$ such that for every $\bar{p}\geq\bar{p}_0$:
		\begin{enumerate}[label=(\roman*)]
			\item \textbf{Cutoff.} A PBE exists, and in every PBE there is $c^*\in(\ell_{\min},\ell_{\max}]$ such that the sender manipulates if $\ell^P<c^*$ and discloses if $\ell^P>c^*$. In particular, full disclosure is not a PBE.
			\item \textbf{Disclosure vanishes.} There exists $\bar{p}_H\in[\bar{p}_0,1)$ such that for $\bar{p}\geq\bar{p}_H$ the unique PBE has $c^*=\ell_{\max}$: every type manipulates.
		\end{enumerate}
	\end{proposition}

	Under disclosure, the receiver's log-odds shift by $\ell^P$: the stronger the sender's evidence, the easier it is for the receiver's belief to cross the threshold $\bar{p}$, so that the receiver chooses $a=1$. Under manipulation, the sender forgoes this shift $\ell^P$, is inferred to hold evidence below the cutoff, and incurs the indirect penalty $-\lambda\Delta$, but gains a heavier right tail of the posterior distribution. The cutoff separates the senders whose evidence is worth more than the tail advantage from those whose evidence is not. There is, however, a second effect: the sender's evidence also moves his own belief about the state, and since manipulation works best in state~$1$, stronger evidence raises the appeal of manipulation as well. Part~(i) shows that for a demanding receiver the effect through the receiver's belief dominates, so the disclosure decision is monotone in the evidence. The cutoff characterization is in line with experiments on narrative persuasion: embedding facts in a story raises persuasion when the facts are weak and lowers it when they are strong \citep{krause2020strategic}, and narrative processing outperforms analytical processing only when the arguments are weak \citep{escalas2007self}. Lastly, full disclosure does not arise for a demanding receiver: the sender with the weakest evidence is at worst inferred as such whether he discloses or not, so manipulation costs him only the indirect effect, which the tail advantage outweighs as shown in Proposition~\ref{prop:general_g}(ii). 

	Part~(ii) is the counterpart of Proposition~\ref{prop:general_g}(ii) in the game. As the receiver becomes more demanding, crossing the threshold requires an extreme posterior, and extreme posteriors come from recalling many favorable signals, which manipulation makes more likely. This advantage compounds with the threshold while the disclosed evidence adds only a bounded shift to the receiver's belief, so eventually even the sender with the strongest evidence prefers manipulation, and no information is transmitted through the sender's action.

	\subsubsection{Receiver's incentive to listen}
	We now extend the game and assume that the receiver can decide whether to listen to the sender. Call the receiver's own reading of her experience her \emph{impression}: a signal $s^R$ drawn from $f_\omega$, independently of her memory. The impression is fleeting: it guides the decision to listen but is no longer available when the receiver acts. The timeline is as before, with stage~3 followed by the following stages in place of stages~4 and~5:
	\begin{enumerate}
		\item[3$'$.] The receiver observes her impression $s^R$, updates her belief by Bayes' rule, and decides whether to listen to the sender; the sender observes the receiver's decision.
		\item[4$'$.] If the receiver listens, the sender discloses or manipulates as in stage~4. If the reciever does not, the sender makes no choice.
		\item[5$'$.] The receiver recalls her memory---with the retrieval probabilities induced by the sender's choice if she listened, and with probability $r$ for every signal otherwise---and updates on her decision to listen, on what she observed from the sender if she listened, and on $\mathscr{M}_r$, but not on $s^R$, which she has forgotten. She then takes action $a=\mathds{1}[p_1\geq\bar{p}]$.
	\end{enumerate}
	The receiver's payoff is $1-\bar{p}$ from $a=1$ in state~1, $-\bar{p}$ from $a=1$ in state~0, and $0$ from $a=0$, so she chooses $a=1$ if and only if her posterior belief is at least $\bar{p}$, as before. Although the receiver forgets her impression $s^R$, her listening decision carries information about it, and she updates on that decision in the same way as on non-disclosure in Equation~\eqref{eq:belief_manipulate}: her log-odds shift, if she listened, by the term
	\begin{equation*}
		\ell_{\text{listen}}\;\equiv\;\log\frac{\Pr(\text{listen}\mid\omega=1)}{\Pr(\text{listen}\mid\omega=0)},
	\end{equation*}
	and if she did not, by the term $\ell_{\text{not}}$, defined in the same way with the event that she does not listen; both are computed from her own strategy. We assume that after an off-path listening decision, she attributes not listening to the lowest impression and listening to the highest, so that $\ell_{\text{not}}=\inf_s\ell(s)$ and $\ell_{\text{listen}}=\sup_s\ell(s)$; these are the limits of the on-path values under a cutoff rule as the cutoff approaches the ends of the impression's range.

	The extension describes a receiver who has some idea of what her experience says, albeit an imperfect and fleeting one, and decides on that basis whether to expose herself to the sender. A voter who senses that her own experience speaks for a policy may seek out the politician's message; one who senses the opposite may tune out. Listening exposes her to both faces of the sender: the evidence he may disclose and the recall he may manipulate.

	\begin{proposition}[Listening and the echo chamber]\label{prop:listen}
		There exists $\bar{p}_E<1$ such that for every $\bar{p}\geq\bar{p}_E$:
		\begin{enumerate}[label=(\roman*)]
			\item \textbf{Existence.} A perfect Bayesian equilibrium exists.
			\item \textbf{Cutoff rule.} In every perfect Bayesian equilibrium, every sender type manipulates whenever the receiver listens, and there exists $\ell^*\in[\inf_s\ell(s),\sup_s\ell(s)]$ such that the receiver listens on $\{\ell(s^R)>\ell^*\}$ and does not listen on $\{\ell(s^R)<\ell^*\}$.
			\item \textbf{Echo chamber.} In every perfect Bayesian equilibrium, $\Pr(\omega=1\mid\emph{listen})\geq\tfrac{1}{2}\geq\Pr(\omega=1\mid\emph{not listen})$, strictly whenever both decisions have positive probability.
		\end{enumerate}
	\end{proposition}

	Part~(i) is existence. With the sender's choice between disclosure and manipulation pinned down by Proposition~\ref{prop:equilibrium_threshold}, the receiver's cutoff is a fixed point of a continuous map from the range of $\ell(s^R)$, a closed interval, into itself, and such a fixed point exists by the intermediate value theorem.

	Part~(ii) is the cutoff rule. For a demanding receiver, every sender type manipulates when listened to, by Proposition~\ref{prop:equilibrium_threshold}, and the heavier right tail then raises the chance of acting in the sender's favor in both states, so the receivers who listen are those whose impressions favor state~1: their interests are aligned with the sender's. Part~(iii) is the echo chamber: favorable impressions are more likely in state~1, so the sender's audience is skewed toward the state he wants the receiver to believe in, and the receivers he reaches are those already inclined to believe him. Since manipulation boosts the recall of favorable signals, which are plentiful in state~1 and scarce in state~0, this is also the audience on which it works best. The echo chamber arises without any taste for confirmation: the receiver is Bayesian and listens only because, given her impression, listening raises her expected payoff. It is consistent with the robust preference for information that supports one's existing views, documented in experiments \citep{hart2009feeling}, and with partisans' choice of news outlets \citep{iyengar2009red}, but it does not require that preference, in line with theories in which Bayesian decision-makers optimally seek information that confirms strong beliefs \citep{che2019optimal}. It also has an empirical implication. Because the audience selects on impressions that are correlated with the state, comparing the actions of listeners and non-listeners overstates the effect of the sender's message, as recognized by studies that rely on exogenous variation in exposure, such as cable channel positions \citep{martin2017bias}.



\section{Conclusion}

This paper studies persuasion that works not by supplying evidence but by changing which evidence comes to mind. A Bayesian decision-maker who understands that recall is manipulated corrects for it, and this correction disciplines what manipulation can achieve: manipulation cannot move the average posterior, and it backfires on individuals with few memories, whose sparse recall reads as evidence against the favored state. But the correction stays bounded while recalled evidence accumulates, so manipulation is effective for large memories, thickens the right tail of beliefs and, under Poisson memory, lowers its hazard rate. In the first application, a monopolist selling an expensive experience good manipulates consumers' recall and, because manipulation makes demand less elastic at high prices, charges a higher price. In the second, a politician facing a demanding voter withholds all but his strongest evidence and manipulates recall instead; when voters choose whether to listen, only those whose own impressions already favor him do, creating an echo chamber. Given the growing evidence that memory shapes belief formation, the model is a first step toward understanding communication that works through recall, and we encourage further theoretical and empirical work on this channel.

\newpage
\bibliographystyle{econ}
\bibliography{reference}

@article{vafeiadis2018narratives,
	title={Narratives in political advertising: An analysis of the political advertisements in the 2014 midterm elections},
	author={Vafeiadis, Michail and Li, Ruobing and Shen, Fuyuan},
	journal={Journal of Broadcasting \& Electronic Media},
	volume={62},
	number={2},
	pages={354--370},
	year={2018},
	publisher={Taylor \& Francis}
}

@article{crawford1982strategic,
	title={Strategic information transmission},
	author={Crawford, Vincent P and Sobel, Joel},
	journal={Econometrica},
	pages={1431--1451},
	year={1982},
	publisher={JSTOR}
}

@article{kamenica2011bayesian,
	title={Bayesian persuasion},
	author={Kamenica, Emir and Gentzkow, Matthew},
	journal={American Economic Review},
	volume={101},
	number={6},
	pages={2590--2615},
	year={2011},
	publisher={American Economic Association}
}

@article{zhao2014remembering,
	title={Remembering the good old days: The moderating role of consumer affective state on the effectiveness of nostalgic advertising},
	author={Zhao, Guangzhi and Muehling, Darrel D and Kareklas, Ioannis},
	journal={Journal of Advertising},
	volume={43},
	number={3},
	pages={244--255},
	year={2014},
	publisher={Taylor \& Francis}
}

@article{thoma2025nostalgic,
	title={Nostalgic Advertising Enhances Brand Name Recall by Reactivating Brand-Related Autobiographical Memories, Especially for Familiar and Personally Relevant Brands},
	author={Thoma, Dieter and Koziak, Jessica},
	journal={Psychology \& Marketing},
	volume={42},
	number={12},
	pages={3306--3318},
	year={2025},
	publisher={Wiley Online Library}
}

@article{fetterman2025topics,
	title={The topics of nostalgic recall: The benefits of nostalgia depend on the topics that one recalls},
	author={Fetterman, Adam K and Evans, Nicholas D and Ravey, Eriksen P and Henderson, Perla R and Tran, Bao Han L and Boyd, Ryan L},
	journal={Social Psychological and Personality Science},
	volume={16},
	number={4},
	pages={444--456},
	year={2025},
	publisher={SAGE Publications Sage CA: Los Angeles, CA}
}

@article{tversky1973availability,
	title={Availability: A heuristic for judging frequency and probability},
	author={Tversky, Amos and Kahneman, Daniel},
	journal={Cognitive psychology},
	volume={5},
	number={2},
	pages={207--232},
	year={1973},
	publisher={Elsevier}
}

@article{schwarz1991ease,
	title={Ease of retrieval as information: Another look at the availability heuristic.},
	author={Schwarz, Norbert and Bless, Herbert and Strack, Fritz and Klumpp, Gisela and Rittenauer-Schatka, Helga and Simons, Annette},
	journal={Journal of Personality and Social psychology},
	volume={61},
	number={2},
	pages={195},
	year={1991},
	publisher={American Psychological Association}
}

@article{kassin1996social,
	title={The social psychology of false confessions: Compliance, internalization, and confabulation},
	author={Kassin, Saul M and Kiechel, Katherine L},
	journal={Psychological science},
	volume={7},
	number={3},
	pages={125--128},
	year={1996},
	publisher={SAGE Publications Sage CA: Los Angeles, CA}
}

@article{garry1996imagination,
	title={Imagination inflation: Imagining a childhood event inflates confidence that it occurred},
	author={Garry, Maryanne and Manning, Charles G and Loftus, Elizabeth F and Sherman, Steven J},
	journal={Psychonomic bulletin \& review},
	volume={3},
	number={2},
	pages={208--214},
	year={1996},
	publisher={Springer}
}

@article{loftus1979malleability,
	title={The malleability of human memory: Information introduced after we view an incident can transform memory},
	author={Loftus, Elizabeth F},
	journal={American scientist},
	volume={67},
	number={3},
	pages={312--320},
	year={1979},
	publisher={JSTOR}
}

@article{liv2020deep,
	title={Deep fakes and memory malleability: False memories in the service of fake news},
	author={Liv, Nadine and Greenbaum, Dov},
	journal={AJOB neuroscience},
	volume={11},
	number={2},
	pages={96--104},
	year={2020},
	publisher={Taylor \& Francis}
}

@techreport{andre2024narratives,
  title={Narratives about the Macroeconomy},
  author={Andre, Peter and Haaland, Ingar and Roth, Christopher and Wiederholt, Mirko and Wohlfart, Johannes},
  year={2024},
  institution={SAFE Working Paper}
}

@techreport{barron2024narrative,
  title={Narrative persuasion},
  author={Barron, Kai and Fries, Tilman},
  year={2024},
  institution={WZB Discussion Paper}
}

@article{schwartzstein2021using,
  title={Using models to persuade},
  author={Schwartzstein, Joshua and Sunderam, Adi},
  journal={American Economic Review},
  volume={111},
  number={1},
  pages={276--323},
  year={2021},
  publisher={American Economic Association 2014 Broadway, Suite 305, Nashville, TN 37203}
}

@article{eliaz2020model,
  title={A model of competing narratives},
  author={Eliaz, Kfir and Spiegler, Ran},
  journal={American Economic Review},
  volume={110},
  number={12},
  pages={3786--3816},
  year={2020},
  publisher={American Economic Association 2014 Broadway, Suite 305, Nashville, TN 37203}
}

@article{green2004transportation,
  title={Transportation into narrative worlds: The role of prior knowledge and perceived realism},
  author={Green, Melanie C},
  journal={Discourse processes},
  volume={38},
  number={2},
  pages={247--266},
  year={2004},
  publisher={Taylor \& Francis}
}

@article{hamby2017reflecting,
  title={Reflecting on the journey: Mechanisms in narrative persuasion},
  author={Hamby, Anne and Brinberg, David and Daniloski, Kim},
  journal={Journal of Consumer Psychology},
  volume={27},
  number={1},
  pages={11--22},
  year={2017},
  publisher={Elsevier}
}

@article{enke2024associative,
  title={Associative memory, beliefs and market interactions},
  author={Enke, Benjamin and Schwerter, Frederik and Zimmermann, Florian},
  journal={Journal of Financial Economics},
  volume={157},
  pages={103853},
  year={2024},
  publisher={Elsevier}
}

@techreport{leung2024simple,
	title={A Simple Model of Memory-Based Belief Formation},
	author={Leung, Benson Tsz Kin},
	year={2023},
	institution={Working Paper}
}

@article{roos2024narratives,
	title={Narratives in economics},
	author={Roos, Michael and Reccius, Matthias},
	journal={Journal of Economic Surveys},
	volume={38},
	number={2},
	pages={303--341},
	year={2024},
	publisher={Wiley Online Library}
}

@article{bordalo2022overreaction,
	title={Overreaction and diagnostic expectations in macroeconomics},
	author={Bordalo, Pedro and Gennaioli, Nicola and Shleifer, Andrei},
	journal={Journal of Economic Perspectives},
	volume={36},
	number={3},
	pages={223--244},
	year={2022},
	publisher={American Economic Association 2014 Broadway, Suite 305, Nashville, TN 37203-2418}
}

@article{bordalo2020belief,
	title={Belief overreaction and stock market puzzles},
	author={Bordalo, Pedro and Gennaioli, Nicola and La Porta, Rafael and Shleifer, Andrei},
	journal={Journal of Political Economy},
	volume={132},
	number={5},
	pages={1450--1484},
	year={2024},
	publisher={The University of Chicago Press}
}

@article{graeber2022stories,
	title={Stories, statistics, and memory},
	author={Graeber, Thomas and Roth, Christopher and Zimmermann, Florian},
	journal={The Quarterly Journal of Economics},
	volume={139},
	number={4},
	pages={2181--2225},
	year={2024},
	publisher={Oxford University Press}
}

@article{bordalo2023imagining,
	title={Imagining the Future: Memory, Simulation, and Beliefs},
	author={Bordalo, Pedro and Burro, Giovanni and Coffman, Katherine and Gennaioli, Nicola and Shleifer, Andrei},
	journal={The Review of Economic Studies},
	volume={92},
	number={3},
	pages={1532--1563},
	year={2025},
	publisher={Oxford University Press}
}

@article{bordalo2023memory,
	title={Memory and probability},
	author={Bordalo, Pedro and Conlon, John J and Gennaioli, Nicola and Kwon, Spencer Y and Shleifer, Andrei},
	journal={The Quarterly Journal of Economics},
	volume={138},
	number={1},
	pages={265--311},
	year={2023},
	publisher={Oxford University Press}
}

@article{shiller2017narrative,
	title={Narrative economics},
	author={Shiller, Robert J},
	journal={American economic review},
	volume={107},
	number={4},
	pages={967--1004},
	year={2017},
	publisher={American Economic Association 2014 Broadway, Suite 305, Nashville, TN 37203}
}

@book{jensen1995saddlepoint,
	title={Saddlepoint Approximations},
	author={Jensen, Jens Ledet},
	year={1995},
	publisher={Oxford University Press},
	series={Oxford Statistical Science Series}
}

@article{myerson1981optimal,
	title={Optimal Auction Design},
	author={Myerson, Roger B.},
	journal={Mathematics of Operations Research},
	volume={6},
	number={1},
	pages={58--73},
	year={1981}
}

@article{bagnoli2005log,
	title={Log-concave Probability and Its Applications},
	author={Bagnoli, Mark and Bergstrom, Ted},
	journal={Economic Theory},
	volume={26},
	number={2},
	pages={445--469},
	year={2005}
}

@book{dembo1998large,
	title={Large Deviations Techniques and Applications},
	author={Dembo, Amir and Zeitouni, Ofer},
	edition={2nd},
	year={1998},
	publisher={Springer},
	address={New York}
}

@article{ofir2000ease,
	title={Ease of Recall vs Recalled Evidence in Judgment: Experts vs Laymen},
	author={Ofir, Chezy},
	journal={Organizational Behavior and Human Decision Processes},
	volume={81},
	number={1},
	pages={28--42},
	year={2000}
}

@article{ofir2008memory,
	title={Memory-Based Store Price Judgments: The Role of Knowledge and Shopping Experience},
	author={Ofir, C. and Raghubir, P. and Brosh, G. and Monroe, K. B. and Heiman, A.},
	journal={Journal of Retailing},
	volume={84},
	number={4},
	pages={414--423},
	year={2008}
}

@article{weingarten2018does,
	title={Does Ease Mediate the Ease-of-Retrieval Effect? A Meta-Analysis},
	author={Weingarten, Evan and Hutchinson, J. Wesley},
	journal={Psychological Bulletin},
	volume={144},
	number={3},
	pages={227--283},
	year={2018}
}

@article{muehling2014exploring,
	title={Exploring the Boundaries of Nostalgic Advertising Effects: A Consideration of Childhood Brand Exposure and Attachment on Consumers' Responses to Nostalgia-Themed Advertisements},
	author={Muehling, Darrel D. and Sprott, David E. and Sultan, Abdullah J.},
	journal={Journal of Advertising},
	volume={43},
	number={1},
	pages={73--84},
	year={2014}
}

@article{lewis2014advertising,
	title={Advertising Effectively Influences Older Users: How Field Experiments Can Improve Measurement and Targeting},
	author={Lewis, Randall A. and Reiley, David H.},
	journal={Review of Industrial Organization},
	volume={44},
	number={2},
	pages={147--159},
	year={2014}
}

@article{krause2020strategic,
	title={Strategic Storytelling: When Narratives Help Versus Hurt the Persuasive Power of Facts},
	author={Krause, Rebecca J. and Rucker, Derek D.},
	journal={Personality and Social Psychology Bulletin},
	volume={46},
	number={2},
	pages={216--227},
	year={2020}
}

@article{escalas2007self,
	title={Self-Referencing and Persuasion: Narrative Transportation versus Analytical Elaboration},
	author={Escalas, Jennifer Edson},
	journal={Journal of Consumer Research},
	volume={33},
	number={4},
	pages={421--429},
	year={2007}
}

@article{daniels1954saddlepoint,
  author = {Daniels, Henry E.},
  title = {Saddlepoint Approximations in Statistics},
  journal = {The Annals of Mathematical Statistics},
  volume = {25},
  number = {4},
  pages = {631--650},
  year = {1954}
}

@article{lugannani1980saddle,
  author = {Lugannani, Robert and Rice, Stephen},
  title = {Saddle Point Approximation for the Distribution of the Sum of Independent Random Variables},
  journal = {Advances in Applied Probability},
  volume = {12},
  number = {2},
  pages = {475--490},
  year = {1980}
}

@article{guitart2021impact,
	title={The Impact of Informational and Emotional Television Ad Content on Online Search and Sales},
	author={Guitart, Ivan A. and Stremersch, Stefan},
	journal={Journal of Marketing Research},
	volume={58},
	number={2},
	pages={299--320},
	year={2021}
}

@article{kim2016narrative,
	title={Narrative-Transportation Storylines in Luxury Brand Advertising: Motivating Consumer Engagement},
	author={Kim, Jae-Eun and Lloyd, Stephen and Cervellon, Marie-C{\'e}cile},
	journal={Journal of Business Research},
	volume={69},
	number={1},
	pages={304--313},
	year={2016}
}

@article{hart2009feeling,
	title={Feeling Validated Versus Being Correct: A Meta-Analysis of Selective Exposure to Information},
	author={Hart, William and Albarrac{\'i}n, Dolores and Eagly, Alice H. and Brechan, Inge and Lindberg, Matthew J. and Merrill, Lisa},
	journal={Psychological Bulletin},
	volume={135},
	number={4},
	pages={555--588},
	year={2009}
}

@article{iyengar2009red,
	title={Red Media, Blue Media: Evidence of Ideological Selectivity in Media Use},
	author={Iyengar, Shanto and Hahn, Kyu S.},
	journal={Journal of Communication},
	volume={59},
	number={1},
	pages={19--39},
	year={2009}
}

@article{martin2017bias,
	title={Bias in Cable News: Persuasion and Polarization},
	author={Martin, Gregory J. and Yurukoglu, Ali},
	journal={American Economic Review},
	volume={107},
	number={9},
	pages={2565--2599},
	year={2017}
}

@article{che2019optimal,
	title={Optimal Dynamic Allocation of Attention},
	author={Che, Yeon-Koo and Mierendorff, Konrad},
	journal={American Economic Review},
	volume={109},
	number={8},
	pages={2993--3029},
	year={2019}
}

\newpage

\appendix
\setcounter{equation}{0}
\renewcommand{\theequation}{\Alph{section}.\arabic{equation}}
\section{Proofs}\label{sec:proofs}

	\subsection{Proof of Lemma~\ref{lemma:martingale}}
	\begin{proof}
		We show that $E[p_1(\mathscr{M}_r)]=\Pr(\omega=1)=\frac{1}{2}$. Let $\mathscr{M}_r$ denote the (random) set of recalled signals. The expectation is taken over three sources of randomness: the true state $\omega$, the memory $\mathscr{M}$ (whose size $M\sim g$ and whose signals are i.i.d.\ from $f_\omega$), and the recall outcome (each signal $s\in\mathscr{M}$ is independently recalled with probability $r(s)$). The posterior $p_1(\mathscr{M}_r)=E[\mathds{1}[\omega=1]\mid\mathscr{M}_r]$, and therefore:
		\begin{equation*}
			E[p_1(\mathscr{M}_r)] = E\!\left[E[\mathds{1}[\omega=1]\mid\mathscr{M}_r]\right] = E[\mathds{1}[\omega=1]] = \Pr(\omega=1) = \frac{1}{2},
		\end{equation*}
		where the first equality is the definition of the posterior, the second applies the law of iterated expectations, and the third is by the prior. The result follows.

	\end{proof}

	\subsection{Proof of Proposition~\ref{prop:general_g}}
	\begin{proof}
		\textbf{Part~(i): Backfire at small $M$ and effective at large $M$.} 
		
		\emph{Backfire at small $M$.} When $M=0$, $\mathscr{M}_r=\emptyset$ deterministically. From Equation~\eqref{eq:memorybelief_complex1}, the log-odds equals $\log\frac{E_n[\overline{R}_1^n]}{E_n[\overline{R}_0^n]}$, where $\overline{R}_\omega=\int_S f_\omega(s)(1-r(s))\,ds$. We show $\overline{R}_1<\overline{R}_0$:
		\begin{equation}\label{eq:p1_Rbar}
			\overline{R}_1-\overline{R}_0=\int(f_1(s)-f_0(s))(1-r(s))\,ds<(1-r)\int(f_1-f_0)\,ds=0.
		\end{equation}
		The strict inequality holds because $(1-r(s))$ covaries negatively with $(f_1(s)-f_0(s))$ under manipulation. Since $0<\overline{R}_1<\overline{R}_0\leq 1$ and $g(n)>0$ for some $n\geq 1$ (by non-degeneracy), we have $E_n[\overline{R}_1^n]<E_n[\overline{R}_0^n]$, so $p_1(\emptyset)<\frac{1}{2}$.

		\emph{Effective at large $M$.} Fix the memory size $M=m$ and a state~$\omega$, and recall from Equation~\eqref{eq:memorybelief_complex1} that the log-odds posterior can be decomposed into two terms capturing the direct and indirect effect, i.e., $L=D+I(k)$ where the \emph{direct effect} $D=\sum_{s\in\mathscr{M}_r}\ell(s)$ sums the log-likelihood ratios of the recalled signals, while the \emph{indirect correction} $I(k)$ depends only on the number of recalled signals, $k\equiv|\mathscr{M}_r|\leq m$. Because each of the $m$ signals is recalled independently, $D$ is a sum of $m$ i.i.d.\ terms $X=\ell(s)\,\mathds{1}[\text{signal }s\text{ recalled}]$, with moment generating function
		\begin{equation*}
			\psi_\omega(t)\;\equiv\;E\!\left[e^{tX}\mid\omega\right]\;=\;E_\omega\!\left[r(s)e^{t\ell(s)}+\bigl(1-r(s)\bigr)\right].
		\end{equation*}
		First, we show that the indirect correction is bounded uniformly in the recalled count, such that $|I(k)|/k\rightarrow 0$ as $k\rightarrow \infty$. Every term in the numerator of
		\begin{equation*}
			I(k)=\log\frac{\sum_{n\geq k}\binom{n}{k}\overline{R}_1^{\,n-k}g(n)}{\sum_{n\geq k}\binom{n}{k}\overline{R}_0^{\,n-k}g(n)}
		\end{equation*}
		is nonnegative, so the numerator is at least the $n=k$ term, $g(k)$. For $k\geq n_0$, writing $n=k+j$ and using $\binom{k+j}{k}=\prod_{i=1}^{j}(k+i)/j!$ and Assumption~\ref{ass:light_tail},
		\begin{equation*}
			\binom{k+j}{k}\frac{g(k+j)}{g(k)}=\frac{1}{j!}\prod_{i=1}^{j}\frac{(k+i)\,g(k+i)}{g(k+i-1)}\leq\frac{\overline{C}^{\,j}}{j!},
		\end{equation*}
		so the denominator inside the log operator of $I(k)$ is at most $g(k)\sum_{j\geq 0}(\overline{C}\,\overline{R}_0)^{j}/j!$, that is, $g(k)\,e^{\overline{C}\,\overline{R}_0}$. Hence, combined with the observation that the numerator is at least $g(k)$, we have
		\begin{equation}\label{eq:p1_Ibound}
			-\overline{C}\,\overline{R}_0\;\leq\;I(k)\;\leq\;0\qquad\text{for every }k\geq n_0\text{ with }g(k)>0
		\end{equation}
		(if $g(k)=0$ for some $k\geq n_0$, the assumption implies $g=0$ beyond $k$, so such $k$ are never observed). The finitely many remaining values $I(k)$, $k<n_0$, are finite as well. The correction is the log of an \emph{average} over the possible numbers of unrecalled signals, $I(k)=\log E_{\pi_k}\big[(\overline{R}_1/\overline{R}_0)^{M-k}\big]$, where $\pi_k(n)\propto\binom{n}{k}\overline{R}_0^{\,n-k}g(n)$ is the DM's posterior over his memory size after recalling $k$ signals in state~0; by Jensen's inequality, $|I(k)|\leq E_{\pi_k}[M-k]\cdot\log(\overline{R}_0/\overline{R}_1)$, the expected number of unrecalled signals times the penalty per unrecalled signal, which is finite because $M$ has finite moments under Assumption~\ref{ass:light_tail} and $0<\overline{R}_1<\overline{R}_0$. Thus $\sup_k|I(k)|<\infty$, and in particular $|I(k)|=o(m)$ uniformly over $k\leq m$.
		
		Next, we show that $E(p_1\mid m)-\tfrac{1}{2}>0$ as $m\rightarrow \infty$. As shown above, $\sup_k|I(k)|<\infty$ allows us to focus on analyzing the direct effect $D$. Writing $A_\omega(m)\equiv E(p_1\mid m,\omega)$ for the average posterior in state~$\omega$ and averaging over the two equally likely states,
		\begin{equation}\label{eq:p1_decomp}
			E(p_1\mid m)-\tfrac{1}{2}=\tfrac{1}{2}\Big[\,\underbrace{A_0(m)}_{\text{state-0 error (believing 1)}}-\underbrace{(1-A_1(m))}_{\text{state-1 error (believing 0)}}\,\Big].
		\end{equation}
		Let $\bar{\ell}_\omega\equiv\psi_\omega'(0)=\int r(s)f_\omega(s)\ell(s)\,ds$ denote the per-signal drift of $D$ in state~$\omega$. In state~1 this drift is positive: writing $r(s)=r+\Delta r(s)$, we have $\Delta r(s)\,\ell(s)\geq0$, so $\bar{\ell}_1=r\!\int f_1\ell\,ds+\int\Delta r\,f_1\ell\,ds\geq r\!\int f_1\ell\,ds>0$. Since $|I(k)|\leq\bar{I}\equiv\sup_k|I(k)|<\infty$ by the first step, $L/m=D/m+I(k)/m\to\bar{\ell}_\omega$ almost surely in state~$\omega$; hence the state-1 belief converges to the truth, $A_1(m)\to1$. If $\bar{\ell}_0>0$ the same argument gives $A_0(m)\to 1$, and if $\bar{\ell}_0=0$ then $D/\sqrt{m}$ is asymptotically normal while $I(k)$ stays bounded, so $A_0(m)\to\tfrac{1}{2}$; in either case $A_0(m)$ stays bounded away from~$0$ and, combined with $A_1(m)\to1$, $E(p_1\mid m)>\tfrac{1}{2}$ for large~$m$ a fortiori. The substantive case is $\bar{\ell}_0<0$, where both errors vanish and which one dominates is decided by their convergent rates. Now we bound their convergent rate. Denote $\sigma(x)\equiv 1/(1+e^{-x})$ the logistic function, so that $p_1=\sigma(L)$, and $\operatorname{logit}(p)\equiv\log\frac{p}{1-p}$ its inverse. Since $-\bar{I}\leq I(k)\leq 0$, the log-odds satisfy $D-\bar{I}\leq L\leq D$. Combining this with $\tfrac{1}{2}\min(1,e^{x})\leq\sigma(x)\leq\min(1,e^{x})$ gives
		\begin{equation*}
			\begin{split}
					\sigma(L)&\;\geq\;\tfrac{1}{2}\mathds{1}[L\geq 0]\;\geq\;\tfrac{1}{2}\mathds{1}[D\geq\bar{I}],\\
				\sigma(L)&\;\leq\;\mathds{1}[L\geq 0]+e^{L}\mathds{1}[L<0]\;\leq\;\mathds{1}[D\geq 0]+e^{D}\mathds{1}[D<\bar{I}].
			\end{split}
		\end{equation*}
		Taking expectations under $\omega=0$,
		\begin{equation}\label{eq:p1_sandwich0}
			\tfrac{1}{2}\Pr\nolimits_0(D\geq\bar{I})\;\leq\;A_0(m)\;\leq\;\Pr\nolimits_0(D\geq 0)+E_0\!\left[e^{D}\mathds{1}[D<\bar{I}]\right],
		\end{equation}
		and symmetrically
		\begin{equation}\label{eq:p1_sandwich1}
			\tfrac{1}{2}\Pr\nolimits_1(D\leq 0)\;\leq\;1-A_1(m)\;\leq\;\Pr\nolimits_1(D\leq\bar{I})+e^{\bar{I}}E_1\!\left[e^{-D}\mathds{1}[D>0]\right].
		\end{equation}
		
		Consider first the state-0 error. For the lower end $\tfrac{1}{2}\Pr\nolimits_0(D\geq\bar{I})$ of \eqref{eq:p1_sandwich0}, Cram\'er's theorem, i.e., the large-deviation principle for sums of i.i.d.\ terms \citep[Ch.~2]{dembo1998large}, gives
		\begin{equation}\label{eq:p1_rate0}
			\tfrac{1}{m}\log\big[\tfrac{1}{2}\Pr\nolimits_0(D\geq\bar{I})\big]\;\to\;-\rho_0,\qquad \rho_0\equiv-\log\min_t\psi_0(t),
		\end{equation}
		where $\rho_0$ is the rate function of $D/m$ at zero; a fixed threshold such as $\bar{I}$ does not affect the rate, and $\rho_0>0$ because $\psi_0(0)=1$ while $\psi_0'(0)=\bar{\ell}_0<0$. For the upper end of \eqref{eq:p1_sandwich0}, the first term $\Pr_0(D\geq 0)$ has the same rate by Cram\'er's theorem, and we bound the rate of the second term as follows. On $\{D<\bar{I}\}$, $e^{D}=e^{tD}e^{(1-t)D}\leq e^{(1-t)\bar{I}}e^{tD}$ for any $t\leq 1$, so $E_0[e^{D}\mathds{1}[D<\bar{I}]]\leq e^{(1-t)\bar{I}}\psi_0(t)^m$ and
		\begin{equation}\label{eq:p1_limsup0}
			\limsup_{m\to\infty}\tfrac{1}{m}\log E_0\!\left[e^{D}\mathds{1}[D<\bar{I}]\right]\;\leq\;\log\psi_0(t)\qquad\text{for every }t\leq 1.
		\end{equation}
		The right-hand side is minimized at $t=t^\ast$, the minimizer of $\psi_0$, where it equals $-\rho_0$; hence, if $t^\ast\leq 1$, \eqref{eq:p1_limsup0} bounds the rate of $E_0\!\left[e^{D}\mathds{1}[D<\bar{I}]\right]$ by $-\rho_0$. We now argue $t^\ast\leq 1$ holds: $\psi_0$ being a mixture of exponentials is strictly convex, so $t^\ast$ is unique; $\psi_0'(0)=\bar{\ell}_0<0$ implies that $t^\ast>0$; and combined with $\psi_0(1)=\int r f_1\,ds+\overline{R}_0=1+\overline{R}_0-\overline{R}_1>1=\psi_0(0)$ (using $f_1=e^{\ell}f_0$) implies that $t^\ast\in(0,1)$. Both terms of the upper end of \eqref{eq:p1_sandwich0} therefore decay at least at rate $\rho_0$, hence so does their sum, and \eqref{eq:p1_sandwich0} yields $\tfrac{1}{m}\log A_0(m)\to-\rho_0$.

		Next consider the state-1 error. For the lower end $\tfrac{1}{2}\Pr\nolimits_1(D\leq 0)$ of \eqref{eq:p1_sandwich1}, Cram\'er's theorem gives
		\begin{equation}\label{eq:p1_rate1}
			\tfrac{1}{m}\log\big[\tfrac{1}{2}\Pr\nolimits_1(D\leq 0)\big]\;\to\;-\rho_1,\qquad \rho_1\equiv-\log\min_t\psi_1(t),
		\end{equation}
		where $\rho_1>0$ because $\psi_1(0)=1$ while $\psi_1'(0)=\bar{\ell}_1>0$. For the upper end of \eqref{eq:p1_sandwich1}, the first term $\Pr_1(D\leq\bar{I})$ has the same rate by Cram\'er's theorem, and we bound the rate of the second term as follows. On $\{D>0\}$, $e^{-D}=e^{tD}e^{-(1+t)D}\leq e^{tD}$ for any $t\geq -1$, so $e^{\bar{I}}E_1[e^{-D}\mathds{1}[D>0]]\leq e^{\bar{I}}\psi_1(t)^m$ and
		\begin{equation}\label{eq:p1_limsup1}
			\limsup_{m\to\infty}\tfrac{1}{m}\log\Big(e^{\bar{I}}E_1\!\left[e^{-D}\mathds{1}[D>0]\right]\Big)\;\leq\;\log\psi_1(t)\qquad\text{for every }t\geq -1.
		\end{equation}
		The right-hand side is minimized at the minimizer of $\psi_1$, where it equals $-\rho_1$; hence, if that minimizer is at least $-1$, \eqref{eq:p1_limsup1} bounds the rate of the second term by $-\rho_1$. We now argue that this holds. The two per-signal transforms under state 0 and 1 are the same function up to a shift: writing $\eta(t)\equiv E_0[r(s)e^{t\ell(s)}]$ and using $f_1=e^{\ell}f_0$ again,
		\begin{equation*}
			\psi_0(t)=\eta(t)+\overline{R}_0,\qquad \psi_1(t)=\eta(t+1)+\overline{R}_1,
		\end{equation*}
		so with $\eta^\ast\equiv\min_t\eta(t)$, $\psi_0$ is minimized at $t^\ast$ with value $\eta^\ast+\overline{R}_0$ and $\psi_1$ at $t^\ast-1$ with value $\eta^\ast+\overline{R}_1$; since $t^\ast\in(0,1)$, the minimizer of $\psi_1$ lies in $(-1,0)$. Both terms of the upper end of \eqref{eq:p1_sandwich1} therefore decay at least at rate $\rho_1$, hence so does their sum, and \eqref{eq:p1_sandwich1} yields $\tfrac{1}{m}\log\big(1-A_1(m)\big)\to-\rho_1$.

		Finally, $\min_t\psi_0-\min_t\psi_1=(\eta^\ast+\overline{R}_0)-(\eta^\ast+\overline{R}_1)=\overline{R}_0-\overline{R}_1>0$, so $\rho_0<\rho_1$: the state-0 error decays strictly \emph{slower}. Therefore $A_0(m)>1-A_1(m)$ for all large $m$, and $E(p_1\mid m)>\tfrac{1}{2}$ for all large $m$ by \eqref{eq:p1_decomp}.

		\emph{Confident false beliefs (the ``moreover'' statement).} We show that sufficiently aggressive manipulation---formally, $\bar{\ell}_0=\int r(s)f_0(s)\ell(s)\,ds>0$---drives the DM to certainty in state~1 even when state~0 is true: $p_1\to 1$ almost surely as $M\to\infty$ conditional on $\omega=0$. Realize the memories on a single probability space by drawing an infinite i.i.d.\ sequence of signals and recall indicators, and let the memory at size $M$ consist of the first $M$ draws. Conditional on $\omega$, the strong law of large numbers gives $D/M\to\bar{\ell}_\omega$ almost surely, and $\bar{\ell}_\omega>0$ in \emph{both} states (by hypothesis in state~0, and by $\bar{\ell}_1>0$ shown above), so $D\to+\infty$ regardless of the true state. Since $|I(k)|\leq\bar{I}$, $L=D+I(k)\geq D-\bar{I}\to+\infty$ almost surely even under $\omega=0$, so $p_1=\sigma(L)\to 1$ almost surely.

		\textbf{Part~(ii): heavier right tail and lighter left tail.}

		\emph{Heavier right tail.} We show that $B^{\text{manip}}(p)<B^{\text{no}}(p)$ for all posterior belief $p$ with $B^{\text{no}}(p)\in(0,1)$ sufficiently close to the top of the posterior support, treating unbounded memory support (Steps~1--2) and bounded memory support (Step~2$'$) separately.

		\emph{Step~1 (unbounded support): reduce to $\omega=1$.} Write $1-B(p) = \frac{1}{2}\Pr(p_1>p\mid\omega=1) + \frac{1}{2}\Pr(p_1>p\mid\omega=0)$, and let $x\equiv\operatorname{logit}(p)$ denote the threshold in log-odds that corresponds to the belief threshold $p$. Since $p_1=\sigma(L)$ is monotone in the log-odds $L$, we have $\Pr(p_1>p\mid\omega)=\Pr(L>x\mid\omega)$. We consider $p>\frac{1}{2}$, so that $x>0$. Since the DM is Bayesian with a uniform prior, his posterior odds equal the likelihood ratio of what he observes, $e^{L}=L_1(\mathscr{M}_r)/L_0(\mathscr{M}_r)$, where $L_{\omega}$  follows Equation~\eqref{eq:Lomega_def} taking into account both the direct effect and indirect effect. Let $P_\omega$ denote the probability distribution of the recalled set $\mathscr{M}_r$ in state~$\omega$; its density is $\Pr(M\geq k)\,L_\omega(\mathscr{M}_r)$ with $k=|\mathscr{M}_r|$, so $dP_0=e^{-L}\,dP_1$. Hence, with and without manipulation,
		\begin{equation}\label{eq:p1_change_measure}
			\begin{split}
				\Pr(L>x\mid\omega=0)&=\int\mathds{1}[L>x]\,dP_0\\
				&=\int\mathds{1}[L>x]\,e^{-L}\,dP_1\\
				&=E\big[e^{-L}\mathds{1}[L>x]\mid\omega=1\big]\\
				&\leq e^{-x}\,E\big[\mathds{1}[L>x]\mid\omega=1\big]\\
				&=e^{-x}\Pr(L>x\mid\omega=1),
			\end{split}
		\end{equation}
		where the second line changes measure from state~0 to state~1 and the inequality uses $e^{-L}<e^{-x}$ on the event $\{L>x\}$. The equation implies that crossing a high threshold is exponentially less likely in state~0. Hence the difference in survival functions ($1$ minus the c.d.f.~of posterior belief) satisfies
		\begin{equation}\label{eq:p1_right_reduce}
			\begin{split}
				&\big[1-B^{\text{manip}}(p)\big]-\big[1-B^{\text{no}}(p)\big]\\
				&\quad=\tfrac{1}{2}\big[\Pr(L^{\text{manip}}>x\mid\omega=1)-\Pr(L^{\text{no}}>x\mid\omega=1)\big]\\
				&\quad\quad+\tfrac{1}{2}\big[\Pr(L^{\text{manip}}>x\mid\omega=0)-\Pr(L^{\text{no}}>x\mid\omega=0)\big]\\
				&\quad\geq\tfrac{1}{2}\big[\Pr(L^{\text{manip}}>x\mid\omega=1)-\Pr(L^{\text{no}}>x\mid\omega=1)\big]-\tfrac{1}{2}\Pr(L^{\text{no}}>x\mid\omega=0)\\
				&\quad\geq\tfrac{1}{2}\big[\Pr(L^{\text{manip}}>x\mid\omega=1)-\Pr(L^{\text{no}}>x\mid\omega=1)\big]-\tfrac{1}{2}e^{-x}\Pr(L^{\text{no}}>x\mid\omega=1)\\
				&\quad=\tfrac{1}{2}\Pr(L^{\text{no}}>x\mid\omega=1)\left[\frac{\Pr(L^{\text{manip}}>x\mid\omega=1)}{\Pr(L^{\text{no}}>x\mid\omega=1)}-1-e^{-x}\right],
			\end{split}
		\end{equation}
		where the first equality splits each survival function by state, the first inequality drops the nonnegative term $\Pr(L^{\text{manip}}>x\mid\omega=0)$, and the second applies \eqref{eq:p1_change_measure} to the benchmark. So for large $p$ the comparison reduces to the state-1 tails $\Pr(L>x\mid\omega=1)$ with and without manipulation: it suffices that their ratio exceed $1+e^{-x}$.

		\emph{Step~2 (unbounded support): the boost to the direct effect compounds.} Given Step~1, we condition on $\omega=1$ throughout this step and show that $\Pr(L^{\text{manip}}>x\mid\omega=1)/\Pr(L^{\text{no}}>x\mid\omega=1)\to\infty$ as $x\to\infty$ and the result follows. Call a signal \emph{favorable} if $\ell(s)>0$, let $F_1^{+}\equiv\Pr_1(\ell(s)>0)$, and let $T_m$ denote the sum of $m$ i.i.d.\ draws of $\ell(s)$ from $f_1$ conditioned on $\ell(s)>0$. Since $\ell$ is bounded (no single signal rules out a state), $T_m\leq m\sup_s\ell(s)$, so $\Pr(T_m>x)=0$ unless $m>x/\sup_s\ell(s)$; in particular, once $x\geq n_0\sup_s\ell(s)$, which we assume from now on, only $m\geq n_0$ contributes to the sums below.

		\emph{Benchmark.} Without manipulation $L^{\text{no}}=D^{\text{no}}$, which is at most the sum of the recalled favorable log-likelihood ratios. Each signal in memory is favorable and recalled with probability $F_1^{+}r$, independently across signals, and the values of the recalled favorable signals are i.i.d.\ draws from $f_1$ conditioned on $\ell>0$. Hence
		\begin{equation}\label{eq:tail_bench}
			\begin{split}
				\Pr(L^{\text{no}}>x\mid\omega=1)&\;\leq\;\sum_{m}\Big[\sum_{n\geq m}g(n)\binom{n}{m}(F_1^{+}r)^m(1-F_1^{+}r)^{n-m}\Big]\Pr(T_m>x)\\
				&\;\leq\;e^{\overline{C}}\sum_{m}g(m)\,(F_1^{+}r)^m\,\Pr(T_m>x),
			\end{split}
		\end{equation}
		where the bracket is the probability that exactly $m$ favorable signals are recalled. The second inequality drops the factor $(1-F_1^{+}r)^{n-m}\leq 1$ and uses, for $m\geq n_0$ and $n=m+u$,
		\begin{equation*}
			\binom{m+u}{m}g(m+u)=\frac{g(m)}{u!}\prod_{i=1}^{u}\frac{(m+i)\,g(m+i)}{g(m+i-1)}\;\leq\;g(m)\,\frac{\overline{C}^{\,u}}{u!}
		\end{equation*}
		by Assumption~\ref{ass:light_tail}, and that $g(m)(F_1^{+}r)^m\sum_{u\geq 0}\overline{C}^{\,u}/u!=g(m)(F_1^{+}r)^m e^{\overline{C}}$.

		\emph{Manipulation.} If memory consists of exactly $m$ signals, all favorable and all recalled, then $D^{\text{manip}}=T_m$ and $L^{\text{manip}}\geq T_m-\bar{I}$ by Part~(i). Each favorable signal is recalled with probability $r(s)\geq r_+$, so
		\begin{equation}\label{eq:tail_manip}
			\Pr(L^{\text{manip}}>x\mid\omega=1)\;\geq\;\sum_m g(m)\,(F_1^{+}r_+)^m\,\Pr(T_m>x+\bar{I}).
		\end{equation}

		\emph{Comparing the benchmark with manipulation.} Fix an integer $j>\bar{I}/\sup_s\ell(s)$, so that $\Pr(T_j>\bar{I})>0$. Since the summands of $T$ are positive and independent, $T_{m+j}$ exceeds $x+\bar{I}$ whenever its first $m$ summands exceed $x$ and its last $j$ summands exceed $\bar{I}$: $\Pr(T_{m+j}>x+\bar{I})\geq\Pr(T_j>\bar{I})\Pr(T_m>x)$. By the lower bound in Assumption~\ref{ass:light_tail},
		\begin{equation}\label{eq:tail_gratio}
			g(m+j)=g(m)\prod_{i=1}^{j}\frac{g(m+i)}{g(m+i-1)}\;\geq\;g(m)\Big(\frac{\underline{C}}{m+j}\Big)^{j}\qquad\text{for }m\geq n_0.
		\end{equation}
		Using \eqref{eq:tail_manip} on the indices $m+j$, then \eqref{eq:tail_gratio} and \eqref{eq:tail_bench}, we have
		\begin{equation}\label{eq:p1_right_ratio}
			\begin{split}
				\Pr(L^{\text{manip}}>x\mid\omega=1)&\;\geq\;\Pr(T_j>\bar{I})\sum_m g(m+j)\,(F_1^{+}r_+)^{m+j}\,\Pr(T_m>x)\\
				&\;\geq\;\Pr(T_j>\bar{I})\,(F_1^{+}r_+)^{j}\min_{m>x/\sup_s\ell(s)}\Big[\Big(\frac{r_+}{r}\Big)^{m}\Big(\frac{\underline{C}}{m+j}\Big)^{j}\Big]\\
				&\qquad\times\sum_m g(m)\,(F_1^{+}r)^m\,\Pr(T_m>x)\\
				&\;\geq\;\Pr(T_j>\bar{I})\,(F_1^{+}r_+)^{j}\,e^{-\overline{C}}\min_{m>x/\sup_s\ell(s)}\Big[\Big(\frac{r_+}{r}\Big)^{m}\Big(\frac{\underline{C}}{m+j}\Big)^{j}\Big]\\
				&\qquad\times\Pr(L^{\text{no}}>x\mid\omega=1).
			\end{split}
		\end{equation}
		The function $m\mapsto(r_+/r)^{m}(m+j)^{-j}$ is eventually increasing and diverges, so the minimum diverges as $x\to\infty$. In particular the state-1 tail ratio $\Pr(L^{\text{manip}}>x\mid\omega=1)/\Pr(L^{\text{no}}>x\mid\omega=1)$ exceeds $1+e^{-x}$ for all large $x$, and \eqref{eq:p1_right_reduce} gives $B^{\text{manip}}(p)<B^{\text{no}}(p)$ for all $p$ close enough to~$1$.

		\emph{Step~2$'$ (bounded support).} Let $\bar{M}\equiv\max\operatorname{supp}g$. Since $I(k)\leq 0$, and $I(\bar{M})=0$ because only the term $n=\bar{M}$ survives in both sums defining $I(\bar{M})$, we have, with and without manipulation,
		\begin{equation*}
			L\;\leq\;D\;\leq\;k\,\sup_s\ell(s),\qquad\text{with }L=D\text{ when }k=\bar{M}.
		\end{equation*}
		Hence for $x\in\big((\bar{M}-1)\sup_s\ell(s),\,\bar{M}\sup_s\ell(s)\big)$ the event $\{L>x\}$ requires $M=\bar{M}$, all $\bar{M}$ signals recalled, and $\sum_{i\leq\bar{M}}\ell(s_i)>x$, which forces $\ell(s_i)>x-(\bar{M}-1)\sup_s\ell(s)>0$ and hence $r(s_i)\geq r_+$ for every $i$. Therefore, for each state $\omega$,
		\begin{equation*}
			\begin{split}
				\Pr\nolimits_\omega(L^{\text{manip}}>x)&=g(\bar{M})\,E_\omega\Big[\prod_{i\leq\bar{M}}r(s_i)\,\mathds{1}\Big[\sum_{i\leq\bar{M}}\ell(s_i)>x\Big]\Big]\\
				&\geq\Big(\frac{r_+}{r}\Big)^{\bar{M}}g(\bar{M})\,r^{\bar{M}}\Pr\nolimits_\omega\Big(\sum_{i\leq\bar{M}}\ell(s_i)>x\Big)\\
				&=\Big(\frac{r_+}{r}\Big)^{\bar{M}}\Pr\nolimits_\omega(L^{\text{no}}>x),
			\end{split}
		\end{equation*}
		so $1-B^{\text{manip}}(p)\geq(r_+/r)^{\bar{M}}[1-B^{\text{no}}(p)]>1-B^{\text{no}}(p)$ for every $p=\sigma(x)$ with $x$ in this range, which is exactly the set of $p$ close enough to the top of the support with $B^{\text{no}}(p)<1$.


		\emph{Lighter left tail.} The proof mirrors the heavier-right-tail argument: Step~1 reduces the comparison to $\omega=0$, and Step~2 bounds the tail with manipulation from above and the tail without it from below, so the roles of the two cases are exchanged throughout.

		\emph{Step~1 (unbounded support): reduce to $\omega=0$.} As before, let $x\equiv\operatorname{logit}(p)$, so that $B(p)=\Pr(L\leq x)$; we now consider $p<\tfrac{1}{2}$, so that $x<0$. The change of measure in \eqref{eq:p1_change_measure}, read in the other direction ($dP_1=e^{L}\,dP_0$), gives, with and without manipulation,
		\begin{equation}\label{eq:p1_change_measure_left}
			\Pr(L\leq x\mid\omega=1)=E\big[e^{L}\mathds{1}[L\leq x]\mid\omega=0\big]\leq e^{x}\Pr(L\leq x\mid\omega=0),
		\end{equation}
		using $e^{L}\leq e^{x}$ on the event $\{L\leq x\}$: falling below a low threshold is exponentially less likely in state~1. Hence, splitting each c.d.f.\ by state, dropping the nonnegative term $\Pr(L^{\text{no}}\leq x\mid\omega=1)$ and applying \eqref{eq:p1_change_measure_left} to the case with manipulation,
		\begin{equation}\label{eq:p1_left_reduce}
			\begin{split}
				&B^{\text{no}}(p)-B^{\text{manip}}(p)\\
				&\quad\geq\;\tfrac{1}{2}\big[\Pr(L^{\text{no}}\leq x\mid\omega=0)-\Pr(L^{\text{manip}}\leq x\mid\omega=0)\big]-\tfrac{1}{2}\Pr(L^{\text{manip}}\leq x\mid\omega=1)\\
				&\quad\geq\;\tfrac{1}{2}\big[\Pr(L^{\text{no}}\leq x\mid\omega=0)-\Pr(L^{\text{manip}}\leq x\mid\omega=0)\big]-\tfrac{1}{2}e^{x}\Pr(L^{\text{manip}}\leq x\mid\omega=0)\\
				&\quad=\;\tfrac{1}{2}\Pr(L^{\text{no}}\leq x\mid\omega=0)\left[1-\big(1+e^{x}\big)\frac{\Pr(L^{\text{manip}}\leq x\mid\omega=0)}{\Pr(L^{\text{no}}\leq x\mid\omega=0)}\right],
			\end{split}
		\end{equation}
		so for small $p$ the comparison reduces to the state-0 tails $\Pr(L\leq x\mid\omega=0)$ with and without manipulation: it suffices that their ratio fall below $1/(1+e^{x})$.

		\emph{Step~2 (unbounded support): the suppression of unfavorable signals compounds.} Given Step~1, we condition on $\omega=0$ throughout this step and show that $\Pr(L^{\text{manip}}\leq x\mid\omega=0)/\Pr(L^{\text{no}}\leq x\mid\omega=0)\to 0$ as $x\to-\infty$ and the result follows. Call a signal \emph{unfavorable} if $\ell(s)<0$, let $F_0^{-}\equiv\Pr_0(\ell(s)<0)$, and let $\hat{T}_m$ denote the sum of $m$ i.i.d.\ draws of $|\ell(s)|$ from $f_0$ conditioned on $\ell(s)<0$. Since $\ell$ is bounded, $\hat{T}_m\leq m|\inf_s\ell(s)|$, so $\Pr(\hat{T}_m\geq y)=0$ unless $m\geq y/|\inf_s\ell(s)|$; in particular, once $x\leq-(n_0|\inf_s\ell(s)|+\bar{I})$, which we assume from now on, only $m\geq n_0$ contributes to the sums below.

		\emph{Manipulation.} With manipulation $L^{\text{manip}}\geq D^{\text{manip}}-\bar{I}$ by Part~(i), and $D^{\text{manip}}$ is at least minus the sum of the recalled unfavorable $|\ell(s)|$, so $\{L^{\text{manip}}\leq x\}$ requires that sum to reach $-x-\bar{I}$. An unfavorable signal is recalled with probability $r(s)\leq r_-$, so the probability that a given set of $m$ signals in memory are unfavorable, recalled, and have $|\ell|$-sum at least $y$ is at most $(F_0^{-}r_-)^m\Pr(\hat{T}_m\geq y)$, while the probability that each remaining signal is not a recalled unfavorable one is at most one. Hence, bounding $\binom{n}{m}g(n)$ as in \eqref{eq:tail_bench},
		\begin{equation}\label{eq:ltail_manip}
			\begin{split}
				\Pr(L^{\text{manip}}\leq x\mid\omega=0)&\;\leq\;\sum_m\Big[\sum_{n\geq m}g(n)\binom{n}{m}(F_0^{-}r_-)^m\Big]\Pr(\hat{T}_m\geq -x-\bar{I})\\
				&\;\leq\;e^{\overline{C}}\sum_m g(m)\,(F_0^{-}r_-)^m\,\Pr(\hat{T}_m\geq -x-\bar{I}).
			\end{split}
		\end{equation}

		\emph{Benchmark.} If memory consists of exactly $m$ signals, all unfavorable and all recalled, then $L^{\text{no}}=D^{\text{no}}=-\hat{T}_m$, so, as in \eqref{eq:tail_manip},
		\begin{equation}\label{eq:ltail_bench}
			\Pr(L^{\text{no}}\leq x\mid\omega=0)\;\geq\;\sum_m g(m)\,(F_0^{-}r)^m\,\Pr(\hat{T}_m\geq -x).
		\end{equation}

		\emph{Comparing the benchmark with manipulation.} Fix an integer $j>\bar{I}/|\inf_s\ell(s)|$, so that $\Pr(\hat{T}_j>\bar{I})>0$. As in the right-tail Step~2, $\Pr(\hat{T}_{m+j}\geq y)\geq\Pr(\hat{T}_j>\bar{I})\Pr(\hat{T}_m\geq y-\bar{I})$, which now bounds the sum with manipulation from above, and \eqref{eq:tail_gratio} reads
		\begin{equation}\label{eq:ltail_gratio}
			g(m)\;\leq\;g(m+j)\Big(\frac{m+j}{\underline{C}}\Big)^{j}\qquad\text{for }m\geq n_0.
		\end{equation}
		Using \eqref{eq:ltail_manip}, then re-indexing by $m\mapsto m-j$ and applying \eqref{eq:ltail_gratio}, and finally \eqref{eq:ltail_bench}, we have
		\begin{equation*}
			\begin{split}
				\Pr(L^{\text{manip}}\leq x\mid\omega=0)&\;\leq\;\frac{e^{\overline{C}}}{\Pr(\hat{T}_j>\bar{I})}\sum_{m}g(m)\,(F_0^{-}r_-)^{m}\,\Pr(\hat{T}_{m+j}\geq -x)\\
				&\;=\;\frac{e^{\overline{C}}}{\Pr(\hat{T}_j>\bar{I})}\sum_{m}g(m-j)\,(F_0^{-}r_-)^{m-j}\,\Pr(\hat{T}_{m}\geq -x)\\
				&\;\leq\;\frac{e^{\overline{C}}}{\Pr(\hat{T}_j>\bar{I})\,(F_0^{-}r_-)^{j}}\max_{m\geq -x/|\inf_s\ell(s)|}\Big[\Big(\frac{r_-}{r}\Big)^{m}\Big(\frac{m}{\underline{C}}\Big)^{j}\Big]\\
				&\qquad\times\sum_m g(m)\,(F_0^{-}r)^m\,\Pr(\hat{T}_m\geq -x)\\
				&\;\leq\;\frac{e^{\overline{C}}}{\Pr(\hat{T}_j>\bar{I})\,(F_0^{-}r_-)^{j}}\max_{m\geq -x/|\inf_s\ell(s)|}\Big[\Big(\frac{r_-}{r}\Big)^{m}\Big(\frac{m}{\underline{C}}\Big)^{j}\Big]\\
				&\qquad\times\Pr(L^{\text{no}}\leq x\mid\omega=0).
			\end{split}
		\end{equation*}
		The function $m\mapsto(r_-/r)^{m}m^{j}$ is eventually decreasing and vanishes, so the maximum vanishes as $x\to-\infty$. In particular the state-0 tail ratio $\Pr(L^{\text{manip}}\leq x\mid\omega=0)/\Pr(L^{\text{no}}\leq x\mid\omega=0)$ falls below $1/(1+e^{x})$ for all sufficiently negative $x$, and \eqref{eq:p1_left_reduce} gives $B^{\text{manip}}(p)<B^{\text{no}}(p)$ for all $p$ close enough to~$0$.

		\emph{Step~2$'$ (bounded support).} Mirror the right-tail Step~2$'$. Since $I(k)\geq-\bar{I}$, $I(k)\leq 0$ and $I(\bar{M})=0$,
		\begin{equation*}
			L^{\text{manip}}\;\geq\;D^{\text{manip}}-\bar{I}\;\geq\;k\,\inf_s\ell(s)-\bar{I}\qquad\text{and}\qquad L^{\text{no}}=D^{\text{no}}\;\geq\;k\,\inf_s\ell(s),
		\end{equation*}
		with $L=D$ in both cases when $k=\bar{M}$.
		Hence for $x\in\big(\bar{M}\inf_s\ell(s),\,(\bar{M}-1)\inf_s\ell(s)-\bar{I}\big)$ the event $\{L\leq x\}$ requires, with and without manipulation, $M=\bar{M}$, all $\bar{M}$ signals recalled, and $\sum_{i\leq\bar{M}}\ell(s_i)\leq x$, which forces $\ell(s_i)\leq x-(\bar{M}-1)\inf_s\ell(s)<0$ and hence $r(s_i)\leq r_-$ for every $i$. Therefore, for each state $\omega$,
		\begin{equation*}
			\Pr\nolimits_\omega(L^{\text{manip}}\leq x)=g(\bar{M})\,E_\omega\Big[\prod_{i\leq\bar{M}}r(s_i)\,\mathds{1}\Big[\sum_{i\leq\bar{M}}\ell(s_i)\leq x\Big]\Big]\;\leq\;\Big(\frac{r_-}{r}\Big)^{\bar{M}}\Pr\nolimits_\omega(L^{\text{no}}\leq x),
		\end{equation*}
		so $B^{\text{manip}}(p)\leq(r_-/r)^{\bar{M}}B^{\text{no}}(p)<B^{\text{no}}(p)$ for every $p=\sigma(x)$ with $x$ in this range, which is exactly the set of $p$ close enough to the bottom of the support with $B^{\text{no}}(p)>0$.

	\end{proof}

	\subsection{Proof of Proposition~\ref{prop:hazard}}
	\begin{proof}
		Let $L$ denote the posterior log-odds and $x\equiv\operatorname{logit}(\bar{p})$, as in the proof of Proposition~\ref{prop:general_g}(ii), so that $1-B(\bar{p})=\Pr(p_1>\bar{p})=\Pr(L>x)$ with and without manipulation, and $x\to\infty$ as $\bar{p}\to 1$. By condition~(ii) in Proposition~\ref{prop:hazard}, $L$ admits a density in each state except for the atom at empty recall, which lies at $L\leq 0$; hence for $\bar{p}>\tfrac{1}{2}$ both c.d.f.s are differentiable and $\Pr(L>x)=\Pr(L\geq x)$. Since $h(\bar{p})=B'(\bar{p})/[1-B(\bar{p})]=-\frac{d}{d\bar{p}}\log[1-B(\bar{p})]$,
		\begin{equation*}
			\frac{d}{d\bar{p}}\log\frac{1-B^{\text{manip}}(\bar{p})}{1-B^{\text{no}}(\bar{p})}=\frac{d}{d\bar{p}}\log\big[1-B^{\text{manip}}(\bar{p})\big]-\frac{d}{d\bar{p}}\log\big[1-B^{\text{no}}(\bar{p})\big]=h^{\text{no}}(\bar{p})-h^{\text{manip}}(\bar{p}),
		\end{equation*}
		and, since $dx/d\bar{p}=1/[\bar{p}(1-\bar{p})]>0$,
		\begin{equation*}
			\frac{d}{d\bar{p}}\log\frac{1-B^{\text{manip}}(\bar{p})}{1-B^{\text{no}}(\bar{p})}=\frac{1}{\bar{p}(1-\bar{p})}\,\frac{d}{dx}\log\frac{\Pr(L^{\text{manip}}\geq x)}{\Pr(L^{\text{no}}\geq x)}\Big|_{x=\operatorname{logit}(\bar{p})}.
		\end{equation*}
		Hence $h^{\text{manip}}(\bar{p})<h^{\text{no}}(\bar{p})$ for all $\bar{p}\in[p^*,1)$ if and only if $\frac{d}{dx}\log[\Pr(L^{\text{manip}}\geq x)/\Pr(L^{\text{no}}\geq x)]>0$ for all $x\geq\operatorname{logit}(p^*)$, which is also the statement that the survival-function ratio is strictly increasing on $[p^*,1)$. Moreover, the survival-function ratio tends to infinity as $\bar{p}\uparrow 1$ if and only if $\Pr(L^{\text{manip}}\geq x)/\Pr(L^{\text{no}}\geq x)\to\infty$ as $x\to\infty$. It therefore suffices to show that $\frac{d}{dx}\log[\Pr(L^{\text{manip}}\geq x)/\Pr(L^{\text{no}}\geq x)]>0$ for all sufficiently large $x$ and that $\Pr(L^{\text{manip}}\geq x)/\Pr(L^{\text{no}}\geq x)\to\infty$. In the following, we first establish both properties for the state-1 ratio $\Pr_1(L^{\text{manip}}\geq x)/\Pr_1(L^{\text{no}}\geq x)$, where $\Pr_1$ and $E_1$ condition on $\omega=1$ (Steps~1--5), and then show that both properties hold for the unconditional ratio $\Pr(L^{\text{manip}}\geq x)/\Pr(L^{\text{no}}\geq x)$ (Step~6).

		\emph{Step~1 (Express $\Pr\nolimits_1(L^{\text{manip}}\geq x)$ through the direct effect).} In Step~1, we show that under Poisson memory the indirect correction is a constant, so that $\Pr\nolimits_1(L^{\text{manip}}\geq x)$ is a tail probability of the direct effect at a shifted threshold.

		Decompose the log-odds as $L=D+I(k)$ as in the proof of Proposition~\ref{prop:general_g}(i): $D=\sum_{s\in\mathscr{M}_r}\ell(s)$ is the direct effect and $I(k)$, $k=|\mathscr{M}_r|$, the indirect correction, which vanishes without manipulation, so $L^{\text{no}}=D^{\text{no}}$. Under Poisson memory, $g(n)=e^{-\lambda}\lambda^n/n!$, so for every $y\geq 0$, substituting $n=k+j$,
		\begin{equation*}
			\sum_{n\geq k}\binom{n}{k}y^{\,n-k}g(n)=e^{-\lambda}\sum_{n\geq k}\frac{n!}{k!\,(n-k)!}\,y^{\,n-k}\,\frac{\lambda^n}{n!}=e^{-\lambda}\,\frac{\lambda^k}{k!}\sum_{j\geq 0}\frac{(\lambda y)^j}{j!}=\frac{\lambda^k}{k!}\,e^{\lambda(y-1)}.
		\end{equation*}
		Applying this with $y=\overline{R}_1$ in the numerator and $y=\overline{R}_0$ in the denominator of $I(k)$, the factors $\lambda^k/k!$ cancel, and
		\begin{equation*}
			I(k)=\log\frac{e^{\lambda(\overline{R}_1-1)}}{e^{\lambda(\overline{R}_0-1)}}=-\lambda(\overline{R}_0-\overline{R}_1)\qquad\text{for every }k,
		\end{equation*}
		which is negative because $\overline{R}_0>\overline{R}_1$ by \eqref{eq:p1_Rbar}. Hence $L^{\text{manip}}=D^{\text{manip}}-\lambda(\overline{R}_0-\overline{R}_1)$ and
		\begin{equation}\label{eq:p2_shift}
			\Pr\nolimits_1\big(L^{\text{manip}}\geq x\big)=\Pr\nolimits_1\big(D^{\text{manip}}\geq x+\lambda(\overline{R}_0-\overline{R}_1)\big).
		\end{equation}

		\emph{Step~2 (Properties of the per-signal moment generating function).} In Step~2, we show that the per-signal moment generating function $\psi_1$ is strictly convex, finite for any real $t$, that $\psi_1(t)\geq 1$ for $t\geq 0$, and that $\psi_1'(t)\to\infty$ as $t\to\infty$. As in the proof of Proposition~\ref{prop:general_g}(i), in state~1 the direct effect $D=\sum_{i=1}^{M}X_i$ is a sum of $M$ i.i.d.\ terms $X_i=\ell(s_i)\,\mathds{1}[\text{signal }i\text{ recalled}]$ with per-signal moment generating function 
		\begin{equation*}
			\psi_1(t)\;=\;E_1\big[r(s)e^{t\ell(s)}+1-r(s)\big]\;=\;E_1\big[r(s)e^{t\ell(s)}\big]+\overline{R}_1.
		\end{equation*}
		Since $\ell$ is bounded, $\psi_1(t)\leq E_1\big[r(s)e^{|t|\sup_s|\ell(s)|}+1-r(s)\big]\leq e^{|t|\sup_s|\ell(s)|}<\infty$ for every real $t$. Differentiating under the expectation (for each $t$ the integrand is bounded), $\psi_1'(t)=E_1\big[r(s)\ell(s)e^{t\ell(s)}\big]$ and
		\begin{equation*}
			\begin{split}
				\psi_1''(t)\;=\;E_1\big[r(s)\ell(s)^{2}e^{t\ell(s)}\big]&\;\geq\;E_1\big[r(s)\ell(s)^{2}e^{t\ell(s)}\mathds{1}[\ell(s)>0]\big]\\
				&\;\geq\;r\,E_1\big[\ell(s)^{2}e^{t\ell(s)}\mathds{1}[\ell(s)>0]\big]\;>\;0,
			\end{split}
		\end{equation*}
		where the first inequality drops nonnegative terms, the second holds because signals with $\ell(s)>0$ are recalled with probability at least $r$ with and without manipulation, and the last holds because $r>0$ and $\Pr_1(\ell>0)>0$. Hence $\psi_1$ is strictly convex. Moreover, $\psi_1(0)=1$ and $\psi_1'(0)=\bar{\ell}_1>0$, where $\bar{\ell}_1$ is the per-signal drift from the proof of Proposition~\ref{prop:general_g}(i), so convexity gives $\psi_1(t)\geq\psi_1(0)+\psi_1'(0)\,t\geq 1$ for $t\geq 0$. Finally, fix $\eta\in(0,\sup_s\ell(s))$. For $t\geq 0$, splitting the signals according to $\ell(s)$,
		\begin{equation}\label{eq:p2_psi_lower}
			\begin{split}
				\psi_1'(t)\;&=\;E_1\big[r(s)\ell(s)e^{t\ell(s)}\mathds{1}[\ell(s)\geq \sup\nolimits_s\ell(s)-\eta]\big]\\
				&\qquad+E_1\big[r(s)\ell(s)e^{t\ell(s)}\mathds{1}[0\leq\ell(s)<\sup\nolimits_s\ell(s)-\eta]\big]\\
				&\qquad+E_1\big[r(s)\ell(s)e^{t\ell(s)}\mathds{1}[\ell(s)<0]\big]\\
				&\geq\;E_1\big[r(s)\ell(s)e^{t\ell(s)}\mathds{1}[\ell(s)\geq \sup\nolimits_s\ell(s)-\eta]\big]+E_1\big[r(s)\ell(s)\mathds{1}[\ell(s)<0]\big]\\
				&\geq\;r\,\big(\sup\nolimits_s\ell(s)-\eta\big)\,e^{t(\sup_s\ell(s)-\eta)}\,\Pr\nolimits_1\big(\ell\geq\sup\nolimits_s\ell(s)-\eta\big)-\sup\nolimits_s|\ell(s)|\\
				&\to\;\infty\qquad\text{as }t\to\infty.
			\end{split}
		\end{equation}
		The first inequality drops the middle term, which is nonnegative, and uses $e^{t\ell(s)}\leq 1$ and $r(s)\ell(s)\leq 0$ when $\ell(s)<0$. The second inequality uses, on $\{\ell(s)\geq\sup_s\ell(s)-\eta\}$, where $\ell(s)>0$, that $r(s)\geq r$ with and without manipulation and $\ell(s)e^{t\ell(s)}\geq(\sup_s\ell(s)-\eta)e^{t(\sup_s\ell(s)-\eta)}$; and, on $\{\ell(s)<0\}$, that $r(s)\ell(s)\geq-\sup_s|\ell(s)|$. The limit holds because $\sup_s\ell(s)-\eta>0$ and, by the definition of the supremum, $\Pr_1(\ell\geq\sup_s\ell(s)-\eta)>0$.

		\emph{Step~3 (Characterize $\Pr\nolimits_1(D\geq x)$ as $x\to\infty$).} In Step~3, we derive an asymptotic expression for the hazard rate of the direct effect in state~$1$, with and without manipulation. Conditional on $M=n$, $D$ is a sum of $n$ of the i.i.d.\ terms of Step~2, so $E_1[e^{tD}\mid M=n]=\psi_1(t)^n$; averaging over $M$ with $g(n)=e^{-\lambda}\lambda^n/n!$, the moment generating function of $D$ is
		\begin{equation*}
			E_1\big[e^{tD}\big]\;=\;\sum_{n\geq 0}e^{-\lambda}\frac{\lambda^n}{n!}\,\psi_1(t)^n\;=\;e^{\lambda(\psi_1(t)-1)},
		\end{equation*}
		which is finite for every real $t$ because $\psi_1(t)$ is, by Step~2. For every $t\geq 0$, Markov's inequality applied to $e^{tD}$ gives the Chernoff bound $\Pr_1(D\geq x)\leq e^{-tx}E_1[e^{tD}]$, and optimizing over $t$,
		\begin{equation*}
			\Pr\nolimits_1(D\geq x)\;\leq\;e^{-\mathcal{I}(x)},\qquad\mathcal{I}(x)\equiv\sup_{t\geq 0}\,\big[tx-\lambda(\psi_1(t)-1)\big],
		\end{equation*}
		where $\mathcal{I}$ is the rate function of $D$, the analogue for the random sum of the rates $\rho_\omega$ in \eqref{eq:p1_rate0} and \eqref{eq:p1_rate1}. By Step~2, $\lambda(\psi_1(t)-1)$ is strictly convex, its derivative $\lambda\psi_1'(t)$ is continuous with $\lambda\psi_1'(0)=\lambda\bar{\ell}_1=E_1[D]>0$, and $\lambda\psi_1'(t)\to\infty$ as $t\to\infty$. Consequently, for every $x>E_1[D]$ the function $t\mapsto tx-\lambda(\psi_1(t)-1)$ is strictly concave with derivative $x-\lambda\psi_1'(t)$, which is positive at $t=0$ and tends to $-\infty$. Thus, the supremum is attained at the unique $t^*(x)>0$ at which
		\begin{equation}\label{eq:p2_saddle}
			\lambda\psi_1'(t^*(x))\;=\;x.
		\end{equation}
		Because $\lambda\psi_1'$ is continuous, strictly increasing and finite at every $t$, $t^*(x)$ is continuous and strictly increasing in $x$, and $t^*(x)\to\infty$ as $x\to\infty$. Moreover, $t^*(x)$ grows at most logarithmically in $x$. To see this, for $\eta\in(0,\sup_s\ell(s))$ as in Step~2, write $c_\eta\equiv\tfrac{1}{2}\lambda r\big(\sup_s\ell(s)-\eta\big)\Pr_1\big(\ell(s)\geq\sup_s\ell(s)-\eta\big)>0$. The lower bound \eqref{eq:p2_psi_lower} on $\psi_1'(t)$, multiplied by $\lambda$, reads $\lambda\psi_1'(t)\geq 2c_\eta e^{t(\sup_s\ell(s)-\eta)}-\lambda\sup_s|\ell(s)|$, and the subtracted constant $\lambda\sup_s|\ell(s)|$ is at most $c_\eta e^{t(\sup_s\ell(s)-\eta)}$ once $t$ is large, so
		\begin{equation*}
			\lambda\psi_1'(t)\;\geq\;c_\eta\,e^{t(\sup_s\ell(s)-\eta)}\qquad\text{for all large }t.
		\end{equation*}
		Evaluating at $t=t^*(x)$, using \eqref{eq:p2_saddle}, and taking logarithms,
		\begin{equation}\label{eq:p2_tstar_log}
			t^*(x)\;\leq\;\frac{\log x-\log c_\eta}{\sup_s\ell(s)-\eta}\qquad\text{for all large }x,
		\end{equation}
	i.e., $t^*(x)$ grows at most logarithmically in $x$. 
	
		Write $\operatorname{Var}_t(D)\equiv\lambda\psi_1''(t)$, the variance of $D$ under the measure with density $e^{tD}/E_1[e^{tD}]$ relative to $\Pr_1$. Under this measure $D$ is still a compound Poisson sum, now with mean count $\lambda\psi_1(t)\to\infty$ as $t\to\infty$ and summands that have a density because $\ell(s)$ does, so the saddlepoint approximations for the tail and the density of $D$ \citep{daniels1954saddlepoint,lugannani1980saddle,jensen1995saddlepoint} give, as $x\to\infty$,
		\begin{equation}\label{eq:p2_expansions}
			\begin{split}
				\Pr\nolimits_1(D\geq x)\;&=\;\frac{e^{-\mathcal{I}(x)}}{t^*(x)\sqrt{2\pi\operatorname{Var}_{t^*(x)}(D)}}\Big[1+O\Big(\frac{1}{\operatorname{Var}_{t^*(x)}(D)}\Big)\Big],\\
				-\frac{d}{dx}\Pr\nolimits_1(D\geq x)\;&=\;\frac{e^{-\mathcal{I}(x)}}{\sqrt{2\pi\operatorname{Var}_{t^*(x)}(D)}}\Big[1+O\Big(\frac{1}{\operatorname{Var}_{t^*(x)}(D)}\Big)\Big].
			\end{split}
		\end{equation}
		Dividing the second line of \eqref{eq:p2_expansions} by the first, and noting that the ratio of two factors of the form $1+O(1/\operatorname{Var})$ is again of that form, the hazard rate of $D$ in state~$1$ is
		\begin{equation}\label{eq:p2_hazard}
			-\frac{d}{dx}\log\Pr\nolimits_1(D\geq x)\;=\;t^*(x)\Big[1+O\Big(\frac{1}{\operatorname{Var}_{t^*(x)}(D)}\Big)\Big]\;=\;t^*(x)+o(1),
		\end{equation}
		where the last equality holds because $t^*(x)$ grows at most logarithmically in $x$, by \eqref{eq:p2_tstar_log}, and $\operatorname{Var}_{t^*(x)}(D)$ at least linearly. To see why $\operatorname{Var}_{t^*(x)}(D)$ grows at least linearly in $x$, note, for any $\ell_0\in(0,\sup_s\ell(s))$ and $t\geq 0$,
		\begin{equation*}
			\begin{split}
				\psi_1''(t)\;&=\;\int r(s)\ell(s)^2e^{t\ell(s)}f_1(s)\,ds\;\geq\;\int_{\ell(s)\geq\ell_0}r(s)\ell(s)^2e^{t\ell(s)}f_1(s)\,ds\\
				&\geq\;\ell_0\int_{\ell(s)\geq\ell_0}r(s)\ell(s)e^{t\ell(s)}f_1(s)\,ds\\
				&=\;\ell_0\Big[\psi_1'(t)-\int_{0\leq\ell(s)<\ell_0}r(s)\ell(s)e^{t\ell(s)}f_1(s)\,ds-\int_{\ell(s)<0}r(s)\ell(s)e^{t\ell(s)}f_1(s)\,ds\Big]\\
				&\geq\;\ell_0\big[\psi_1'(t)-\ell_0e^{t\ell_0}\big].
			\end{split}
		\end{equation*}
		The first inequality drops the nonnegative integrand on $\{\ell(s)<\ell_0\}$, and the second uses $\ell(s)\geq\ell_0>0$ on the remaining region. The equality splits $\psi_1'(t)=\int r(s)\ell(s)e^{t\ell(s)}f_1(s)\,ds$ into the contributions of $\{\ell(s)\geq\ell_0\}$, $\{0\leq\ell(s)<\ell_0\}$ and $\{\ell(s)<0\}$. The last inequality uses that the integrand on $\{\ell(s)<0\}$ is negative, and that on $\{0\leq\ell(s)<\ell_0\}$, $r(s)\leq 1$ and $\ell(s)e^{t\ell(s)}\leq\ell_0e^{t\ell_0}$ for $t\geq 0$, so that integral is at most $\ell_0e^{t\ell_0}$. At $t=t^*(x)$, multiplying by $\lambda$ and using \eqref{eq:p2_saddle} gives $\operatorname{Var}_{t^*(x)}(D)\geq\ell_0\big[x-\lambda\ell_0e^{t^*(x)\ell_0}\big]$, and $e^{t^*(x)\ell_0}=o(x)$ by \eqref{eq:p2_tstar_log}, since $e^{t^*(x)\ell_0}\leq(x/c_\eta)^{\ell_0/(\sup_s\ell(s)-\eta)}$ and $\eta$ can be taken below $\sup_s\ell(s)-\ell_0$. Thus, $\operatorname{Var}_{t^*(x)}(D)$ grows at least linearly in $x$. 
		
		Finally, differentiating \eqref{eq:p2_saddle} gives $dt^*(y)/dy=1/\operatorname{Var}_{t^*(y)}(D)$, so for any fixed shift $a>0$,
		\begin{equation}\label{eq:p2_tstar_shift}
			t^*(x+a)-t^*(x)\;=\;\int_x^{x+a}\frac{dy}{\operatorname{Var}_{t^*(y)}(D)}\;\to\;0\qquad\text{as }x\to\infty.
		\end{equation}
		In summary, this step established that the saddlepoint $t^*(x)$ is the increasing solution of \eqref{eq:p2_saddle}, that it grows at most logarithmically in $x$ by \eqref{eq:p2_tstar_log}, that the hazard rate of $D$ in state~$1$ equals $t^*(x)+o(1)$ by \eqref{eq:p2_hazard}, and that a fixed shift of $x$ moves $t^*(x)$ by $o(1)$ by \eqref{eq:p2_tstar_shift}; Steps~5 and~6 use these four facts.

		\emph{Step~4 (Comparison of the derivatives of the moment generating functions).} In Step~4, we show that, for large $t$, the derivative of the per-signal moment generating function with manipulation exceeds the one without by a factor of at least $r_+/r$, up to a vanishing term. Fix a cutoff $\ell_0\in(0,\sup_s\ell(s))$, which exists because $f_1\neq f_0$ implies $\sup_s\ell(s)>0$, and split $\psi_1'(t)=\int r(s)\ell(s)e^{t\ell(s)}f_1(s)\,ds$ over the regions $\{\ell\geq\ell_0\}$, $\{0\leq\ell<\ell_0\}$ and $\{\ell<0\}$. With manipulation, for $t>0$,
		\begin{equation}\label{eq:p2_deriv_manip}
			\begin{split}
				\psi_1'^{\,\text{manip}}(t)\;&=\;\int_{\ell\geq\ell_0}r(s)\ell(s)e^{t\ell(s)}f_1(s)\,ds+\int_{0\leq\ell<\ell_0}r(s)\ell(s)e^{t\ell(s)}f_1(s)\,ds\\
				&\qquad+\int_{\ell<0}r(s)\ell(s)e^{t\ell(s)}f_1(s)\,ds\\
				&\geq\;r_+\int_{\ell\geq\ell_0}\ell(s)e^{t\ell(s)}f_1(s)\,ds+\int_{\ell<0}r(s)\ell(s)e^{t\ell(s)}f_1(s)\,ds\\
				&\geq\;r_+\int_{\ell\geq\ell_0}\ell(s)e^{t\ell(s)}f_1(s)\,ds-\int_{\ell<0}|\ell(s)|e^{t\ell(s)}f_1(s)\,ds\\
				&\geq\;r_+\int_{\ell\geq\ell_0}\ell(s)e^{t\ell(s)}f_1(s)\,ds-(et)^{-1}.
			\end{split}
		\end{equation}
		The equality splits $\psi_1'^{\,\text{manip}}(t)$ over the three regions. The first inequality uses $r(s)\geq r_+$ on $\{\ell\geq\ell_0\}$, where $\ell(s)>0$, and drops the nonnegative integrand on $\{0\leq\ell<\ell_0\}$. The second uses $r(s)\leq 1$ and $\ell(s)=-|\ell(s)|$ on $\{\ell<0\}$. The last uses $|\ell(s)|e^{t\ell(s)}\leq(et)^{-1}$ on $\{\ell<0\}$, the maximum of $ue^{-tu}$ over $u>0$, and $\Pr_1(\ell(s)<0)\leq 1$. Without manipulation, for $t\geq 0$,
		\begin{equation}\label{eq:p2_deriv_no}
			\begin{split}
				\psi_1'^{\,\text{no}}(t)\;&=\;r\int_{\ell\geq\ell_0}\ell(s)e^{t\ell(s)}f_1(s)\,ds+r\int_{0\leq\ell<\ell_0}\ell(s)e^{t\ell(s)}f_1(s)\,ds+r\int_{\ell<0}\ell(s)e^{t\ell(s)}f_1(s)\,ds\\
				&\leq\;r\int_{\ell\geq\ell_0}\ell(s)e^{t\ell(s)}f_1(s)\,ds+r\int_{0\leq\ell<\ell_0}\ell(s)e^{t\ell(s)}f_1(s)\,ds\\
				&\leq\;r\int_{\ell\geq\ell_0}\ell(s)e^{t\ell(s)}f_1(s)\,ds+r\ell_0e^{t\ell_0}\Pr\nolimits_1\big(0\leq\ell(s)<\ell_0\big)\\
				&\leq\;r\int_{\ell\geq\ell_0}\ell(s)e^{t\ell(s)}f_1(s)\,ds+r\ell_0e^{t\ell_0}.
			\end{split}
		\end{equation}
		The equality uses $r(s)=r$ for every $s$ without manipulation and splits over the three regions. The first inequality drops the negative integrand on $\{\ell<0\}$. The second uses $\ell(s)e^{t\ell(s)}\leq\ell_0e^{t\ell_0}$ on $\{0\leq\ell<\ell_0\}$ for $t\geq 0$. The last uses $\Pr_1(0\leq\ell(s)<\ell_0)\leq 1$. Combining \eqref{eq:p2_deriv_manip} and \eqref{eq:p2_deriv_no},
		\begin{equation}\label{eq:p2_deriv_comp}
			\psi_1'^{\,\text{manip}}(t)\;\geq\;\frac{r_+}{r}\,\psi_1'^{\,\text{no}}(t)-r_+\ell_0e^{t\ell_0}-(et)^{-1}\;=\;\frac{r_+}{r}\,\psi_1'^{\,\text{no}}(t)\,\bigl[1-o(1)\bigr]\qquad\text{as }t\to\infty.
		\end{equation}
		The equality holds because both subtracted terms are $o(\psi_1'^{\,\text{no}}(t))$. For the term $(et)^{-1}$, this is because $\psi_1'^{\,\text{no}}(t)\to\infty$ by \eqref{eq:p2_psi_lower}. For the term $r_+\ell_0e^{t\ell_0}$, the same reasoning as for \eqref{eq:p2_deriv_manip}, with $r$ in place of $r_+$, gives $\psi_1'^{\,\text{no}}(t)\geq r\int_{\ell\geq\ell_0}\ell(s)e^{t\ell(s)}f_1(s)\,ds-(et)^{-1}$, and since $\ell_0<\sup_s\ell(s)$, the signals with $\ell(s)\geq(\ell_0+\sup_s\ell(s))/2$ have positive $f_1$-mass, so that
		\begin{equation*}
			\frac{\int_{\ell\geq\ell_0}\ell(s)e^{t\ell(s)}f_1(s)\,ds}{e^{t\ell_0}}\;\geq\;\ell_0\,e^{t(\sup_s\ell(s)-\ell_0)/2}\,\Pr\nolimits_1\big(\ell(s)\geq(\ell_0+\sup\nolimits_s\ell(s))/2\big)\to\infty\quad\text{as }t\to\infty,
		\end{equation*}
		where the inequality restricts the integral to $\{\ell(s)\geq(\ell_0+\sup_s\ell(s))/2\}$, on which $\ell(s)\geq\ell_0$ and $e^{t\ell(s)}\geq e^{t(\ell_0+\sup_s\ell(s))/2}$. Therefore, we conclude that $\psi_1'^{\,\text{manip}}(t)>\psi_1'^{\,\text{no}}(t)$ for all sufficiently large $t$.

		\emph{Step~5 (Comparison of the hazard rates).} In Step~5, we show that the hazard rate of $D^{\text{no}}$ at $x$ exceeds the hazard rate of $D^{\text{manip}}$ at $x+\lambda(\overline{R}_0-\overline{R}_1)$ by a margin bounded away from zero for all large $x$. This gives the two properties of the state-1 ratio $\Pr_1(L^{\text{manip}}\geq x)/\Pr_1(L^{\text{no}}\geq x)$ that we need: it is strictly increasing for all large $x$, and it tends to infinity as $x\to\infty$.

		By \eqref{eq:p2_deriv_comp}, $\psi_1'^{\,\text{manip}}(t)>\psi_1'^{\,\text{no}}(t)$ for all large $t$. The maximizers defined by \eqref{eq:p2_saddle} satisfy
		\begin{equation*}
			\lambda\psi_1'^{\,\text{manip}}\big(t^*_{\text{manip}}(x)\big)\;=\;x\;=\;\lambda\psi_1'^{\,\text{no}}\big(t^*_{\text{no}}(x)\big)\;<\;\lambda\psi_1'^{\,\text{manip}}\big(t^*_{\text{no}}(x)\big)
		\end{equation*}
		for all large $x$, and since $\lambda\psi_1'^{\,\text{manip}}$ is increasing, $t^*_{\text{manip}}(x)<t^*_{\text{no}}(x)$. We now show that the gap $t^*_{\text{no}}(x)-t^*_{\text{manip}}(x)$ is bounded away from zero. By \eqref{eq:p2_deriv_comp} at $t=t^*_{\text{manip}}(x)$ and \eqref{eq:p2_saddle},
		\begin{equation}\label{eq:p2_x_lower}
			x\;=\;\lambda\psi_1'^{\,\text{manip}}\big(t^*_{\text{manip}}(x)\big)\;\geq\;\frac{r_+}{r}\,[1-o(1)]\,\lambda\psi_1'^{\,\text{no}}\big(t^*_{\text{manip}}(x)\big),
		\end{equation}
		while for every $\eta\geq 0$ and $t>0$,
		\begin{equation}\label{eq:p2_eta_shift}
			\begin{split}
				\psi_1'^{\,\text{no}}(t+\eta)\;&=\;\int_{\ell(s)>0}r\,\ell(s)e^{(t+\eta)\ell(s)}f_1(s)\,ds+\int_{\ell(s)<0}r\,\ell(s)e^{(t+\eta)\ell(s)}f_1(s)\,ds\\
				&\leq\;e^{\eta\sup_s\ell(s)}\int_{\ell(s)>0}r\,\ell(s)e^{t\ell(s)}f_1(s)\,ds\\
				&=\;e^{\eta\sup_s\ell(s)}\Big[\psi_1'^{\,\text{no}}(t)+\int_{\ell(s)<0}r\,|\ell(s)|e^{t\ell(s)}f_1(s)\,ds\Big]\\
				&\leq\;e^{\eta\sup_s\ell(s)}\big[\psi_1'^{\,\text{no}}(t)+r(et)^{-1}\big]\;=\;e^{\eta\sup_s\ell(s)}\,\psi_1'^{\,\text{no}}(t)\,[1+o(1)]\qquad\text{as }t\to\infty.
			\end{split}
		\end{equation}
		The first equality splits $\psi_1'^{\,\text{no}}(t+\eta)$ into the signals with $\ell(s)>0$ and $\ell(s)<0$. The first inequality drops the second integral, which is negative, and uses $e^{\eta\ell(s)}\leq e^{\eta\sup_s\ell(s)}$ on $\{\ell(s)>0\}$. The second equality writes the integral over $\{\ell(s)>0\}$ as $\psi_1'^{\,\text{no}}(t)$ minus the integral over $\{\ell(s)<0\}$, where $\ell(s)=-|\ell(s)|$. The second inequality uses $|\ell(s)|e^{t\ell(s)}\leq(et)^{-1}$ on $\{\ell(s)<0\}$, as in \eqref{eq:p2_deriv_manip}, and $\Pr_1(\ell(s)<0)\leq 1$. The last equality holds because $\psi_1'^{\,\text{no}}(t)\to\infty$ by \eqref{eq:p2_psi_lower}, and the $o(1)$ term does not depend on $\eta$. Apply \eqref{eq:p2_eta_shift} with $t=t^*_{\text{manip}}(x)$ and $\eta=t^*_{\text{no}}(x)-t^*_{\text{manip}}(x)\geq 0$, for which $\lambda\psi_1'^{\,\text{no}}(t+\eta)=\lambda\psi_1'^{\,\text{no}}(t^*_{\text{no}}(x))=x$. Together with \eqref{eq:p2_x_lower}, this bounds $x$ from both sides:
		\begin{equation*}
			\frac{r_+}{r}\,[1-o(1)]\,\lambda\psi_1'^{\,\text{no}}\big(t^*_{\text{manip}}(x)\big)\;\leq\;x\;\leq\;e^{[t^*_{\text{no}}(x)-t^*_{\text{manip}}(x)]\sup_s\ell(s)}\,[1+o(1)]\,\lambda\psi_1'^{\,\text{no}}\big(t^*_{\text{manip}}(x)\big),
		\end{equation*}
		where the $o(1)$ terms are as $x\to\infty$ because $t^*_{\text{manip}}(x)\to\infty$. Dividing by $\lambda\psi_1'^{\,\text{no}}(t^*_{\text{manip}}(x))>0$ and taking logarithms, we have
		\begin{equation}\label{eq:p2_gap}
			t^*_{\text{no}}(x)-t^*_{\text{manip}}(x)\;\geq\;\frac{\log(r_+/r)-o(1)}{\sup_s\ell(s)}.
		\end{equation}
		By \eqref{eq:p2_shift} and the hazard expansion \eqref{eq:p2_hazard}, applied to $D^{\text{no}}$ at $x$ and to $D^{\text{manip}}$ at $x+\lambda(\overline{R}_0-\overline{R}_1)$,
		\begin{equation}\label{eq:p2_state1_deriv}
			\begin{split}
				\frac{d}{dx}\log\frac{\Pr_1(L^{\text{manip}}\geq x)}{\Pr_1(L^{\text{no}}\geq x)}\;&=\;-\frac{d}{dx}\log\Pr\nolimits_1\big(D^{\text{no}}\geq x\big)+\frac{d}{dx}\log\Pr\nolimits_1\big(D^{\text{manip}}\geq x+\lambda(\overline{R}_0-\overline{R}_1)\big)\\
				&=\;t^*_{\text{no}}(x)-t^*_{\text{manip}}\big(x+\lambda(\overline{R}_0-\overline{R}_1)\big)+o(1)\\
				&=\;t^*_{\text{no}}(x)-t^*_{\text{manip}}(x)-\Big[t^*_{\text{manip}}\big(x+\lambda(\overline{R}_0-\overline{R}_1)\big)-t^*_{\text{manip}}(x)\Big]+o(1)\\
				&\geq\;\frac{\log(r_+/r)}{\sup_s\ell(s)}-o(1),
			\end{split}
		\end{equation}
		where the last line uses \eqref{eq:p2_gap} and \eqref{eq:p2_tstar_shift} with $a=\lambda(\overline{R}_0-\overline{R}_1)$. Hence the derivative of $\log[\Pr_1(L^{\text{manip}}\geq x)/\Pr_1(L^{\text{no}}\geq x)]$ exceeds a positive constant for all large $x$: the state-1 ratio is strictly increasing for all large $x$, and, integrating, it tends to infinity.

		\emph{Step~6 (From state~1 to the unconditional ratio).} By the change of measure \eqref{eq:p1_change_measure}, $\Pr(L\geq x\mid\omega=0)=E_1\big[e^{-L}\mathds{1}[L\geq x]\big]$. Hence, with and without manipulation,
		\begin{equation}\label{eq:p2_uncond}
			\Pr(L\geq x)=\tfrac{1}{2}\Pr\nolimits_1(L\geq x)\Big[1+E_1\big[e^{-L}\mid L\geq x\big]\Big],\qquad 0<E_1\big[e^{-L}\mid L\geq x\big]\leq e^{-x},
		\end{equation}
		so
		\begin{equation*}
			\frac{1}{1+e^{-x}}\,\frac{\Pr_1(L^{\text{manip}}\geq x)}{\Pr_1(L^{\text{no}}\geq x)}\;\leq\;\frac{\Pr(L^{\text{manip}}\geq x)}{\Pr(L^{\text{no}}\geq x)}\;\leq\;(1+e^{-x})\,\frac{\Pr_1(L^{\text{manip}}\geq x)}{\Pr_1(L^{\text{no}}\geq x)},
		\end{equation*}
		and divergence of the state-1 ratio implies that the unconditional ratio tends to infinity. 
		
		For monotonicity, note that for $x>0$, $L$ admits a density in state~1 by condition~(ii) (its only atom, at empty recall, lies below~$0$). 	Write $h_1(x)\equiv-\frac{d}{dx}\log\Pr_1(L\geq x)$ for its state-1 hazard rate, so that the density of $L$ at $x$ equals $h_1(x)\Pr_1(L\geq x)$, $\frac{d}{dx}\Pr_1(L\geq x)=-h_1(x)\Pr_1(L\geq x)$ and $\frac{d}{dx}E_1\big[e^{-L}\mathds{1}[L\geq x]\big]=-e^{-x}h_1(x)\Pr_1(L\geq x)$. We then have
		\begin{equation*}
			\begin{split}
				\frac{d}{dx}E_1\big[e^{-L}\mid L\geq x\big]&=\frac{d}{dx}\,\frac{E_1\big[e^{-L}\mathds{1}[L\geq x]\big]}{\Pr_1(L\geq x)}\\
				&=\frac{-e^{-x}h_1(x)\Pr_1(L\geq x)^2+E_1\big[e^{-L}\mathds{1}[L\geq x]\big]\,h_1(x)\Pr_1(L\geq x)}{\Pr_1(L\geq x)^2}\\
				&=h_1(x)\Big[E_1\big[e^{-L}\mid L\geq x\big]-e^{-x}\Big].
			\end{split}
		\end{equation*}
		Since $0<E_1[e^{-L}\mid L\geq x]\leq e^{-x}$, it follows that, with and without manipulation,
		\begin{equation}\label{eq:p2_corr_bound}
			\Big|\frac{d}{dx}\log\Big(1+E_1\big[e^{-L}\mid L\geq x\big]\Big)\Big|\;=\;\frac{h_1(x)\,\big|E_1\big[e^{-L}\mid L\geq x\big]-e^{-x}\big|}{1+E_1\big[e^{-L}\mid L\geq x\big]}\;\leq\;h_1(x)\,e^{-x}.
		\end{equation}
		By \eqref{eq:p2_uncond},
		\begin{equation*}
			\frac{\Pr(L^{\text{manip}}\geq x)}{\Pr(L^{\text{no}}\geq x)}\;=\;\frac{\Pr_1(L^{\text{manip}}\geq x)}{\Pr_1(L^{\text{no}}\geq x)}\cdot\frac{1+E_1\big[e^{-L^{\text{manip}}}\mid L^{\text{manip}}\geq x\big]}{1+E_1\big[e^{-L^{\text{no}}}\mid L^{\text{no}}\geq x\big]},
		\end{equation*}
		so taking logarithms, differentiating, and applying \eqref{eq:p2_corr_bound} with and without manipulation,
		\begin{equation*}
			\frac{d}{dx}\log\frac{\Pr(L^{\text{manip}}\geq x)}{\Pr(L^{\text{no}}\geq x)}\;\geq\;\frac{d}{dx}\log\frac{\Pr_1(L^{\text{manip}}\geq x)}{\Pr_1(L^{\text{no}}\geq x)}-\big[h_1^{\text{manip}}(x)+h_1^{\text{no}}(x)\big]\,e^{-x}.
		\end{equation*}
		By the definition of $h_1$, $h_1^{\text{no}}(x)-h_1^{\text{manip}}(x)=\frac{d}{dx}\log[\Pr_1(L^{\text{manip}}\geq x)/\Pr_1(L^{\text{no}}\geq x)]$, which is positive and bounded away from zero for large $x$ by \eqref{eq:p2_state1_deriv}; hence $h_1^{\text{manip}}(x)<h_1^{\text{no}}(x)$ for large $x$. Without manipulation $L^{\text{no}}=D^{\text{no}}$, so $h_1^{\text{no}}$ is the hazard rate of $D^{\text{no}}$ in state~1, which equals $t^*_{\text{no}}(x)+o(1)$ by \eqref{eq:p2_hazard} and grows at most logarithmically in $x$ by \eqref{eq:p2_tstar_log}. The subtracted term is therefore at most $2h_1^{\text{no}}(x)\,e^{-x}\to 0$, while $h_1^{\text{no}}(x)-h_1^{\text{manip}}(x)$ stays bounded away from zero. Hence $\frac{d}{dx}\log[\Pr(L^{\text{manip}}\geq x)/\Pr(L^{\text{no}}\geq x)]>0$ for all large $x$, and $\Pr(L^{\text{manip}}\geq x)/\Pr(L^{\text{no}}\geq x)\to\infty$.
	\end{proof}

	\subsection{Proof of Proposition~\ref{prop:crowding}}
	\begin{proof}
		Write $x=\operatorname{logit}(\bar{p})>0$, let $d_{TV}$ denote total-variation distance, and let $\varepsilon_\gamma\equiv\int\min\{f_0^{\gamma}(s),f_1^{\gamma}(s)\}\,ds$ denote the overlap of the two signal densities.

		\emph{Part~(i).} Let $P_\omega^{\gamma}$ denote the probability distribution of $\mathscr{M}_r$ given $\omega$ and $\bar{P}=\tfrac{1}{2}(P_1^{\gamma}+P_0^{\gamma})$ the unconditional distribution. Since $P_1^{\gamma}\leq 2\bar{P}$, $P_1^{\gamma}$ is absolutely continuous with respect to $\bar{P}$, so by the Radon--Nikodym theorem it has a density $dP_1^{\gamma}/d\bar{P}$ with respect to $\bar{P}$, with values in $[0,2]$. With equal priors, Bayes' rule gives $p_1=\tfrac{1}{2}\,dP_1^{\gamma}/d\bar{P}$, so $p_1-\tfrac{1}{2}=\tfrac{1}{4}\,d(P_1^{\gamma}-P_0^{\gamma})/d\bar{P}$, and integrating the absolute value against $\bar{P}$ yields
		\begin{equation*}
			E\big|p_1-\tfrac{1}{2}\big|\;=\;\tfrac{1}{2}\,d_{TV}\big(P_1^{\gamma},P_0^{\gamma}\big).
		\end{equation*}
		Conditional on $M=m$, the recalled set is a function of $m$ i.i.d.\ (signal, recall-indicator) pairs, and the recall indicator is drawn state-independently given the signal, so the per-pair distributions differ in total variation by exactly $d_{TV}(f_1^{\gamma},f_0^{\gamma})$. Sub-additivity of $d_{TV}$ over the $m$ pairs, and its convexity over the mixing distribution $g$ (common to both states), give $d_{TV}(P_1^{\gamma},P_0^{\gamma})\leq E[M]\,d_{TV}(f_1^{\gamma},f_0^{\gamma})$, where $E[M]<\infty$ by Assumption~\ref{ass:light_tail}. As $\gamma\to 0$, $d_{TV}(f_1^{\gamma},f_0^{\gamma})\to 0$ by continuity, so Markov's inequality gives $\Pr(p_1\geq\bar{p})\leq E|p_1-\tfrac{1}{2}|/(\bar{p}-\tfrac{1}{2})\to 0$ with and without manipulation; hence $\Gamma_\gamma(\bar{p})\to 0$.

		\emph{Part~(ii).} Continuity at the perfectly revealing endpoint forces the overlap to vanish: $\varepsilon_\gamma\leq\int|f_0^{\gamma}-f_0^{1}|+\int|f_1^{\gamma}-f_1^{1}|\to 0$ as $\gamma\to 1$, since $f_0^{1}$ and $f_1^{1}$ have disjoint supports. Two observations follow. First, for any $T\geq 0$, a state-$1$ draw has $\ell_\gamma(s)\leq T$ with probability at most $e^{T}\varepsilon_\gamma$, because $f_1^{\gamma}\leq e^{T}\min\{f_0^{\gamma},f_1^{\gamma}\}$ on that event; symmetrically, a state-$0$ draw has $\ell_\gamma(s)>0$ with probability at most $\varepsilon_\gamma$. Second, taking the first observation with $T=0$, a state-$1$ draw has $\ell_\gamma(s)\leq 0$ with probability at most $\varepsilon_\gamma$. Under manipulation, a draw with $\ell_\gamma(s)>0$ is recalled with probability $\overline{r}$ and any other draw with probability at most $1$, so the probability that a state-$1$ draw is recalled lies in $[(1-\varepsilon_\gamma)\overline{r},\,\overline{r}+\varepsilon_\gamma]$; under the benchmark it equals $r$. Hence, under manipulation, the probability that a state-$1$ draw is not recalled, $\overline{R}_1(\gamma)\equiv\int f_1^{\gamma}(s)\big(1-r(s)\big)\,ds$, satisfies $1-\overline{r}-\varepsilon_\gamma\leq\overline{R}_1(\gamma)\leq 1-(1-\varepsilon_\gamma)\overline{r}$, so $\overline{R}_1(\gamma)\to 1-\overline{r}$ as $\gamma\to 1$, because $\varepsilon_\gamma\to 0$. Symmetrically, a state-$0$ draw has $\ell_\gamma(s)\geq 0$ with probability at most $\varepsilon_\gamma$, because $f_0^{\gamma}=\min\{f_0^{\gamma},f_1^{\gamma}\}$ on that event, and a draw with $\ell_\gamma(s)<0$ is recalled with probability $\underline{r}$; hence $\overline{R}_0(\gamma)\to 1-\underline{r}$. Without manipulation, $\overline{R}_1(\gamma)=\overline{R}_0(\gamma)=1-r$ for every $\gamma$. Given $M=n$, each of the $n$ signals is independently left unrecalled with probability $\overline{R}_1(\gamma)$ in state~$1$, so the probability of recalling nothing in state~$1$ is
		\begin{equation*}
			\sum_{n\geq 0}g(n)\overline{R}_1(\gamma)^{n}\;=\;Q_g\big(\overline{R}_1(\gamma)\big),
		\end{equation*}
		where $Q_g(y)=E[y^M]=\sum_{n\geq 0}g(n)y^n$ is the probability generating function of $M$ from Proposition~\ref{prop:crowding}. The function $Q_g$ is continuous on $[0,1]$, since its series has nonnegative coefficients summing to one and so converges uniformly there. Hence, as $\gamma\to 1$, the probability of recalling nothing in state~$1$ converges to $Q_g(1-\overline{r})$ under manipulation, and it equals $Q_g(1-r)$ without manipulation.

		We first focus on $\omega=1$. Fix $\delta>0$, choose $m_0$ with $\Pr(M>m_0)<\delta$, and consider the event
		\begin{equation}\label{eq:p3_event}
			A_\gamma(T)\;\equiv\;\{M\leq m_0\}\;\cap\;\{\text{every signal in }\mathscr{M}\text{ has }\ell_\gamma(s)>T\},
		\end{equation}
		whose complement has probability at most $\delta+E[M]\,e^{T}\varepsilon_\gamma$ by our first observation above. On $A_\gamma(T)$ with $k\geq 1$ signals recalled, the direct effect exceeds $T$, while the indirect correction $I_\gamma(k)$, the second term of Equation~\eqref{eq:memorybelief_complex1} under the signal structure $\gamma$, is bounded below: for $k\geq n_0$, the bound \eqref{eq:p1_Ibound} gives $I_\gamma(k)\geq -\overline{C}\,\overline{R}_0(\gamma)\geq -\overline{C}$, with $\overline{C}$ the constant of Assumption~\ref{ass:light_tail}, while for the finitely many $k<n_0$, $I_\gamma(k)$ converges as $\gamma\to 1$ to the finite limit $\log\big[\sum_{n\geq k}\binom{n}{k}(1-\overline{r})^{n-k}g(n)\big/\sum_{n\geq k}\binom{n}{k}(1-\underline{r})^{n-k}g(n)\big]$;\footnote{The limit is finite because both sums are positive and finite. Each sum is at most $\sum_{n\geq k}\binom{n}{k}g(n)=E\big[\binom{M}{k}\big]$, which is finite by Assumption~\ref{ass:light_tail}. The denominator is positive whenever $\Pr(M\geq k)>0$, since $1-\underline{r}>0$, and so is the numerator if $\overline{r}<1$. If $\overline{r}=1$, only the term $n=k$ of the numerator survives (with $0^0=1$), so the numerator equals $g(k)$; on $A_\gamma(T)$ every signal is recalled, so $k=M$ and $g(k)=g(M)>0$.} hence there is a constant $C_0\geq 0$ such that $I_\gamma(k)\geq -C_0$ for every realizable $k\geq 1$ and every $\gamma$ close enough to~$1$. Taking $T=x+C_0$, every outcome in $A_\gamma(T)$ with at least one recalled signal has $L>T-C_0=x$ and thus posterior crosses $\bar{p}$. Combining the two inclusions $\{p_1\geq\bar{p}\}\subseteq\{\text{at least one signal recalled}\}$ and $\{\text{at least one signal recalled}\}\cap A_\gamma(T)\subseteq\{p_1\geq\bar{p}\}$,
		\begin{equation*}
			\begin{split}
				0\;&\leq\;\Pr(\text{at least one signal recalled}\mid\omega=1)-\Pr(p_1\geq\bar{p}\mid\omega=1)\\
				&\leq\;\Pr\big(A_\gamma(T)^{c}\mid\omega=1\big)\;\leq\;\delta+E[M]\,e^{T}\varepsilon_\gamma.
			\end{split}
		\end{equation*}
		The second inequality holds because, by the first inclusion, the difference is the probability of recalling a signal without crossing~$\bar{p}$, and by the second inclusion this event lies outside $A_\gamma(T)$. The third is the bound on $\Pr(A_\gamma(T)^{c}\mid\omega=1)$ stated after \eqref{eq:p3_event}. Since $\delta$ is arbitrary, $\Pr(p_1\geq\bar{p}\mid\omega=1)\to 1-Q_g(1-\overline{r})$ under manipulation and $\to 1-Q_g(1-r)$ under the benchmark.

		Now we turn to $\omega=0$. Posterior crossing $\bar{p}$ requires a strictly positive direct effect because the indirect effect is negative with manipulation and is $0$ without manipuation. Hence it requires recalling a draw with $\ell_\gamma>0$, an event of probability at most $E[M]\,\varepsilon_\gamma\to 0$ with and without manipulation. Averaging over the equal prior,
		\begin{equation*}
			\Gamma_\gamma(\bar{p})\;\to\;\tfrac{1}{2}\big[Q_g(1-r)-Q_g(1-\overline{r})\big],
		\end{equation*}
		strictly positive because $\Pr(M\geq 1)>0$ makes $Q_g$ strictly increasing and $1-\overline{r}<1-r$.

		\emph{Part~(iii).} For any $y\in(0,1)$ and any $K$, $Q_g(y)=E[y^M]\leq\Pr(M\leq K)+y^{K}$. If $M\to\infty$ in probability (e.g., Poisson memory with $\lambda\to\infty$), both terms become arbitrarily small (taking $K$ large, then passing along the sequence), so $Q_g(y)\to 0$ for every $y\in(0,1)$ and the limit in~(ii) vanishes.
	\end{proof}

	\subsection{Proof of Corollary~\ref{cor:monopoly_manipulation}}
	\begin{proof}
		Poisson memory satisfies the hypotheses of Proposition~\ref{prop:general_g}: $g(0)=e^{-\lambda}>0$, $g$ is non-degenerate, and Assumption~\ref{ass:light_tail} holds with $\underline{C}=\overline{C}=\lambda$. The manipulation $r^*$ belongs to the manipulation class with $r_+=\overline{r}$ and $r_-=\underline{r}$. Moreover, since every memory size has positive probability, the consumer recalls arbitrarily many favorable signals, or arbitrarily many unfavorable ones, with positive probability, so $B^{\text{no}}(p)\in(0,1)$ for every $p\in(0,1)$. Let $P\in(\max\{\hat{P},c\},v_H)$. Since $\bar{p}$ is strictly increasing with $\bar{p}(\hat{P})=\hat{p}_H$ and $\bar{p}(v_H)=1$, we have $\bar{p}(P)\in(\hat{p}_H,1)$, and Proposition~\ref{prop:general_g}(ii), in the form for $\Pr(p_1\geq\bar{p})$ noted before Corollary~\ref{cor:monopoly_manipulation}, gives $\Pr(p_1^{\text{manip}}\geq\bar{p}(P))>\Pr(p_1^{\text{no}}\geq\bar{p}(P))$. Multiplying by the margin $P-c>0$ preserves the inequality, so $\Pi^{\text{manip}}(P,r^*)>\Pi^{\text{no}}(P)$.
	\end{proof}

	\subsection{Proof of Corollary~\ref{cor:complementarity}}
	\begin{proof}
		Let $p^*$ be the threshold of Proposition~\ref{prop:hazard}, increased if necessary so that $p^*>\tfrac{1}{2}$; the hazard-rate dominance of Proposition~\ref{prop:hazard} continues to hold on the smaller interval $[p^*,1)$. Condition~\eqref{eq:high_cost} is equivalent to $\bar{p}_c\geq p^*$. Since $P\in[c,v_H]$ implies $\bar{p}(P)\in[\bar{p}_c,1]\subseteq[p^*,1]$, the monopolist's entire feasible price range lies in the right-tail region in which Proposition~\ref{prop:hazard} applies. Moreover, since $\ell$ has a density, the posterior has a density on $(\tfrac{1}{2},1)$, with and without manipulation: its only atom, at empty recall, lies at or below $\tfrac{1}{2}$, as in the proof of Proposition~\ref{prop:hazard}. Hence, for $\bar{p}\geq p^*>\tfrac{1}{2}$, demand is $\Pr(p_1\geq\bar{p})=1-B(\bar{p})$, and $B$ is differentiable at $\bar{p}$. The proof proceeds in two steps.

		\emph{Step~1 (Hazard rate comparison on $[p^*,1)$).} The hazard-rate dominance $h^{\text{manip}}(\bar{p})<h^{\text{no}}(\bar{p})$ on $[p^*,1)$ follows from Proposition~\ref{prop:hazard}, whose hypotheses are satisfied here: memory is Poisson, the bang-bang manipulation belongs to the manipulation class with $r_+=\overline{r}$ and $r_-=\underline{r}$, and $\ell$ has a density.

		\emph{Step~2 (Complementarity via the FOC).} Without manipulation, profit is zero at $P=c$ and at $P=v_H$, and it is positive in between, because $B^{\text{no}}(\bar{p})<1$ for $\bar{p}<1$ under Poisson memory. Profit is continuous on $[c,v_H]$ by the previous paragraph, so a maximizer $P^{\text{no}}$ exists and is interior. Since $B^{\text{no}}$ is differentiable at $\bar{p}^{\text{no}}\equiv\bar{p}(P^{\text{no}})$, the first-order condition~\eqref{eq:monopoly_foc} holds there: $(\bar{p}^{\text{no}}-\bar{p}_c)\,h^{\text{no}}(\bar{p}^{\text{no}})=1$. By Condition~\eqref{eq:high_cost}, $\bar{p}^{\text{no}}\geq\bar{p}_c\geq p^*$, so Step~1 applies at the no-manipulation optimum: $h^{\text{manip}}(\bar{p}^{\text{no}})<h^{\text{no}}(\bar{p}^{\text{no}})$. Hence
		\[
			\left.\frac{d\Pi^{\text{manip}}}{dP}\right|_{P=P^{\text{no}}}=[1-B^{\text{manip}}(\bar{p}^{\text{no}})]\cdot\bigl[1-(\bar{p}^{\text{no}}-\bar{p}_c)\,h^{\text{manip}}(\bar{p}^{\text{no}})\bigr]>0.
		\]
		Finally, every maximizer $P^{\text{manip},*}$ of the quasi-concave function $\Pi^{\text{manip}}$ lies strictly to the right of $P^{\text{no}}$. To see that, first, it cannot equal $P^{\text{no}}$, where the derivative is positive. Second, suppose $P^{\text{manip},*}<P^{\text{no}}$, and take $P''>P^{\text{no}}$ close enough that $\Pi^{\text{manip}}(P'')>\Pi^{\text{manip}}(P^{\text{no}})$. Quasi-concavity on $[P^{\text{manip},*},P'']$ gives $\Pi^{\text{manip}}(P^{\text{no}})\geq\min\{\Pi^{\text{manip}}(P^{\text{manip},*}),\Pi^{\text{manip}}(P'')\}=\Pi^{\text{manip}}(P'')$, a contradiction.
	\end{proof}


	\subsection{Proof of Proposition~\ref{prop:equilibrium_threshold}}
	\begin{proof}
		Write $x\equiv\operatorname{logit}(\bar{p})>0$. With equal priors, a sender whose signal has log-likelihood ratio $\ell^P$ assigns probability $\sigma(\ell^P)$ to state~$1$. Let $B^{\text{no}}_\omega$ and $B^{\text{manip}}_\omega$ denote the distribution functions of the posterior $p_1$ conditional on the state, without and with manipulation, so that $B=\tfrac{1}{2}(B_1+B_0)$ in each case. Under Poisson memory the posterior log-odds from recalled signals are $L^{\text{no}}=D$ without manipulation and $L^{\text{manip}}=D^{\text{manip}}-\lambda\Delta$ with manipulation; since $p_1=\sigma(L)$ and neither has an atom above~$0$, for every $y>0$
		\begin{equation*}
			\Pr\nolimits_\omega(L^{\text{no}}\geq y)=1-B^{\text{no}}_\omega(\sigma(y)),\qquad\Pr\nolimits_\omega(L^{\text{manip}}\geq y)=1-B^{\text{manip}}_\omega(\sigma(y)),
		\end{equation*}
		where $\Pr_\omega$ conditions on the state. Suppose $x>\ell_{\max}$. By Equation~\eqref{eq:belief_disclose}, a sender of type $\ell^P$ who discloses obtains
		\begin{equation}\label{eq:p4_UD}
			U^{D}(\ell^P)\;=\;\sigma(\ell^P)\big[1-B^{\text{no}}_1(\sigma(x-\ell^P))\big]+\big(1-\sigma(\ell^P)\big)\big[1-B^{\text{no}}_0(\sigma(x-\ell^P))\big].
		\end{equation}
		By Equation~\eqref{eq:belief_manipulate}, if he manipulates and the receiver attaches the shift $\ell_{\text{inf}}$ to non-disclosure, he obtains
		\begin{equation}\label{eq:p4_UM}
			U^{M}(\ell^P;\ell_{\text{inf}})\;=\;\sigma(\ell^P)\big[1-B^{\text{manip}}_1(\sigma(x-\ell_{\text{inf}}))\big]+\big(1-\sigma(\ell^P)\big)\big[1-B^{\text{manip}}_0(\sigma(x-\ell_{\text{inf}}))\big],
		\end{equation}
		which is increasing in $\ell_{\text{inf}}$. As specified before Proposition~\ref{prop:equilibrium_threshold}, $\ell_{\text{inf}}\in[\ell_{\min},\ell_{\max}]$ in any PBE, on and off the path.

		\emph{Step~1 (The best response to any inference is a cutoff).} Fix $\ell_{\text{inf}}\in[\ell_{\min},\ell_{\max}]$. Multiplying the gain from disclosure, $U^{D}(\ell^P)-U^{M}(\ell^P;\ell_{\text{inf}})$ from \eqref{eq:p4_UD} and \eqref{eq:p4_UM}, by $1+e^{\ell^P}>0$, and using $\sigma(\ell^P)=e^{\ell^P}/(1+e^{\ell^P})$ and $1-\sigma(\ell^P)=1/(1+e^{\ell^P})$,
		\begin{equation}\label{eq:p4_G}
			\begin{split}
				G(\ell^P)\;\equiv\;\big(1+e^{\ell^P}\big)\big[U^{D}(\ell^P)-U^{M}(\ell^P;\ell_{\text{inf}})\big]\;&=\;e^{\ell^P}\big[B^{\text{manip}}_1(\sigma(x-\ell_{\text{inf}}))-B^{\text{no}}_1(\sigma(x-\ell^P))\big]\\
				&\quad+\big[B^{\text{manip}}_0(\sigma(x-\ell_{\text{inf}}))-B^{\text{no}}_0(\sigma(x-\ell^P))\big].
			\end{split}
		\end{equation}
		On $(0,\infty)$, $L^{\text{no}}$ has a continuous density in each state, because $\ell(s)$ has a bounded continuous density. As in the proof of Proposition~\ref{prop:hazard}, write $h_1(y)\equiv-\frac{d}{dy}\log\big[1-B^{\text{no}}_1(\sigma(y))\big]$ for its hazard rate in state~$1$, so that its density in state~$1$ at $y$ is $h_1(y)\big[1-B^{\text{no}}_1(\sigma(y))\big]$. By the change of measure \eqref{eq:p1_change_measure}, $\Pr_0(L^{\text{no}}\in A)=E_1\big[e^{-L^{\text{no}}}\mathds{1}[L^{\text{no}}\in A]\big]$ for every event $A$, so its density in state~$0$ at $y$ is $e^{-y}$ times its density in state~$1$, and, for every $y>0$,
		\begin{equation}\label{eq:p4_cm_bound}
			1-B^{\text{no}}_0(\sigma(y))\;\leq\;e^{-y}\big[1-B^{\text{no}}_1(\sigma(y))\big].
		\end{equation}
		Since $1-B^{\text{no}}_\omega(\sigma(x-\ell^P))=\Pr_\omega(L^{\text{no}}\geq x-\ell^P)$, its derivative with respect to $\ell^P$ is the density of $L^{\text{no}}$ in state~$\omega$ at $x-\ell^P$. Differentiating,
		\begin{equation}\label{eq:p4_Gprime}
			\begin{split}
				G'(\ell^P)\;&=\;e^{\ell^P}\big[B^{\text{manip}}_1(\sigma(x-\ell_{\text{inf}}))-B^{\text{no}}_1(\sigma(x-\ell^P))\big]\\
				&\quad+\big(e^{\ell^P}+e^{-(x-\ell^P)}\big)h_1(x-\ell^P)\big[1-B^{\text{no}}_1(\sigma(x-\ell^P))\big]\\
				&=\;e^{\ell^P}\Big[B^{\text{manip}}_1(\sigma(x-\ell_{\text{inf}}))-B^{\text{no}}_1(\sigma(x-\ell^P))\\
				&\qquad\quad+\big(1+e^{-x}\big)h_1(x-\ell^P)\big[1-B^{\text{no}}_1(\sigma(x-\ell^P))\big]\Big].
			\end{split}
		\end{equation}
		Take a type $\ell_0$ such that $G(\ell_0)=0$, and write $y_0\equiv x-\ell_0\in[x-\ell_{\max},x-\ell_{\min}]$. We now show $G'(\ell_0)>0$. From $G(\ell_0)=0$,
		\begin{equation}\label{eq:p4_indiff}
			\begin{split}
				e^{\ell_0}\big[B^{\text{manip}}_1(\sigma(x-\ell_{\text{inf}}))-B^{\text{no}}_1(\sigma(y_0))\big]\;&=\;B^{\text{no}}_0(\sigma(y_0))-B^{\text{manip}}_0(\sigma(x-\ell_{\text{inf}}))\\
				&\geq\;-\big[1-B^{\text{no}}_0(\sigma(y_0))\big]\;\geq\;-e^{\ell_0-x}\big[1-B^{\text{no}}_1(\sigma(y_0))\big].
			\end{split}
		\end{equation}
		The equality rearranges $G(\ell_0)=0$. The first inequality uses $B^{\text{manip}}_0(\sigma(x-\ell_{\text{inf}}))\leq 1$, and the second is \eqref{eq:p4_cm_bound} at $y=y_0$, with $e^{-y_0}=e^{\ell_0-x}$. Substituting \eqref{eq:p4_indiff} for the first term in \eqref{eq:p4_Gprime} at $\ell^P=\ell_0$,
		\begin{equation*}
			\begin{split}
				G'(\ell_0)\;&\geq\;e^{\ell_0}\Big[\big(1+e^{-x}\big)h_1(y_0)-e^{-x}\Big]\big[1-B^{\text{no}}_1(\sigma(y_0))\big]\\
				&=\;e^{\ell_0}\big(1+e^{-x}\big)\big[1-B^{\text{no}}_1(\sigma(y_0))\big]\big[h_1(y_0)-(1-\bar{p})\big],
			\end{split}
		\end{equation*}
		using $e^{-x}/(1+e^{-x})=1-\bar{p}$. By \eqref{eq:p2_hazard}, applied to $D=L^{\text{no}}$ in state~$1$, $h_1(y)=t^*(y)+o(1)$ tends to infinity as $y\to\infty$; let $\check{y}$ be such that $h_1(y)\geq 1$ for all $y\geq\check{y}$. If $x\geq\max\{\ell_{\max},\check{y}+\ell_{\max}\}$, then $y_0\geq\check{y}$ and $G'(\ell_0)>0$ at every indifferent type $\ell_0$. Hence, for such $x$ and every $\ell_{\text{inf}}$, there is $c(\ell_{\text{inf}})\in[\ell_{\min},\ell_{\max}]$ such that types $\ell^P<c(\ell_{\text{inf}})$ strictly prefer to manipulate and types $\ell^P>c(\ell_{\text{inf}})$ strictly prefer to disclose, with $c(\ell_{\text{inf}})=\ell_{\min}$ if $G>0$ on the support and $c(\ell_{\text{inf}})=\ell_{\max}$ if $G<0$ on the support.

		\emph{Step~2 (The lowest type manipulates).} Since $U^{M}$ is increasing in $\ell_{\text{inf}}$ and $\ell_{\text{inf}}\geq\ell_{\min}$ in any PBE, it suffices to show $U^{D}(\ell_{\min})<U^{M}(\ell_{\min};\ell_{\min})$ for large $x$. Both payoffs evaluate the same threshold $y\equiv x-\ell_{\min}$. Multiplying their difference by $1+e^{\ell_{\min}}$ as in \eqref{eq:p4_G},
		\begin{equation}\label{eq:p4_lowest}
			\begin{split}
				&\big(1+e^{\ell_{\min}}\big)\big[U^{D}(\ell_{\min})-U^{M}(\ell_{\min};\ell_{\min})\big]\\
				&\qquad=\;e^{\ell_{\min}}\big[B^{\text{manip}}_1(\sigma(y))-B^{\text{no}}_1(\sigma(y))\big]+\big[B^{\text{manip}}_0(\sigma(y))-B^{\text{no}}_0(\sigma(y))\big]\\
				&\qquad\leq\;e^{\ell_{\min}}\big[B^{\text{manip}}_1(\sigma(y))-B^{\text{no}}_1(\sigma(y))\big]+\big[1-B^{\text{no}}_0(\sigma(y))\big]\\
				&\qquad\leq\;e^{\ell_{\min}}\big[B^{\text{manip}}_1(\sigma(y))-B^{\text{no}}_1(\sigma(y))\big]+e^{-y}\big[1-B^{\text{no}}_1(\sigma(y))\big]\\
				&\qquad=\;e^{\ell_{\min}}\big[1-B^{\text{no}}_1(\sigma(y))\big]\Big[1+e^{-\ell_{\min}-y}-\frac{1-B^{\text{manip}}_1(\sigma(y))}{1-B^{\text{no}}_1(\sigma(y))}\Big].
			\end{split}
		\end{equation}
		The first equality is \eqref{eq:p4_G} with $\ell^P=\ell_{\text{inf}}=\ell_{\min}$. The first inequality uses $B^{\text{manip}}_0(\sigma(y))\leq 1$, the second is \eqref{eq:p4_cm_bound}, and the last equality writes $B^{\text{manip}}_1(\sigma(y))-B^{\text{no}}_1(\sigma(y))$ as $[1-B^{\text{no}}_1(\sigma(y))]-[1-B^{\text{manip}}_1(\sigma(y))]$ and factors out $e^{\ell_{\min}}[1-B^{\text{no}}_1(\sigma(y))]$. The right-hand side is negative for all large $y$ because the last ratio equals $\Pr_1(L^{\text{manip}}\geq y)/\Pr_1(L^{\text{no}}\geq y)$, which tends to infinity by \eqref{eq:p1_right_ratio}. Let $\bar{p}_0<1$ be large enough that, for every $\bar{p}\geq\bar{p}_0$, the condition on $x$ in Step~1 holds and the right-hand side of \eqref{eq:p4_lowest} is negative at $y=x-\ell_{\min}$. For $\bar{p}\geq\bar{p}_0$, type $\ell_{\min}$ manipulates under every inference, so full disclosure is not a PBE, and every cutoff of Step~1 satisfies $c(\ell_{\text{inf}})>\ell_{\min}$.

		\emph{Step~3 (Every PBE is a cutoff equilibrium, and one exists).} Let $\bar{p}\geq\bar{p}_0$. In a PBE the sender's strategy is a best response to the equilibrium inference, hence by Steps~1 and~2 a cutoff strategy with cutoff $c^*\in(\ell_{\min},\ell_{\max}]$, and the inference is consistent with it: $\ell_{\text{inf}}=\ell_{\text{inf}}(c^*)=\log[F^P_1(c^*)/F^P_0(c^*)]$. This proves Part~(i) except for existence. Conversely, a cutoff $c^*\in(\ell_{\min},\ell_{\max}]$ is a PBE if the cutoff strategy is a best response to $\ell_{\text{inf}}(c^*)$: by Step~1, either $c^*=\ell_{\max}$ and $G(\ell_{\max})\leq 0$ under $\ell_{\text{inf}}=\ell_{\text{inf}}(\ell_{\max})=0$, the inference when every type manipulates and non-disclosure is uninformative, so that every type weakly prefers to manipulate, or $c^*<\ell_{\max}$ and $G(c^*)=0$ under $\ell_{\text{inf}}(c^*)$. Define $\varphi(c)\equiv(1+e^{c})[U^{D}(c)-U^{M}(c;\ell_{\text{inf}}(c))]$ on $(\ell_{\min},\ell_{\max}]$. It is continuous, because $\ell_{\text{inf}}(c)$ is continuous and $L^{\text{no}}$ and $L^{\text{manip}}$ have no atoms above $0$; and $\varphi(c)\to(1+e^{\ell_{\min}})[U^{D}(\ell_{\min})-U^{M}(\ell_{\min};\ell_{\min})]<0$ as $c\downarrow\ell_{\min}$ by \eqref{eq:p4_lowest}, since $\ell_{\text{inf}}(c)\to\ell_{\min}$. If $\varphi(\ell_{\max})\leq 0$, then $c^*=\ell_{\max}$ is a PBE. Otherwise $\varphi(\ell_{\max})>0$, and by the intermediate value theorem there is $c^*\in(\ell_{\min},\ell_{\max})$ with $\varphi(c^*)=0$, which is a PBE.

		\emph{Step~4 (Every type manipulates for $\bar{p}$ close to $1$).} For any type $\ell^P$ and any inference $\ell_{\text{inf}}\geq\ell_{\min}$, we have $(x-\ell_{\text{inf}})-(x-\ell^P)=\ell^P-\ell_{\text{inf}}\leq\ell_{\max}-\ell_{\min}$, so $1-B^{\text{manip}}_1(\sigma(x-\ell_{\text{inf}}))\geq 1-B^{\text{manip}}_1(\sigma(x-\ell^P+\ell_{\max}-\ell_{\min}))$. As in \eqref{eq:p4_lowest},
		\begin{equation}\label{eq:p4_all_manip}
			\begin{split}
				&\big(1+e^{\ell^P}\big)\big[U^{D}(\ell^P)-U^{M}(\ell^P;\ell_{\text{inf}})\big]\\
				&\qquad=\;e^{\ell^P}\big[B^{\text{manip}}_1(\sigma(x-\ell_{\text{inf}}))-B^{\text{no}}_1(\sigma(x-\ell^P))\big]+\big[B^{\text{manip}}_0(\sigma(x-\ell_{\text{inf}}))-B^{\text{no}}_0(\sigma(x-\ell^P))\big]\\
				&\qquad\leq\;e^{\ell^P}\big[B^{\text{manip}}_1(\sigma(x-\ell_{\text{inf}}))-B^{\text{no}}_1(\sigma(x-\ell^P))\big]+\big[1-B^{\text{no}}_0(\sigma(x-\ell^P))\big]\\
				&\qquad\leq\;e^{\ell^P}\big[B^{\text{manip}}_1(\sigma(x-\ell_{\text{inf}}))-B^{\text{no}}_1(\sigma(x-\ell^P))\big]+e^{-(x-\ell^P)}\big[1-B^{\text{no}}_1(\sigma(x-\ell^P))\big]\\
				&\qquad\leq\;e^{\ell^P}\big[B^{\text{manip}}_1(\sigma(x-\ell^P+\ell_{\max}-\ell_{\min}))-B^{\text{no}}_1(\sigma(x-\ell^P))\big]\\
				&\qquad\qquad+e^{-(x-\ell^P)}\big[1-B^{\text{no}}_1(\sigma(x-\ell^P))\big]\\
				&\qquad=\;e^{\ell^P}\big[1-B^{\text{no}}_1(\sigma(x-\ell^P))\big]\Big[1+e^{-x}-\frac{1-B^{\text{manip}}_1(\sigma(x-\ell^P+\ell_{\max}-\ell_{\min}))}{1-B^{\text{no}}_1(\sigma(x-\ell^P))}\Big].
			\end{split}
		\end{equation}
		The first equality is \eqref{eq:p4_G}. The first inequality uses $B^{\text{manip}}_0(\sigma(x-\ell_{\text{inf}}))\leq 1$, the second is \eqref{eq:p4_cm_bound} at $y=x-\ell^P$, and the third is the inequality between the manipulation terms noted above. The last equality writes $e^{-(x-\ell^P)}=e^{\ell^P}e^{-x}$ and factors out $e^{\ell^P}[1-B^{\text{no}}_1(\sigma(x-\ell^P))]$. The last ratio equals $\Pr_1(L^{\text{manip}}\geq y+\ell_{\max}-\ell_{\min})/\Pr_1(L^{\text{no}}\geq y)$ at $y=x-\ell^P$, and \eqref{eq:p1_right_ratio}, with $\bar{I}$ replaced by $\bar{I}+\ell_{\max}-\ell_{\min}$, shows that it tends to infinity as $y\to\infty$. Since $x-\ell^P\geq x-\ell_{\max}$ for every type, there is $\bar{p}_H\in[\bar{p}_0,1)$ such that for $\bar{p}\geq\bar{p}_H$ the right-hand side of \eqref{eq:p4_all_manip} is negative for every type and every inference. Then every type strictly prefers to manipulate in any PBE, the inference is $\ell_{\text{inf}}(\ell_{\max})=0$, and the unique PBE has $c^*=\ell_{\max}$, which is Part~(ii).
	\end{proof}

	\subsection{Proof of Proposition~\ref{prop:listen}}
	\begin{proof}
		Write $x\equiv\operatorname{logit}\bar{p}$. Every inference in the game is bounded: $\ell_{\text{listen}}$ and $\ell_{\text{not}}$, defined in the setup of the extension, are log-likelihood ratios of events in $\ell(s^R)$, on and off the path, so they lie in $[\inf_s\ell(s),\sup_s\ell(s)]$, which is a bounded interval; and the shift $\ell_{\text{inf}}$ she attaches to non-disclosure lies in $[\ell_{\min},\ell_{\max}]$, as specified before Proposition~\ref{prop:equilibrium_threshold}. As in that proof, let $B^{\text{no}}_\omega$ and $B^{\text{manip}}_\omega$ denote the distribution functions of the posterior conditional on the state, so that $\Pr_\omega(L^{\text{no}}\geq y)=1-B^{\text{no}}_\omega(\sigma(y))$ and $\Pr_\omega(L^{\text{manip}}\geq y)=1-B^{\text{manip}}_\omega(\sigma(y))$ for every $y>0$; we assume $x>\sup_s\ell(s)+\ell_{\max}$.

		\emph{Part~(i).} We proceed in four steps.

		\emph{Step~1 (Every sender type manipulates when listened to).} Step~4 of the proof of Proposition~\ref{prop:equilibrium_threshold} gives $\bar{x}$ such that every type strictly prefers to manipulate under every receiver's inference whenever the threshold he faces is at least $\bar{x}$. After the receiver listens, he makes additional inferences on the impression that he had, and he faces the threshold $x-\ell_{\text{listen}}\geq x-\sup_s\ell(s)$, so the property holds if $x\geq\bar{x}+\sup_s\ell(s)$. We therefore fix $\bar{p}$ high enough such that the property holds. The inference from non-disclosure is then $0$ whenever listening has positive probability.

		\emph{Step~2 (Listening raises $\Pr(a=1\mid\omega)$ in both states).} For each decision $E\in\{\text{listen},\text{not}\}$ write $A_\omega(E)\equiv\Pr(a=1\mid\omega,E)$. By Step~1, $A_\omega(\text{listen})=1-B^{\text{manip}}_\omega(\sigma(x-\ell_{\text{listen}}))$ and $A_\omega(\text{not})=1-B^{\text{no}}_\omega(\sigma(x-\ell_{\text{not}}))$. Since $\ell_{\text{listen}}\geq\inf_s\ell(s)$ and $\ell_{\text{not}}\leq\sup_s\ell(s)$,
		\begin{equation}\label{eq:p5_A_bounds}
			A_\omega(\text{listen})\;\geq\;1-B^{\text{manip}}_\omega(\sigma(x-\inf_s\ell(s))),\qquad A_\omega(\text{not})\;\leq\;1-B^{\text{no}}_\omega(\sigma(x-\sup_s\ell(s))),
		\end{equation}
		whatever the inferences. It therefore suffices to show that, in each state, the ratio of the two right-hand sides of \eqref{eq:p5_A_bounds} tends to infinity as $x\to\infty$. In state~$1$, \eqref{eq:p1_right_ratio}, with $\bar{I}$ replaced by $\bar{I}+\sup_s\ell(s)-\inf_s\ell(s)$, gives
		\begin{equation}\label{eq:p5_state1}
			\frac{1-B^{\text{manip}}_1(\sigma(x-\inf_s\ell(s)))}{1-B^{\text{no}}_1(\sigma(x-\sup_s\ell(s)))}\;\to\;\infty\qquad\text{as }x\to\infty.
		\end{equation}
		In state~$0$, the change of measure \eqref{eq:p1_change_measure}, $\Pr_0(L\geq t)=E_1[e^{-L}\mathds{1}[L\geq t]]$ for the log-odds $L$ with or without manipulation, gives two bounds for every threshold $t>0$. Bounding $e^{-L}$ above by $e^{-t}$ on $\{L\geq t\}$ gives $1-B^{\text{no}}_0(\sigma(t))\leq e^{-t}[1-B^{\text{no}}_1(\sigma(t))]$, which is \eqref{eq:p4_cm_bound} at $y=t$. Keeping only the part of the tail within one unit of the threshold, where $e^{-L}$ exceeds $e^{-(t+1)}$, gives
		\begin{equation}\label{eq:p5_window}
			\begin{split}
				1-B^{\text{manip}}_0(\sigma(t))\;&=\;E_1\big[e^{-L^{\text{manip}}}\mathds{1}[L^{\text{manip}}\geq t]\big]\\
				&\geq\;E_1\big[e^{-L^{\text{manip}}}\mathds{1}[t\leq L^{\text{manip}}<t+1]\big]\;\geq\;e^{-(t+1)}\Pr\nolimits_1(t\leq L^{\text{manip}}<t+1).
			\end{split}
		\end{equation}
		By \eqref{eq:p2_hazard}, applied to $D^{\text{manip}}$ in state~$1$, the hazard rate of $L^{\text{manip}}$ in state~$1$ tends to infinity, so it is at least $1$ on $[t,\infty)$ for all large $t$, and then $\Pr_1(t\leq L^{\text{manip}}<t+1)=B^{\text{manip}}_1(\sigma(t+1))-B^{\text{manip}}_1(\sigma(t))\geq(1-e^{-1})[1-B^{\text{manip}}_1(\sigma(t))]$. Hence, for all large $x$,
		\begin{equation*}
			\begin{split}
				&\frac{1-B^{\text{manip}}_0(\sigma(x-\inf_s\ell(s)))}{1-B^{\text{no}}_0(\sigma(x-\sup_s\ell(s)))}\\
				&\qquad\geq\;\frac{e^{-(x-\inf_s\ell(s)+1)}\Pr_1(x-\inf_s\ell(s)\leq L^{\text{manip}}<x-\inf_s\ell(s)+1)}{e^{-(x-\sup_s\ell(s))}\big[1-B^{\text{no}}_1(\sigma(x-\sup_s\ell(s)))\big]}\\
				&\qquad=\;e^{-1-(\sup_s\ell(s)-\inf_s\ell(s))}\,\frac{\Pr_1(x-\inf_s\ell(s)\leq L^{\text{manip}}<x-\inf_s\ell(s)+1)}{1-B^{\text{no}}_1(\sigma(x-\sup_s\ell(s)))}\\
				&\qquad\geq\;e^{-1-(\sup_s\ell(s)-\inf_s\ell(s))}\big(1-e^{-1}\big)\,\frac{1-B^{\text{manip}}_1(\sigma(x-\inf_s\ell(s)))}{1-B^{\text{no}}_1(\sigma(x-\sup_s\ell(s)))}\;\to\;\infty\qquad\text{as }x\to\infty.
			\end{split}
		\end{equation*}
		The first inequality applies \eqref{eq:p5_window} at the threshold $x-\inf_s\ell(s)$ and \eqref{eq:p4_cm_bound} at $x-\sup_s\ell(s)$; the equality collects the exponentials; the second inequality applies the bound on the unit window at $x-\inf_s\ell(s)$; and the limit is \eqref{eq:p5_state1}. Both ratios therefore exceed $1$ once $x$ is large enough, so that $A_\omega(\text{listen})>A_\omega(\text{not})$ in both states, whatever the receiver's inferences based on her listening decisions. Let $\bar{p}_E<1$ be such that, for every $\bar{p}\geq\bar{p}_E$, $x\geq\bar{x}+\sup_s\ell(s)$ and $x$ is large enough that $A_\omega(\text{listen})>A_\omega(\text{not})$ in both states. Then Steps~1 and~2 hold in every PBE.

		\emph{Step~3 (The receiver's listening decision).} At stage~3$'$ the receiver's belief is $q\equiv\Pr(\omega=1\mid s^R)=\sigma(\ell(s^R))$, increasing in $\ell(s^R)$. Her action at stage~5$'$ depends on her listening decision, on the sender's message and on $\mathscr{M}_r$, but not on $s^R$; and given $\omega$, the sender's signal and her memory are independent of $s^R$. Hence the probabilities $A_\omega(E)$ are fixed by the equilibrium strategies and do not vary with $s^R$, and her expected payoff from $E$ is
		\begin{equation}\label{eq:p5_V}
			V_E(q)=q\,(1-\bar{p})A_1(E)-(1-q)\,\bar{p}A_0(E).
		\end{equation}
		The difference $V_{\text{listen}}(q)-V_{\text{not}}(q)$ is affine in $q$ with slope $(1-\bar{p})[A_1(\text{listen})-A_1(\text{not})]+\bar{p}[A_0(\text{listen})-A_0(\text{not})]$, which is positive by Step~2. The difference is negative at $q=0$ and positive at $q=1$, so it has a unique root $q^*\in(0,1)$, and she listens if and only if $q>q^*$. Since $q=\sigma(\ell(s^R))$ with $\ell(s^R)\in[\inf_s\ell(s),\sup_s\ell(s)]$, this is the cutoff rule with $\ell^*=\operatorname{logit}(q^*)$ if $q^*$ lies in $[\sigma(\inf_s\ell(s)),\sigma(\sup_s\ell(s))]$; if $q^*$ lies below that interval she listens after every impression, which is the cutoff rule with $\ell^*=\inf_s\ell(s)$, and if it lies above she never listens, which is the cutoff rule with $\ell^*=\sup_s\ell(s)$.

		\emph{Step~4 (A fixed point).} Suppose the receiver listens if and only if $\ell(s^R)>\ell^*$, where $\ell^*$ lies in the interval $[\inf_s\ell(s),\sup_s\ell(s)]$, and let
		\begin{equation*}
			\ell_{\text{listen}}(\ell^*)\;\equiv\;\log\frac{\Pr_1(\ell(s^R)>\ell^*)}{\Pr_0(\ell(s^R)>\ell^*)},\qquad\ell_{\text{not}}(\ell^*)\;\equiv\;\log\frac{\Pr_1(\ell(s^R)\leq\ell^*)}{\Pr_0(\ell(s^R)\leq\ell^*)}
		\end{equation*}
		be her inferences. Since the density of $\ell(s^R)$ in state~$1$ is $e^{\ell}$ times its density in state~$0$, $\Pr_1(\ell(s^R)\in A)=E_0\big[e^{\ell(s^R)}\mathds{1}[\ell(s^R)\in A]\big]$ for every event $A$, so
		\begin{equation*}
			\begin{split}
				e^{\ell_{\text{listen}}(\ell^*)}\;&=\;\frac{E_0\big[e^{\ell(s^R)}\mathds{1}[\ell(s^R)>\ell^*]\big]}{\Pr_0(\ell(s^R)>\ell^*)}\;=\;E_0\big[e^{\ell(s^R)}\,\big|\,\ell(s^R)>\ell^*\big],\\
				e^{\ell_{\text{not}}(\ell^*)}\;&=\;\frac{E_0\big[e^{\ell(s^R)}\mathds{1}[\ell(s^R)\leq\ell^*]\big]}{\Pr_0(\ell(s^R)\leq\ell^*)}\;=\;E_0\big[e^{\ell(s^R)}\,\big|\,\ell(s^R)\leq\ell^*\big].
			\end{split}
		\end{equation*}
		Thus, both $e^{\ell_{\text{listen}}(\ell^*)}$ and $	e^{\ell_{\text{not}}(\ell^*)}$ are conditional means of $e^{\ell(s^R)}$, and thus lies between the values of $e^{\ell}$ at the ends of the interval and moves continuously with them, because $\ell(s^R)$ has a density. Both inferences are therefore continuous on the open interval, and they extend continuously to its endpoints: $\ell_{\text{listen}}(\inf_s\ell(s))=0$ and $\ell_{\text{not}}(\sup_s\ell(s))=0$, since the conditioning event then has probability one; and $\ell_{\text{not}}(\ell^*)\to\inf_s\ell(s)$ as $\ell^*\downarrow\inf_s\ell(s)$ and $\ell_{\text{listen}}(\ell^*)\to\sup_s\ell(s)$ as $\ell^*\uparrow\sup_s\ell(s)$, since the conditioning interval then shrinks to an endpoint. These limits are the off-path beliefs specified in the setup. By Steps~2 and~3, her best response to the cutoff $\ell^*$ is to listen if and only if $q>q^*(\ell^*)$, where
		\begin{equation*}
			q^*(\ell^*)\;\equiv\;\frac{\bar{p}\,[A_0(\text{listen})-A_0(\text{not})]}{(1-\bar{p})[A_1(\text{listen})-A_1(\text{not})]+\bar{p}\,[A_0(\text{listen})-A_0(\text{not})]}\;\in\;(0,1)
		\end{equation*}
		is the root of $V_{\text{listen}}(q)=V_{\text{not}}(q)$ with $V_E$ as in \eqref{eq:p5_V}.  It is continuous in $\ell^*$, because the inferences are continuous and $L^{\text{no}}$ and $L^{\text{manip}}$ have no atoms above~$0$. Let $\tau(\ell^*)\equiv\operatorname{logit}q^*(\ell^*)$ and let $\tilde{\tau}(\ell^*)$ be $\tau(\ell^*)$ truncated to $[\inf_s\ell(s),\sup_s\ell(s)]$. Then $\tilde{\tau}(\ell^*)-\ell^*$ is continuous, nonnegative at $\inf_s\ell(s)$ and nonpositive at $\sup_s\ell(s)$, so it has a root $\hat{\ell}$. By Step~3, the receiver's best response to the inferences $\ell_{\text{listen}}(\hat{\ell})$ and $\ell_{\text{not}}(\hat{\ell})$ is the cutoff rule whose cutoff is $\operatorname{logit}q^*(\hat{\ell})$ truncated to $[\inf_s\ell(s),\sup_s\ell(s)]$, that is, $\tilde{\tau}(\hat{\ell})=\hat{\ell}$. The cutoff $\hat{\ell}$ with these inferences is therefore a PBE: if $\hat{\ell}$ is interior, both decisions are on the path and the inferences are Bayesian; if $\hat{\ell}$ is an endpoint, the decision that no impression takes is off the path and its inference is the belief specified in the setup.

		\emph{Part~(ii).} Step~1 of Part~(i) is the first claim. For the second, Step~3 of Part~(i) applies in any PBE with the equilibrium inferences, so she listens if and only if $\ell(s^R)>\ell^*$ for some $\ell^*$ in $[\inf_s\ell(s),\sup_s\ell(s)]$.

		\emph{Part~(iii).} By Part~(ii), the receiver listens on $\{\ell(s^R)>\ell^*\}$. The likelihood ratio of the impression, $e^{\ell(s^R)}$, is increasing in $\ell(s^R)$, so the distribution of $\ell(s^R)$ in state~$1$ dominates its distribution in state~$0$ in the likelihood-ratio order and hence in the sense of first-order stochastic dominance: $\Pr_1(\ell(s^R)>\ell^*)\geq\Pr_0(\ell(s^R)>\ell^*)$, strictly whenever both this event and its complement have positive probability. With the prior $\tfrac{1}{2}$, $\Pr(\omega=1\mid\text{listen})=\Pr_1(\text{listen})/[\Pr_1(\text{listen})+\Pr_0(\text{listen})]\geq\tfrac{1}{2}$, strictly when $0<\Pr(\text{listen})<1$. The inequality for not listening follows similarly.

	\end{proof}

\end{document}